\documentclass[aps,prx,twocolumn,showpacs,notitlepage,superscriptaddress,longbibliography]{revtex4-2}
\usepackage{graphicx}
\usepackage[justification=raggedright]{caption}
\usepackage{xcolor}
\usepackage{enumerate}
\usepackage{amsmath, amssymb}
\usepackage{stmaryrd}
\usepackage{multirow}
\usepackage{float}
\usepackage{subcaption}
\usepackage{dsfont}
\usepackage{amssymb}
\usepackage{placeins}
\usepackage[pdftex,colorlinks=true,linkcolor=blue,citecolor=blue]{hyperref}

\usepackage{algorithm2e}
\RestyleAlgo{ruled}

\usepackage{soul}
\setstcolor{red}

\DeclareMathOperator*{\argmax}{arg\,max}
\DeclareMathOperator*{\argmin}{arg\,min}

\usepackage[normalem]{ulem}
\newcommand{\stkout}[1]{\ifmmode\text{\sout{\ensuremath{#1}}}\else\sout{#1}\fi}

\DeclareMathAlphabet\mathbfcal{OMS}{cmsy}{b}{n}
\newcommand\undermat[2]{% http://tex.stackexchange.com/a/102468/5764
  \makebox[0pt][l]{$\smash{\underbrace{\phantom{%
    \begin{matrix}#2\end{matrix}}}_{\text{$#1$}}}$}#2}

\newcommand*{\mybox}[2][5cm]{%
  \makebox[#1][c]{#2}}

\begin{document}
% \title{Minimum energy decoding for multimode Gottesman-Kitaev-Preskill codes}
\title{Closest lattice point decoding for multimode Gottesman-Kitaev-Preskill codes}
\author{Mao Lin}
% \affiliation{AWS Quantum Technologies, Seattle, WA, USA, 98170, USA}%
\affiliation{Amazon Braket, Seattle, WA, USA, 98170, USA}%
\author{Christopher Chamberland}
\affiliation{AWS Center for Quantum Computing, Pasadena, CA 91125, USA}
\affiliation{IQIM, California Institute of Technology, Pasadena, CA 91125, USA}
\author{Kyungjoo Noh}
\affiliation{AWS Center for Quantum Computing, Pasadena, CA 91125, USA}
\affiliation{IQIM, California Institute of Technology, Pasadena, CA 91125, USA}

\date{\today}

\begin{abstract}
    Quantum error correction (QEC) plays an essential role in fault-tolerantly realizing quantum algorithms of practical interest. Among different approaches to QEC, encoding logical quantum information in harmonic oscillator modes has been shown to be promising and hardware efficient. {In this work, we study multimode Gottesman-Kitaev-Preskill (GKP) codes, encoding a qubit in many oscillators, through a lattice perspective. In particular, we implement a closest point decoding strategy for correcting random Gaussian shift errors. For decoding a generic multimode GKP code, we first identify its corresponding lattice followed by finding the closest lattice point in its symplectic dual lattice to a candidate shift error compatible with the error syndrome.}
    We use this method to characterize the error correction capabilities of several known multimode GKP codes, including their code distances and fidelities.
    We also perform numerical optimization of {multimode} GKP codes up to ten modes and find {three} instances (with three{, seven} and nine modes) with better code distances and fidelities compared to the known GKP codes with the same number of modes.
    While exact closest point decoding incurs exponential time cost in the number of modes for general unstructured GKP codes, we give several examples of structured GKP codes {(i.e., of the repetition-rectangular GKP code types)} where the closest point decoding can be performed exactly in linear time.
    For the surface-GKP code, we show that the closest point decoding can be performed exactly in {polynomial time} with the help of a minimum-weight-perfect-matching algorithm (MWPM). {We show that this MWPM closest point decoder improves both the fidelity and the noise threshold of the surface-GKP code to 0.602 compared to the previously studied MWPM decoder assisted by log-likelihood analog information which yields a noise threshold of 0.599}. 
\end{abstract}

\maketitle

\tableofcontents

\section{Introduction}

% Decoder plays an essential role in QEC
Quantum computers hold the promises to solve certain families of problems with significant speedups compared to its classical counterparts \cite{nielsen2002quantum}.
However, due to the ubiquitous noise in the physical systems that are used to build the quantum computers \cite{preskill2018quantum}, quantum error correction (QEC) is essential to protect quantum information from errors due to decoherence and other quantum noise \cite{gottesman2010introduction}. 
The idea behind QEC is to encode a logical qubit onto several physical qubits that are highly entangled \cite{shor1995scheme,steane1996multiple,bennett1996mixed,calderbank1996good,steane1996error}. 
A widely used family of QEC code is the stabilizer code where the logical information is stored in the $+1$ eigenstates of a set of commuting Pauli operators, known as stabilizers \cite{gottesman1996class,gottesman1997stabilizer}. 
The syndrome measurements of the stabilizers provide information on the location and nature of the possible errors. Before attempting to correct the errors, a classical decoding algorithm is typically used to analyze the results of the syndrome measurements to determine the most likely errors.
During the development of fault-tolerant quantum computing, a lot of efforts have been put into creating better ways of encoding the logical qubits or reducing the noise of physical qubits. 
However, given its essential role in QEC, devising classical decoding algorithms that can reduce the effect of noise in a fast time scale is an equally important problem \cite{terhal2015quantum}.

% Among many platforms for QEC, CV, bosonic GKP are promising
Among different platforms for quantum computers, bosonic systems have became increasingly promising because, thanks to the infinite dimensional Hilbert spaces of the bosonic modes, QEC can be implemented in a hardware efficient way \cite{albert2018performance,joshi2021quantum,cai2021bosonic,noh2020quantum}.
For example, two-component cat codes, which can be realized in circuit QED and trapped ion, naturally realize noise-biased qubits where the phase-flip error is more prominent compared to the bit-flip error \cite{cochrane1999macroscopically,jeong2002efficient,mirrahimi2014dynamically,leghtas2015confining,lescanne2020exponential}.
With that, a bias preserved CNOT gate can be realized with two noise-biased cat qubits \cite{guillaud2019repetition,puri2020bias,guillaud2021error,chamberland2022building}, which is however not possible with conventional two level systems \cite{aliferis2008fault}.
Such unique feature can be used to significantly
reduce the required resource overheads for implementing
fault-tolerant quantum computation \cite{aliferis2008fault,aliferis2009fault,webster2015reducing,robertson2017tailored,tuckett2019tailoring,bonilla2021xzzx,tuckett2018ultrahigh,darmawan2021practical}.
The Gottesman-Kitaev-Preskill (GKP) qubit is another example of bosonic qubit that has unique feature unattainable to two level qubits \cite{gottesman2001encoding}. The main novelty of the GKP encoding is that it is designed to protect against small errors on all qubits, which is in contrast to the conventional encoding that corrects errors of arbitrary amplitude for only a subset of qubits \cite{grimsmo2021quantum}. 
Hence the GKP encoding is more resilient to errors in the phase space that shift the values of the canonical variables $\hat{q}$ and $\hat{p}$ of the quantum system.
GKP qubit with a single mode has been realized in various platforms \cite{de2022error, fluhmann2019encoding,campagne2020quantum, sivak2022real}, and shown to suppress errors from photon losses and dissipation processes. 
Unfortunately, GKP code with a single mode cannot correct random shift errors with size large than certain critical value, thus the logical error rate cannot be suppressed to arbitrarily small value.
To improve the QEC properties of the GKP code, or increase the critical value of shift errors, one approach is to consider multimode GKP codes. For instance, one could concatenate a single-mode GKP code with a conventional multi-qubit code, such as the repetition code or the surface code \cite{noh2022low,noh2020fault}, which is generally referred to as {the concatenated GKP} codes \cite{noh2020encoding}. 
For this family of codes, the standard {decoding} techniques of {the multi-qubit stabilizer codes}, such as minimum-weight perfect matching (MWPM) \cite{fowler2013minimum, higgott2023sparse}, can be used for error correction. 
Importantly, the accuracy of the decoder can be significantly enhanced by using the analog information from the homodyne measurements of the GKP qubits \cite{fukui2017analog,vuillot2019quantum,noh2020fault,terhal2020towards,noh2022low,fukui2018high,fukui2018tracking,fukui2019high}. 

The QEC property of multimode GKP codes can be understood in terms of lattices in the phase space. 
In the original proposal \cite{gottesman2001encoding}, it has been shown that the stabilizer group elements of an $N$-mode GKP code are in one-to-one correspondence to the points of a $2N$-dimensional lattice in the phase space. It follows from the commutation relation between the canonical variables that the lattice has to be symplectic integral. 
The logical operators of the GKP codes, which commute with the stabilizers but not in the stabilizer group, correspond to the symplectic dual lattice quotient by the original lattice. 
Although the lattices for single-mode GKP codes have been often used for illustration purpose, there are only a few examples in the literature \cite{Harrington2001_achievable,harrington2004analysis,hanggli2020enhanced,hanggli2021oscillator,conrad2022gottesman,royer2022encoding, conrad2023good} that attempt to use the lattice structure to better understand the properties of such codes, {especially in high (i.e., greater than two) dimensions}, let alone devise a lattice-based decoder.

In this work, we {numerically implement} an exact {closest point decoder} for the multimode GKP codes that are based on the lattice structures in the phase space. 
% \footnote{The Julia code used in this work can be found in the repository \href{https://github.com/maolinml/LatticeAlgorithms.jl}{LatticeAlgorithms.jl}.}.
%
For a given GKP code, we first identify the lattice $\Lambda$ that is isomorphic to its stabilizer group, and its symplectic dual $\Lambda^\perp$ that consists both the stabilizers and logical operators {of the GKP code}.
{In the absence of noise, the outcome of the syndrome measurement corresponds to a lattice point in $\Lambda\subset\Lambda^\perp$, the identity operator in the code space. A random shift errors on the canonical variables of the GKP code will shift the syndrome away from the lattice points in $\Lambda^\perp$.
Since the actual shift error is unknown a priori, the goal of the decoding algorithm is to find a candidate error that has the shortest length and being the most likely shift error compatible with the syndrome measurement. It has been known that this is equivalent to finding the closest lattice point in the symplectic dual lattice $\Lambda^\perp$ of the GKP code for the given syndrome, which is also known as the closest point search problem in the mathematical literature \cite{conway2013sphere}.
}

The general closest point problem, however, is known to be NP-hard \cite{van1981another,micciancio2001hardness} thus solving the problem, or even finding an approximate solution \cite{arora1997hardness,dinur2000improved,micciancio2001shortest}, requires runtime that is exponential in the dimension of the lattice.
But intuitively we expect that the closest point of a lattice can be found much faster if the lattice has certain structure \cite{conway1982fast,conway1986soft,be1989fast,van2016cryptographic}.
One trivial example is the integer lattice $Z_n$: in order to find the closest lattice point to an arbitrary real-valued vector,
we simply round each component to its nearest integer. 
Hence a GKP code based on $Z_n$ lattice can be decoded with runtime that scales linearly with the number of modes.
Building on that, root lattices, such as checkerboard lattices $D_n$ and their Euclidean dual lattices, also admit decoding algorithms with runtime scaling linearly with the dimensionality of the lattices \cite{conway1982fast}. 
More complex lattices can be built by taking the direct sum of lattices, or a union of cosets for certain lattice $\Lambda$. Instead of decoding the lattice as a whole, in these cases, one could decode different components separately followed by assembling the result together. Aided with these strategies, we show that {certain concatenated GKP codes}, which correspond to glue lattices, can be decoded more efficiently by decoding different cosets separately followed by selecting the result with the shortest distance to the input vector. We apply these techniques to {generalizations of the tesseract and $D_{4}$ codes as well as to the surface-GKP codes}, and show that they can be decoded in linear and polynomial time respectively.

The remainder of the paper is organized as follows.
In Sec.~\ref{sec:Preliminary and notations}, we provide the necessary background information on the GKP codes and show that every GKP code can be viewed as a symplectic integral lattice.
In Sec.~\ref{sec:Closest point decoder for the GKP codes}, we formulate the problem of decoding GKP codes as finding the closest lattice point in its sympletic dual lattice. A general algorithm is presented for solving the problem in the unstructured lattices.
In Sec.~\ref{sec:Search for optimized GKP codes}, we utilize the algorithm to analyze the error correction properties, including code distance and fidelity, for several known GKP codes, as well as {generalizations of the tesseract and $D_{4}$ codes}. 
We provide a proof-of-concept demonstration that the lattice perspective of the GKP code allows one to numerically search for optimized GKP codes. The {numerically} found GKP codes, despite being not optimal, exhibit distances and fidelities that are comparable or better than {those of the best known GKP codes}.
However, solving the closest point problem for unstructured lattices incurs exponential time cost in the number of modes. In Sec.~\ref{sec:Efficient closest point decoder for structured GKP codes}, we present several strategies to decode structured GKP lattices.
The strategies play an important role in decoding the $D_n$ lattices, as shown in Sec.~\ref{sec:Linear time decoder for Dn}, which in turn serve as building block for more sophisticated GKP codes.
In Sec.~\ref{sec:generalized_codes_intro}, we show that {the generalizations of the tesseract and $D_{4}$ codes} can be decoded with runtime that scales linearly with respect to the size of the codes, which enables us to benchmark the error correction capabilities of {these code families at scale}.
In Sec.~\ref{sec:Polynomial time closest point decoder for surface-GKP code}, we {present an exact and polynomial-time closest point decoder for the surface-GKP codes based on MWPM}. We show that this decoder improves both the fidelity and the noise threshold of the surface-GKP code, {compared to the previously studied MWPM decoder assisted by log-likelihood analog information. }
We conclude and discuss the future directions in Sec.~\ref{sec:Discussion and conclusion}. 
We provide more technical details in the appendices. 

\section{Summary of main contributions}

{Here we summarize the three key contributions of this work. 
First, we numerically demonstrate a closest point searching algorithm for decoding general multimode GKP codes. Since the initial proposal \cite{gottesman2001encoding}, decoding a GKP code is known to be related to finding the Voronoi cell of its dual lattice, the cell containing all the points that are closer to the origin than to any other lattice site. In a recent publication \cite{conrad2022gottesman}, it is shown that the optimal maximum likelihood decoding strategy for multimode GKP codes can be approximated by the closest point decoder. 
In these prior works, however, there has not been numerical implementations of the exact closest point decoder for general GKP codes. Here we present a self-contained introduction of a closest point searching algorithm for general lattices \cite{agrell2002closest}. The source code and data used in this work are available through the package LatticeAlgorithms.jl \cite{latticealgorithms}. 

Closest point searching is a well known NP-hard problem, and decoding a generic unstructured GKP code generally takes exponential time cost in the number of modes. Hence, there is no a priori known evidence that it is practical to decode large instances of multimode GKP codes via the closest point searching strategy.
The second contribution of this work is to show that certain structured GKP codes can be decoded efficiently with the closest point decoder. We present a set of tools to decode structured GKP codes, with which linear time closest point decoders are constructed for two families of repetition-GKP codes. 
More remarkably, we demonstrate an efficient closest point decoder for the surface-GKP code, which improves both the fidelity and threshold of the surface-GKP code while has \emph{exactly the same time complexity} as the commonly used MWPM decoder.
Our finding suggests that efficient closest point decoding strategy may exist for other commonly studied structured GKP codes.
}

{
The third contribution of our work is we find three instances of GKP codes that outperform the known structured GKP codes in terms of code distances and fidelity. In particular, with numerical optimization of multimode GKP codes up to ten modes, we find GKP codes with three, seven and nine modes that have better performance than the known GKP codes of the same modes, including repetition-GKP code, $[[7,1,3]]$-hexagonal GKP code and surface-GKP code.
For GKP codes with even number of modes, the distances of the optimized codes are smaller than the YY-repetition-GKP codes, the concatenation of two copies of repetition-GKP codes with the YY stabilizer. Despite that, we find that the fidelity of the optimized codes are the same or better than that of the YY-repetition-GKP codes with the same number of modes.
The detailed study of these new GKP codes are interesting future research topics. 
}

\section{Preliminary and notations}
\label{sec:Preliminary and notations}

\subsection{Displacement operators and Gaussian unitaries}

In this work, we will work {with} the quantum Hilbert space of $N$ harmonic oscillator modes. Let $\hat{a}_j$ and $\hat{a}^\dagger_j$ denote the creation and annihilation operators for the $j$-th mode, we have the commutation relation $[\hat{a}_j,\hat{a}_k^\dagger]=\delta_{jk}$ where we have set $\hbar=1$. {Since we use Gaussian operations and related concepts in many places}, it proves convenient to introduce the quadrature operator $\hat{\boldsymbol{x}} \equiv (\hat{x}_{1} , \hat{x}_{2} , \cdots , \hat{x}_{2N} )^{T} \equiv (\hat{q}_{1} , \hat{p}_{1} , \cdots \hat{q}_{N} , \hat{p}_{N} )^{T}$ where
\begin{align}
    \hat{q}_j=(\hat{a}_j+\hat{a}_j^\dagger)/\sqrt{2}, \quad \hat{p}_j=-i(\hat{a}_j-\hat{a}_j^\dagger)/\sqrt{2}.
\end{align}
Then, we have
\begin{align}
\label{eq:quad_commutator}
    [\hat{x}_{j} , \hat{x}_{k} ] = i\Omega_{jk} , 
\end{align}
where the symplectic form $\Omega$ is {a $2N\times 2N$ matrix and is} given by 
\begin{align}
\label{eq:def_Omega}
    \Omega 
    % =  \begin{bmatrix} \omega & 0_{2\times 2}  & \cdots &  0_{2\times 2} \\   0_{2\times 2} & \omega  & \cdots & 0_{2\times 2}  \\  \vdots & \vdots & \ddots  & \vdots \\ 0_{2\times 2} & 0_{2\times 2} & \cdots & \omega \end{bmatrix} 
    = I_N \otimes \omega 
    = I_N \otimes \begin{bmatrix} 0 & 1 \\ -1 & 0\end{bmatrix} = \begin{bmatrix}
    \omega & 0 & \cdots & 0\\
    0 & \omega & \cdots & 0 \\
    \vdots & \vdots & \ddots & \vdots\\
    0 & 0 & \cdots & \omega
    \end{bmatrix} .
\end{align}
Here $I_N$ is the $N\times N$ identity matrix, and we have denoted operators with a hat and ({column}) vectors in bold fonts. We note that $\Omega^{-1} = \Omega^{T} =-\Omega$. 

We remark that the choice of the ordering of the position and momentum operators in $\hat{\boldsymbol{x}}$ is not unique. We refer to the ordering convention chosen above as the \texttt{qpqp} ordering. Occasionally it is convenient to work with {a different ordering convention} where $\hat{\boldsymbol{x}}  \equiv (\hat{q}_{1} , \cdots,\hat{q}_{N} , \hat{p}_{1} , \cdots, \hat{p}_{N} )^{T}$, which is referred as the \texttt{qqpp} convention. In {the latter} case, the symplectic form $\Omega$ reads
\begin{align}
\label{eq:Omega_CSS}
    \Omega^\text{(\texttt{qqpp})}
    = \begin{bmatrix} 0_N & I_N \\ -I_N & 0_N\end{bmatrix}.    
\end{align}
In this paper, we will mostly work with the {\texttt{qpqp} ordering} as in Eq.~\ref{eq:def_Omega} unless {we explicitly state that the \texttt{qqpp} ordering is used instead. }

The quadrature operators can be thought as the generators of the translation in the $2N$ dimensional phase space of the $N$ oscillator modes. Specifically, let $\boldsymbol{u} = (u_{q}^{(1)}, u_{p}^{(1)}, \cdots, u_{q}^{(N)}, u_{p}^{(N)} )^{T}\in \mathbb{R}^{2N}$ be a vector in the phase space. Then, the displacement operator $\hat{D}(\boldsymbol{u})$ is defined as
\begin{align} 
    \hat{D}(\boldsymbol{u}) \equiv \exp[ i\boldsymbol{u}^{T} \Omega^{-1} \hat{\boldsymbol{x}} ] = \exp[ -i \boldsymbol{u}^{T} \Omega \hat{\boldsymbol{x}} ]. 
\end{align}
This is a displacement operator in the sense that 
\begin{align}
\label{eq:displace_x}
    \hat{D}^{\dagger}(\boldsymbol{u}) \hat{x}_{j} \hat{D}(\boldsymbol{u}) &= \hat{x}_{j} + [ -i \boldsymbol{u}^{T} \Omega^{-1} \hat{\boldsymbol{x}} , \hat{x}_{j} ] = \hat{x}_{j} + u_{j},
\end{align}
or equivalently, $\hat{D}^{\dagger}(\boldsymbol{u}) \hat{\boldsymbol{x}} \hat{D}(\boldsymbol{u})=\hat{\boldsymbol{x}}+\boldsymbol{u}$, which shifts the quadrature operators $\hat{\boldsymbol{x}}$ by an amount of $\boldsymbol{u}$. With that, we have the following commutation relation for the displacement operators
\begin{align}
    \hat{D}(\boldsymbol{u}) \hat{D}(\boldsymbol{v}) =  \hat{D}(\boldsymbol{v}) \hat{D}(\boldsymbol{u})\exp[i\boldsymbol{u}^{T}\Omega \boldsymbol{v} ]. 
\end{align}
Hence the two displacements associated with $\boldsymbol{u}$ and $\boldsymbol{v}$ commute if and only if their symplectic product $\boldsymbol{u}^{T}\Omega \boldsymbol{v}$ is an integer {multiple of} $2\pi$. 
% $\boldsymbol{u}^{T}\Omega \boldsymbol{v}\in\mathbb{Z}$. 

Displacement operator is an example of Gaussian unitary operators that preserve the symplectic form $\Omega$: it is clear that {the commutation relation in} Eq.~\ref{eq:quad_commutator} is invariant under the translation in Eq.~\ref{eq:displace_x}. More generally, one could consider a Gaussian unitary $\hat{U}$ that transforms the quadrature operator $\hat{\boldsymbol{x}}$ into $S \hat{\boldsymbol{x}}+\boldsymbol{u}$ \cite{weedbrook2012gaussian},
\begin{align}
    \hat{\boldsymbol{x}}' \equiv \hat{U}^{\dagger} \hat{\boldsymbol{x}} \hat{U} = S\hat{\boldsymbol{x}}+\boldsymbol{u}. 
\end{align}
If the symplectic form $\Omega$ is invariant under transformation, i.e., 
\begin{align}
[ \hat{{x}}'_{j} ,  \hat{{x}}'_{k} ] = i\Omega_{jk},
\end{align}
then from
\begin{align}
    [ \hat{{x}}'_{j} ,  \hat{{x}}'_{k} ] =\sum_{l,m} [  S_{jl}\hat{{x}}_{l}+u_j , S_{km}\hat{{x}}_{m}+u_k  ]  = i (S\Omega S^{T})_{jk},\nonumber
\end{align}
we conclude that $S$ is a $2N\times 2N$ symplectic matrix, i.e.,
\begin{align}
\label{eq:def_symplectic}
    S\Omega S^{T} = \Omega. 
\end{align}
Hence a Gaussian unitary operator is fully characterized by {a symplectic matrix} $S$ and $\boldsymbol{u}$. {In this work, it suffices to consider Gaussian operations with $\boldsymbol{u}=0$. T}hose with $\boldsymbol{u}\neq0$ and $S=I_{2N}$ are referred to as the displacement operator{s} as above.

\subsection{Multimode GKP code}
\label{sec:Multimode GKP code}

A GKP code with $N$ modes is stabilized by $2N$ independent stabilizer {generators}. Each stabilizer generator is given by a displacement in the $2N$-dimensional phase space
\begin{align}
\label{eq:def_S_j}
    \hat{{S}}_{j} = \hat{D}(\sqrt{2\pi}\boldsymbol{v}_{j}) = \exp[i\hat{g}_j], 
\end{align}
where $j\in\left\{1,...,2N\right\}$ and $\boldsymbol{v}_j$ is a translation vector {corresponding to the $j$-th stabilizer generator}.  
We {also introduce a vector of operators} $\hat{\boldsymbol{g}}\equiv(\hat{g}_1, ..., \hat{g}_{2N})^T$ {such that} 
$
    \hat{g}_i = \sqrt{2\pi}\boldsymbol{v}^T_j\Omega^{-1}\hat{\boldsymbol{x}},
$
or more compactly 
\begin{align}
    \hat{\boldsymbol{g}} = \sqrt{2\pi}M\Omega^{-1}\hat{\boldsymbol{x}}, 
\end{align}
{where} $M$ is a $2N\times 2N$ matrix with the $j$-th row given by the {row} vector $\boldsymbol{v}_j^T$. 
The full stabilizer group is given by
\begin{align}
    \mathcal{S} = \lbrace \hat{S} = \hat{{S}}_{1}^{a_1}\cdots \hat{{S}}_{2N}^{a_{2N}} ~|~ \boldsymbol{a} = (a_{1},\cdots, a_{2N})^{T} \in \mathbb{Z}^{2N} \rbrace,
\end{align}
and since the stabilizer generators commute with each other, a generic stabilizer group element reads
\begin{align}
\label{eq:stabilizer_as_lattice_point}
    \hat{S} &= \exp[ i \boldsymbol{a}^T\hat{\boldsymbol{g}} ] =\exp[ i  \sqrt{2\pi} ( \boldsymbol{a}^{T} M )  \Omega^{-1} \hat{\boldsymbol{x}}  ] = \hat{D}(\sqrt{2\pi}M^T\boldsymbol{a}) . 
\end{align}
Eq.~\ref{eq:stabilizer_as_lattice_point} establishes an isomorphism between the stabilizer group and a lattice with the generator matrix $M$
\begin{align}
    \Lambda(M) \equiv \lbrace \boldsymbol{a}^{T} M : a = (a_{1},\cdots, a_{2N})^{T} \in \mathbb{Z}^{2N} \rbrace,
\end{align}
where the stabilizer group element $\hat{S}$ is mapped to the lattice point $\sqrt{2\pi}M^T\boldsymbol{a}$. Since the stabilizers form an Abelian group, we require that $\hat{D}(\sqrt{2\pi}M^T\boldsymbol{a})$ commute with $\hat{D}(\sqrt{2\pi}M^T\boldsymbol{b})$ for arbitrary $\boldsymbol{a}, \boldsymbol{b}\in\mathbb{Z}^{2N}$. Equivalently, it is required that the symplectic Gram matrix
\begin{align}
\label{eq:def_A}
    A\equiv M\Omega M^T
\end{align}
has only integer entries. Lattices with this property is called symplectic integral lattices \cite{conway2013sphere}. 

Hereafter we shall use $n$ for the dimension of a general lattice, and $N$ for the number of modes of a GKP code. {We refer to the matrix $M$ in $\hat{\boldsymbol{g}} = \sqrt{2\pi}M\Omega^{-1}\hat{\boldsymbol{x}}$ as the generator matrix of the GKP code (or sometimes simply the GKP generator matrix). We will also use}
% We shall use ``GKP code'' and ``GKP lattice'' interchangeably 
$M_\Lambda$ to denote the generator matrix of a lattice $\Lambda$. 

\subsection{{Canonical generator matrix of a GKP code}}
\label{sec:The canonical basis for the GKP codes}

Much like the set of generators of a stabilizer group is not unique, one could pick a different basis, and hence {a different but equivalent} generator matrix, for the same lattice. For instance, a lattice $\Lambda(M)$ can {also} be generated by
\begin{align}
\label{eq:unimodular_transformation}
    M' = RM,
\end{align}
where $R$ is a unimodular matrix, which is {an integer matrix} with $|\text{det}(R)|=1$. The corresponding symplectic Gram matrices are related via 
\begin{align}
    A' = RAR^T.
\end{align}
Since $A$ is integer-valued, and {anti-symmetric}, i.e., 
\begin{align}
    A^{T} = M \Omega^{T} M^{T} = -M\Omega M^{T} = -A,
\end{align}
it is possible to find a unimodular matrix $R$ such that
\begin{align}
\label{eq:canonize_A}
    RA R^{T} 
    &= \textrm{diag}(d_{1},\cdots,d_{N}) \otimes \omega ,
\end{align}
where $\textrm{diag}(d_{1},\cdots,d_{N})$ (or $\textrm{diag}(\boldsymbol{d})$ in short) is a diagonal matrix {whose elements are natural numbers, i.e.,} $\boldsymbol{d}\in\mathbb{N}^N$.
Eq.~\ref{eq:canonize_A} means that for any valid {generator matrix $M$ of a GKP code}, it is possible to find a unimodular matrix $R$ such that $M' = RM$ satisfies 
\begin{align}
\label{eq:symplectic_gram_matrix}
    A' \equiv M' \Omega M'^{T}  = \textrm{diag}(\boldsymbol{d}) \otimes \omega . 
\end{align}
Here, $d_{k}$ can be interpreted as the number of states encoded in the $k$-th ``mode''. {We say that a generator matrix $M$ of a GKP code is in the canonical form when its symplectic Gram matrix $A = M\Omega M^{T}$ is given by $\textrm{diag}(\boldsymbol{d}) \otimes \omega$ for some $\boldsymbol{d}\in\mathbb{N}^N$. In App.~\ref{app:canonize_A}, we provide more details on how to find an appropriate unimodular matrix $R$ that converts a valid GKP generator matrix $M$ into a canonical GKP generator matrix $M' = RM$. }

\subsection{The logical operators of a GKP code}

Displacement operators that preserve the GKP code subspace form the normalizer group of the code, which consists of phase space translation that commute with the stabilizer group
\begin{align}
    \mathcal{S}^\perp = \left\{\hat{S}^\perp|[\hat{S}^\perp, \hat{S}]=0 ~ \forall \hat{S}\in\mathcal{S}\right\}.
\end{align}
By definition, for an arbitrary element $\hat{S}^\perp\equiv\hat{D}(\boldsymbol{u})\in\mathcal{S}$, we have $\boldsymbol{u}^T\Omega \boldsymbol{v}\in\mathbb{Z}$ for all $\boldsymbol{v}\in\Lambda$. This gives precisely the symplectic dual lattice for $\Lambda$
\begin{align}
    \Lambda^\perp \equiv \Lambda(M^{\perp}) = \left\{\boldsymbol{u} ~|~ \boldsymbol{u}^T\Omega\boldsymbol{v}\in\mathbb{Z} ~ \forall \boldsymbol{v}\in\Lambda\right\},
\end{align}
which could be generated by 
\begin{align}
\label{eq:def_M_perp}
    M^{\perp} = \Omega A^{-1}M. 
\end{align}
Indeed, as one can check, $\boldsymbol{b}^TM^{\perp}\Omega M^T\boldsymbol{a}=\boldsymbol{b}^T\Omega\boldsymbol{a}$ is an integer for all $\boldsymbol{a}, \boldsymbol{b}\in\mathbb{Z}^{2N}$. {In the literature, there is an alternative definition $M^{\perp} = A^{-1}M$ which differs from our definition in Eq.~\ref{eq:def_M_perp} only by a unimodular matrix $\Omega$ multiplied from the left. Hence the two definitions generate the same lattice.}
We note that since the stabilizer group is Abelian and all elements commute with each other, it implies that $\mathcal{S}\subset\mathcal{S}^\perp$. 
Since the logical operators of a QEC code leave the stabilizer group invariant, analogously, we can associate all translations in the dual lattice to the logical operators defined as
\begin{align}
    \label{eq:logical_operators}
    \hat{L}_{j} \equiv \exp[ i \sqrt{2\pi} \boldsymbol{w}_{j}^{T} \Omega^{-1} \hat{\boldsymbol{x}}  ], 
\end{align}
where $j\in \lbrace 1,\cdots, 2N \rbrace$ and $\boldsymbol{w}_{j}^{T}$ is the $j$-th row of $M^{\perp}$. 
The $2N$ logical operators of the GKP code are, however, not independent to each other, because logical operators differ by a stabilizer are indistinguishable in the code subspace.
This corresponds to the fact that the logical information of the GKP code is encoded in the quotient group $\Lambda^\perp/\Lambda$, and the number of distinct logical operators is equal to the number of elements in the quotient group, or \cite{royer2022encoding,conrad2022gottesman}
\begin{align}
\label{eq:num_logicals}
    |\text{det}(M)/\text{det}(M^\perp)| = |\text{det}(M)|^2.
\end{align}
{We have used Eq.~\ref{eq:def_A} and \ref{eq:def_M_perp} to derive Eq.~\ref{eq:num_logicals}}.
Hence the number of states encoded in the GKP lattice is given by $|\text{det}(M)|^{2}$, the {square of the} determinant of the GKP lattice generator matrix.
The generator of the symplectic dual lattice takes a particularly simple form when $M$ is in the canonical form, i.e., $A = \textrm{diag}(\boldsymbol{d}) \otimes \omega$ because 
\begin{align}
\label{eq:logical_canonical}
    M^{\perp} = \Omega A^{-1}M
    % = ( I_{N\times N} \otimes \omega )  (\textrm{diag}(d)^{-1} \otimes \omega^{-1})  M 
    = (\textrm{diag}(\boldsymbol{d})^{-1} \otimes I_2 )  M.
\end{align}
Hence the logical operators are simply the stabilizers {divided} by the corresponding integers $d_i$ in the canonical basis.
We note that sometimes it may be more convenient to use the identity
\begin{align}
\label{eq:identity_1}
    M^{\perp} = \Omega A^{-1}M = \Omega ( M \Omega M^{T})^{-1} M = \Omega (M^{T})^{-1}\Omega^{-1}. 
\end{align}

\subsection{Code distances of a GKP code}

In order to quantify the error correction {capability of a GKP code}, we will need several metrics for evaluating GKP codes. For standard qubit-based stabilizer codes, one such metric is the distance of the code, defined as the weight of the shortest nontrivial logical operator \cite{nielsen2002quantum}. 
Motivated by that, we can define the distance of a GKP code as the Euclidean length of the shortest non-trivial logical operator \cite{conrad2022gottesman}
\begin{align}
\label{eq:distance}
    % d = \min_{\boldsymbol{0}\neq \boldsymbol{x}\in\Lambda^\perp/\Lambda}||\boldsymbol{x}||.
    d = \min_{\boldsymbol{u}\in\Lambda^\perp,~ \boldsymbol{u}\notin\Lambda}\sqrt{2\pi}||\boldsymbol{u}||,
\end{align}
where the factor of $\sqrt{2\pi}$ comes from the definitioin in Eq.~\ref{eq:logical_operators}, and the minimum is taken over the lattice vectors that are in the symplectic dual lattice $\Lambda^\perp = \Lambda(M^{\perp})$ but not in the original lattice $\Lambda = \Lambda(M)$. 

To be more concrete, let us consider a GKP code that encodes a qubit into $N$ modes. Since Euclidean distance is {independent to the basis vectors of the lattice}, we assume that the generator matrix $M$ of the GKP code is canonized with $d_1=2$ and $d_2=...=d_N=1$. From Eq.~\ref{eq:logical_canonical}, we can notice that the symplectic dual lattice $\Lambda^\perp = \Lambda(M^{\perp})$ is spanned by the same set of basis vectors as $\Lambda= \Lambda(M)$, except that the first two basis vectors of $\Lambda^\perp$ are only \emph{half} of those for $\Lambda$. 
Hence the quotient group $\Lambda^\perp/\Lambda$ is {generated} by $\left\{\boldsymbol{w}_1^T, \boldsymbol{w}_2^T\right\}$, where $\boldsymbol{w}_{1,2}$ correspond to the logical operators of the encoded qubit.
Since the logical operators are indistinguishable if they differ by a stabilizer, we can identify the following set of vectors
\begin{align}
\label{eq:omega_1_2_u}
    \left\{\boldsymbol{w}_1+\boldsymbol{u},~\forall~ \boldsymbol{u}\in\Lambda\right\}, \nonumber\\
    \left\{\boldsymbol{w}_2+\boldsymbol{u},~\forall~ \boldsymbol{u}\in\Lambda\right\}, \\
    \left\{\boldsymbol{w}_1+\boldsymbol{w}_2+\boldsymbol{u},~\forall~ \boldsymbol{u}\in\Lambda\right\},\nonumber
\end{align}
to the logical $\bar{X}$, $\bar{Z}$ and $\bar{Y}$ operators respectively. 
{Note that the logical identity operator $\bar{I}$ is simply all the lattice vectors in $\Lambda$, i.e., $\left\{\boldsymbol{u},~\forall~\boldsymbol{u}\in\Lambda\right\}$.}
We can define the distances of the different logical operators as the minimum length of the corresponding set of vectors
\begin{align}
\label{eq:def_distances}
    d_{X} &= \min_{ \boldsymbol{b} \in \mathbb{Z}^{2N} } \sqrt{2\pi}|| \boldsymbol{w}_{1} - M^T\boldsymbol{b} ||, 
    \nonumber\\
    d_{Z} &= \min_{ \boldsymbol{b} \in \mathbb{Z}^{2N} } \sqrt{2\pi}|| \boldsymbol{w}_{2} - M^T\boldsymbol{b} ||,\\ 
    d_{Y} &= \min_{ \boldsymbol{b} \in \mathbb{Z}^{2N} } \sqrt{2\pi}|| \boldsymbol{w}_{1} + \boldsymbol{w}_{2} - M^T\boldsymbol{b} ||,\nonumber 
\end{align}
and the distance is given by
\begin{align}
\label{eq:def_distance}
    d = \min(d_{X},d_{Y},d_{Z}). 
\end{align}
Eq.~\ref{eq:def_distance} is a special case of Eq.~\ref{eq:distance} for GKP codes that are in the canonical basis and encode a single qubit, because the vectors in the three summands of Eq.~\ref{eq:def_distances} are guaranteed to lie in $\Lambda^\perp$ but not $\Lambda$ by construction.

\subsection{Transformation between GKP codes}
\label{sec:Transformation between GKP codes}
Recall a Gaussian unitary $\hat{U}$ transforms the quadrature operator $\hat{\boldsymbol{x}}$ into $S \hat{\boldsymbol{x}}$ where $S$ is a symplectic matrix. If we apply the Gaussian unitary to a stabilizer group element $\hat{S}=\hat{D}(M^T\boldsymbol{a})$, as defined in Eq.~\ref{eq:stabilizer_as_lattice_point}, the new GKP code is then stabilized by the stabilizer
\begin{eqnarray}
    \begin{aligned}
        \hat{S} &\equiv \hat{U} \hat{S} \hat{U}^{\dagger} \\
        &= \exp[ i  \sqrt{2\pi} ( \boldsymbol{a}^{T} M )  \Omega^{-1} \hat{U}\hat{\boldsymbol{x}} \hat{U}^\dagger ]\\
        % &= \exp[ i  \sqrt{2\pi} ( \boldsymbol{a}^{T} M )  \Omega^{-1} S^{-1} \hat{\boldsymbol{x}} ]\\
        &= \exp[ i \sqrt{2\pi} ( \boldsymbol{a}^{T} M )  \Omega^{-1} (\Omega S^{T}\Omega^{-1}) \hat{\boldsymbol{x}}  ] \\
        % &= \exp[ i \sqrt{2\pi} ( \boldsymbol{a}^{T} M S^T)   \Omega^{-1} \hat{\boldsymbol{x}}  ] \\
        &= \hat{D}(\sqrt{2\pi}(MS^T)^T\boldsymbol{a}), 
    \end{aligned}
\end{eqnarray}
where we used the fact that $S^{-1} =\Omega S^{T}\Omega^{-1}$ because $S (\Omega S^{T}\Omega^{-1})  = I$. Thus the new GKP lattice has generator matrix 
\begin{align}
\label{eq:lattice_transformation}
    M' = MS^T.
\end{align}
With Eq.~\ref{eq:lattice_transformation}, it allows us to realize any GKP code {by applying a Gaussian unitary operator to an $N$-mode square lattice GKP code which is} generated by
\begin{align}
\label{eq:def_M_sq}
    M_{\textrm{sq}}(\boldsymbol{d}) \equiv \textrm{diag}(\sqrt{d_{1}}, \cdots, \sqrt{d_{N}} ) \otimes I_2. 
\end{align}
% with 
% \begin{align}
%     \prod_{i=1}^Nd_i=\text{
% \end{align}
To see that, consider a general GKP code in its canonical basis, we can always decompose $M$ into $M = M_{\textrm{sq}}(\boldsymbol{d}) S^{T}$ with $S^{T} =M_{\textrm{sq}}(\boldsymbol{d})^{-1} M $, which is a symplectic matrix because
\begin{align}
    S^{T} \Omega (S^{T})^{T}  &= (M_{sq}(\boldsymbol{d})^{-1} M ) \Omega   (M_{sq}(\boldsymbol{d})^{-1} M )^{T}    
    % = M_{sq}(d)^{-1} \Omega(d) M_{sq}(d)^{-1} 
    = \Omega.
\end{align}
If $S^T$ is symplectic, then $S$ is also symplectic because
\begin{align}
    S\Omega S^T = S\Omega \Omega (\Omega S)^{-1} = -S S^{-1} \Omega^{-1} = \Omega.
\end{align}
Thus we see that any GKP code can be understood as a code that results from applying a Gaussian unitary operator $\hat{U}$, with a corresponding symplectic matrix $S=(M_\text{sq}(\boldsymbol{d})M)^T$, to a square lattice GKP code. The corresponding stabilizers are then given by 
\begin{align}
    \hat{{S}}_{2k-1} &= \hat{D}(\sqrt{2\pi d_{k}} s_{2k-1}), \quad \hat{{S}}_{2k} = \hat{D}(\sqrt{2\pi d_{k}} s_{2k}),
\end{align}
for $k\in\lbrace 1,\cdots, N \rbrace$, where $s_{1},\cdots,s_{2N}$ are the columns of the symplectic matrix $S$.  

\subsection{The {concatenated GKP} code}
\label{sec:The concatenated GKP code}

Here we describe how to concatenate a GKP code with another qubit stabilizer code. We assume that the base GKP code
% , denoted as $M_\text{base}$ 
encodes a single qubit {in each mode and we prepare $N$ such GKP qubits with $N$ modes. Then we consider a standard qubit-based stabilizer code $[[N, k, d_{0}]]$, where $d_{0}$ is the distance of the qubit stabilizer code (not to be confused with the distance of the resulting GKP code). Then, the} resultant {concatenated GKP} code encodes $k$ qubits in $N$ modes. 

For {the qubit stabilizer code $[[N, k, d_{0}]]$}, recall that each stabilizer generator corresponds to a binary vector with $2N$ components, where the odd-numbered and even-numbered entries represent the presence of an $X$ and $Z$
% $Z$ and $X$ 
operators respectively. Hence we shall use $\left\{\boldsymbol{g}_j, j=1,...,N-k\right\}$ to denote the set of binary vectors for the stabilizer generators. 
For simplicity, we start with the square lattice as the base GKP code, and form a separable lattice generated by $N$ copies of the square GKP code $M^{\text{(sq)}}=M_\text{sq}^{\oplus N}$. 
From Eq.~\ref{eq:def_M_sq}, we see that 
\begin{align}
\label{eq:M^sq}
    M^{\text{(sq)}} = \sqrt{2}I_{2N} 
\end{align}
in an appropriately chosen basis. {Here the prefactor $\sqrt{2}$ is due to the fact that each base GKP code encodes one qubit (i.e., two states) in a mode.}
In order to obtain the lattice corresponding to the concatenated code, we replace $N-k$ rows in $M^{\text{(sq)}}$
% the first summand of Eq.~\ref{eq:M^sq} 
by the following set of vectors 
\begin{align}
    \label{eq:basis_construction_A_lattice}
    \left\{\frac{1}{\sqrt{2}}\boldsymbol{g}^T_j, ~\text{for}~ j=1,...,N-k\right\},
\end{align}
such that the resultant matrix, denoted by $M_\text{conc}^{\text{(sq)}}$, remains full-rank. 
In App.~\ref{app:Details of constructing lattices for concatenated GKP codes}, we will show more details on how to arrive at $M_\text{conc}^{\text{(sq)}}$, and that 
$\det(M_\text{conc}^{\text{(sq)}})=2^{k}$ which indicates that the resultant lattice indeed encodes $k$ qubits as desired. 
This process of deriving a lattice from a binary code is known as \emph{Construction A} in Ref.~\cite{conway2013sphere}. 

The Construction A procedure allows us to concatenate generic base GKP code to a stabilizer code. Let $M_\text{base}$ {be the generator matrix of a generic qubit-into-an-oscillator GKP code and assume that $M_\text{base}$ is in the canonical form.} From the discussion in Sec.~\ref{sec:Transformation between GKP codes}, we can always find a symplectic matrix $S_\text{base}$ such that $M_\text{base}=M_\text{sq}S_\text{base}^T$. Hence, the {generator matrix of the concatenated GKP code is given by}
\begin{align}
\label{eq:concatenated_GKP}
    M_\text{conc} = M_\text{conc}^{\text{(sq)}} (S_\text{base}^T)^{\oplus N} . 
\end{align}

\subsection{{Examples of symplectic lattices and GKP codes} }
\label{sec:Lattice examples}

The {error-correcting capabilities} of a GKP code is strongly tied to the properties of the underlying {symplectic lattice of the GKP code}. In this section, we review several {relevant symplectic} lattices. 

\emph{Z-type lattice} - The simplest way to encode a \emph{state} into a multimode GKP code is to use the hypercubic lattice denoted as the $Z_{2N}$ generated by $M_{Z_{2N}}\equiv I_{2N}$, the $2N\times 2N$ identity matrix. The resulting stabilized state is given by a tensor product of $2N$ GKP qunaught states \cite{larsen2021fault,walshe2020continuous,noh2022low}.

One could scale the lattice spacings along different axes for encoding multiple states and qubits in the lattice, as shown in Eq.~\ref{eq:def_M_sq}.
For instance, one way to encode a qubit into a single mode GKP code is to use the two dimensional rectangular lattice given by
\begin{align}
    \label{eq:2D_rec}
    M_\text{rec} = \begin{bmatrix}
    \sqrt{2}\eta & 0\\
    0 & \sqrt{2}/\eta
    \end{bmatrix}
\end{align}
where $\eta>0$ is the {square root of the} aspect ratio between the two axes. 
The rectangular lattice can be obtained from the square lattice $M_\text{sq}(2)$ via the transformation
\begin{align}
    M_\text{rec} = M_\text{sq}(2) S_\text{rec}^T = M_\text{sq}(2) \begin{bmatrix}
    \eta & 0\\
    0 & \eta^{-1}
    \end{bmatrix}.
\end{align}
Here the symplectic matrix $S_\text{rec}$ corresponds to a one-mode squeezing operation. Similarly, an $N$-mode hyper-rectangular GKP code can be obtained by {applying the tensor product of $N$ one-mode squeezing operations to an $N$-mode hypercubic
% hyper-square 
GKP code generated by a scaled $Z_{2N}$ lattice (by a factor of $\sqrt{2}$). }

The logical operators of the code can be deduced from Eq.~\ref{eq:logical_operators}. From Eq.~\ref{eq:def_M_perp}, we has $\boldsymbol{w}_1=(\eta/\sqrt{2}, 0)^T$, $\boldsymbol{w}_2=(0, \sqrt{2}/\eta)^T$ such that \cite{gottesman2001encoding}
\begin{align}
    \bar{X} \equiv \hat{L}_1=\exp\left[-i\sqrt{\pi}\eta\hat{p}\right],\quad \bar{Z} \equiv \hat{L}_2=\exp\left[i\frac{\sqrt{\pi}}{\eta}\hat{q}\right].
\end{align}
Upon solving Eq.~\ref{eq:def_distances}, we have the code distance for the rectangular code as
\begin{align}
    \label{eq:def_distances_rec}
    d^{\text{rec}}_X = \eta\sqrt{\pi}, \quad d^{\text{rec}}_Y = \sqrt{(\eta^2+\eta^{-2})\pi}, \quad d^{\text{rec}}_Z = \eta^{-1}\sqrt{\pi}
\end{align}

\emph{$D$-type lattice} - The $D$-type lattice, denoted as $D_n$, is an $n$-dimensional sublattice of the $Z_n$ lattice such that the sum of the components of the lattice points is even \cite{conway2013sphere}. Formally, 
\begin{align}
    D_n = \left\{(x_1, .., x_n)\in \mathbb{Z}^n: \sum_{k=1}^nx_k \text{ is even}\right\}.
\end{align}
In other words, the $D_n$ lattice can be obtained by coloring the $Z_n$ lattice in a checkerboard pattern, hence the $D_n$ lattice is also called the checkerboard lattice. 
The simpliest example is the $D_2$ lattice given by 
\begin{align}
\label{eq:D_2}
    M_{D_2} = \begin{bmatrix}
    1 & 1\\
    1 & -1
    \end{bmatrix},
\end{align}
which is nothing but a rotated square lattice {(by 45 degrees)}, also known as the diamond code. 
An important feature of the $D_n$ lattice is that the volume of its fundamental parallelotope is always $2$, i.e., $\det(M_{D_n})=2$. {Combining this with the fact that the symplectic gram matrix of $M_{D_n}$ is an integer matrix, we find that the $D_n$ lattice can be used to define a GKP code that encode one qubit in $N$ modes for even valued $n=2N$.}
In fact, {it was shown in Ref.~\cite{royer2022encoding} that the GKP code defined with the lattice $D_{2N}$} can be viewed as an $N$-qubit repetition code {(with $Y$-type stabilizers)} concatenated with the diamond GKP code defined in Eq.~\ref{eq:D_2}. 

The fact that the $D_{2N}$ lattice can be viewed as a repetition code allows us to infer its code distances straightforwardly. 
Because the diamond code in Eq.~\ref{eq:D_2} is simply a rotated square lattice, its code distances are the same as those given in Eq.~\ref{eq:def_distances_rec} with $\eta=1$. Since the YY stabilizers in the repetition code would detect the $X$ and $Z$ errors but not the $Y$ errors, the Euclidean distance for the logical $\bar{Y}$ operator remains the same as that for the diamond code.
Further, because the $\bar{X}$ operator for the YY repetition code is a tensor product of Pauli $X$ operators, and the $\bar{X}$ operator for the diamond code corresponds to $\boldsymbol{w}_1=\frac{1}{2}(1, 1)^T$, one can show that the $\bar{X}$ operator for the $D_{2N}$ code corresponds to an $2N$-component vector with all entries equal to $1/2$ \cite{royer2022encoding}.
This vector has the minimum length of $\sqrt{N/2}$ among $\left\{\boldsymbol{w}_1+\boldsymbol{u},~\forall~ \boldsymbol{u}\in\Lambda\right\}$ because $\Lambda$ is an integral lattice.
As a result, we have $d^{D_{2N}}_X=\sqrt{2\pi}\sqrt{N/2}=\sqrt{N\pi}$. Because of the symmetry between the logical $\bar{Z}$ and $\bar{X}$ operators, we conclude that
\begin{align}
    \label{eq:def_distances_D_2N}
    d^{D_{2N}}_Z = \sqrt{N\pi}, \quad d^{D_{2N}}_Y = \sqrt{2\pi}, \quad d^{D_{2N}}_X = \sqrt{N\pi}
\end{align}
We see that $d^{D_{2N}}_X=d^{D_{2N}}_Y=d^{D_{2N}}_Z$ if and only if $N=2$ for the $D_4$ lattice. We will call a code with equal {$\bar{X},\bar{Y},\bar{Z}$} logical distances a \emph{balanced code}.

\emph{Tesseract lattice} - The tesseract lattice is a four-dimensional analogue of a cube with the following generator matrix
\begin{align}
\label{eq:M_tess}
    M_\text{tess} = 2^\frac{1}{4}\begin{bmatrix}
    \frac{1}{\sqrt{2}} & 0 & \frac{1}{\sqrt{2}} & 0\\
    0 & 1 & 0 & 0\\
    \frac{1}{\sqrt{2}} & 0 & -\frac{1}{\sqrt{2}} & 0\\
    0 & 0 & 0 & 1
    \end{bmatrix}.    
\end{align}
We can notice that this is a direct sum of two sublattices, where $\hat{q}_1$ and $\hat{q}_2$ quadratures form a $D_2$ lattice, and $\hat{p}_1$ and $\hat{p}_2$ form a $Z_2$ lattice. 
More importantly, the tessearct code can be viewed as the rectangular GKP code concatenated with the $2$-qubit repetition code. To see that, we consider two copies of the rectangular GKP codes in Eq.~\ref{eq:2D_rec} with $\eta=2^{1/4}$ and the $2$-qubit repetition code with a $XX$ stabilizer. Following the approach above, we arrive at the following generator matrix
\begin{eqnarray}
    \begin{aligned}
    M_\text{tess}' &=
    \begin{bmatrix}
    \frac{1}{\sqrt{2}} & 0 & \frac{1}{\sqrt{2}} & 0\\
    0 & \sqrt{2} & 0 & 0\\
    0 & 0 & \sqrt{2} & 0\\
    0 & 0 & 0 & \sqrt{2}\\
    \end{bmatrix}
    \begin{bmatrix}
    2^{\frac{1}{4}} & 0 & 0 & 0\\
    0 & 2^{-\frac{1}{4}} & 0 & 0 \\
    0 & 0 & 2^{\frac{1}{4}} & 0 \\
    0 & 0 & 0 & 2^{-\frac{1}{4}}
    \end{bmatrix} \\
    & = 2^{\frac{1}{4}}\begin{bmatrix}
    2^{-\frac{1}{2}} & 0 & 2^{-\frac{1}{2}} & 0\\
    0 & 1 & 0 & 0 \\
    0 & 0 & 2^{\frac{1}{2}} & 0 \\
    0 & 0 & 0 & 1
    \end{bmatrix} \\
    &= RM_\text{tess},    
    \end{aligned}
\end{eqnarray}
where $M_\text{tess}$ is given in Eq.~\ref{eq:M_tess} and 
\begin{align}
    R = \begin{bmatrix}
    1 & 0 & 0 & 0\\
    0 & 1 & 0 & 0\\
    1 & 0 & -1 & 0\\
    0 & 0 & 0 & 1
    \end{bmatrix}.    
\end{align}
Since $M_\text{tess}'$ differ from $M_\text{tess}$ only by a unimodular matrix, they are different basis for the same lattice, namely the tesseract lattice. 

For the distance of the tesseract code, since the XX stabilizer cannot detect the logical $\bar{X}$ errors, it follows that $d^{\text{tess}}_X$ is the same as $d^{\text{rec}}_X$ with $\eta=2^{1/4}$, i.e., $d^{\text{tess}}_X = 2^\frac{1}{4} \sqrt{\pi}$. On the other hand, we expect that both distances for the logical $\bar{Y}$ and $\bar{Z}$ will be improved due to the XX stabilizer. Upon explicitly solving Eq.~\ref{eq:def_distances} for $M_\text{tess}^\perp$, we have
\begin{align}
    \label{eq:def_distances_tess}
    d^{\text{tess}}_X = 2^\frac{1}{4} \sqrt{\pi}, \quad d^{\text{tess}}_Y =2^\frac{1}{4} \sqrt{2\pi}, \quad d^{\text{tess}}_Z = 2^\frac{1}{4} \sqrt{\pi}.
\end{align}

We see that for the case of $N=2$, the $D_4$ lattice has distance $d^{D_4}=\sqrt{2\pi}$ which is greater than $d^{\text{tess}}=2^\frac{1}{4}\sqrt{\pi}$ for the tesseract code. 
However, one problem for the $D_{2N}$ lattice is that its distance remains the same for all values of $N$ because the distance of the logical $\bar{Y}$ operator do not scale with the number of modes. 
In Sec.~\ref{sec:generalized_codes_intro}, we will introduce two generalizations of the tesseract {and $D_{4}$} codes that has larger distances than the $D_{2N}$ lattice for $N>2$.

\section{Closest point decoder for the GKP codes}
\label{sec:Closest point decoder for the GKP codes}

In this section, we introduce the closest point decoder for the GKP code,  which is based on the lattice structure of the code.
We begin by {following Ref.~\cite{conrad2022gottesman} to formulate} the decoding problem as a lattice problem.

\subsection{Error syndrome for GKP code}
\label{sec:Error syndrome for GKP code}

Suppose that we have a GKP code that encodes one qubit in $N$ oscillators. 
The defined GKP code can be used to correct small shift errors, and we assume that the oscillator quadratures undergo independent and identically distributed (iid) additive errors 
\begin{align}
\label{eq:def_shift_errors}
    \hat{\boldsymbol{x}}\rightarrow\hat{\boldsymbol{x}}' \equiv \hat{\boldsymbol{x}} + \boldsymbol{\xi},
\end{align}
where $\boldsymbol{\xi}\equiv(\xi_q^{(1)}, \xi_q^{(1)}, ..., \xi_q^{(N)}, \xi_q^{(N)})\sim_\text{iid}\mathcal{N}(0,\sigma^2)$ are the random shifts that follow the Gaussian distribution $\mathcal{N}(0,\sigma^2)$. 
The errors can be modeled by applying the displacement operator $\hat{D}(\boldsymbol{\xi})$ onto the GKP code. 
Our goal is to apply another displacement operator $\hat{D}(-\boldsymbol{\xi}^*)$ onto the errant GKP code to minimize the chance of getting a logical error. 

The shift error $\boldsymbol{\xi}$ is not known a priori, and we have only the information from the {stabilizer} measurements. 
Recall that the stabilizers of a GKP code are given by 
$\hat{S}_{j} = \exp[ i\sqrt{2\pi} \boldsymbol{v}_{j}^{T} \Omega^{-1}\hat{\boldsymbol{x}} ]$, %
where $v_{j}^{T}$ is the $j$-th row of $M$ (c.f. Eq.~\ref{eq:stabilizer_as_lattice_point}). 
Because the stabilizers commute with each other, they can be measured simultaneously. This is equivalent to measuring the exponents $ i\sqrt{2\pi} \boldsymbol{v}_{j}^{T} \Omega^{-1}\hat{\boldsymbol{x}}$ modulo $2\pi i$. 
Let $s_j$ denote error syndrome from the homodyne measurements, then it differs from $ \sqrt{2\pi} \boldsymbol{v}_{j}^{T} \Omega^{-1}{\boldsymbol{\xi}}$ by an integer multiple of $2\pi$, i.e.,
\begin{align}
\label{eq:def_syndrome}
    % \sqrt{2\pi} M \Omega^{-1}\boldsymbol{\xi} = \boldsymbol{\xi}^{(m)} + 2\pi \boldsymbol{a}
    \boldsymbol{s} \equiv \sqrt{2\pi} M \Omega^{-1}\boldsymbol{\xi} ~\mod~ 2\pi,
\end{align}
where the modulo operation is applied element-wise. 
In other words, the shift error is related to the syndrome via $\boldsymbol{\xi} = \frac{1}{\sqrt{2\pi}}\Omega M^{-1}(\boldsymbol{s} + 2\pi \boldsymbol{a})$ for certain integer valued vector $\boldsymbol{a}$. For the purpose of decoding, we can write
\begin{eqnarray}
    \begin{aligned}
    \boldsymbol{\xi} 
    &=\frac{1}{\sqrt{2\pi}}\Omega (\Omega M^TA^{-1})(\boldsymbol{s} + 2\pi \boldsymbol{a})\\
    &=\frac{{-1}}{\sqrt{2\pi}}(\Omega M^\perp)^T(\boldsymbol{s} + 2\pi \boldsymbol{a})\\
    &=\boldsymbol{\eta}(\boldsymbol{s}) - \sqrt{2\pi}(M^\perp)^T\boldsymbol{b},
    \end{aligned}
\end{eqnarray}
where we have used the identity in Eq.~\ref{eq:identity_1} and introduced the integer valued vector $\boldsymbol{b}\equiv {-}\Omega\boldsymbol{a}$ and 
\begin{align}
    \label{eq:def_eta_s}
    \boldsymbol{\eta}(\boldsymbol{s})\equiv\frac{{-1}}{\sqrt{2\pi}}(\Omega M^\perp)^T\boldsymbol{s}.
\end{align}
Thus, we learn about the shift $\boldsymbol{\xi}$ only modulo the lattice generated by $\sqrt{2\pi}M^{\perp}$, {i.e., a lattice of logical operators}.  
Condition on the error syndrome $\boldsymbol{s}$ obtained from the homodyne measurement, we are looking for a {shift} $\boldsymbol{\xi}^*$ that has the shortest length and being the most likely shift error compatible with the measured stabilizer values. Thus we need to solve the following problem
\begin{align}
\label{eq:closest_point_problem}
    \boldsymbol{b}^*=\argmin_{\boldsymbol{b}\in\mathbb{Z}^{2N}}|\boldsymbol{\eta}(\boldsymbol{s}) - \sqrt{2\pi}(M^\perp)^T\boldsymbol{b}|.
\end{align}
With that we will apply the counter displacement $\hat{D}(-\boldsymbol{\xi}^*)$ with $\boldsymbol{\xi}^*\equiv \boldsymbol{\eta}(\boldsymbol{s}) - \sqrt{2\pi}(M^\perp)^T\boldsymbol{b}^*$. 
After the correction, the initial state is translated by $\hat{D}(-\boldsymbol{\xi}^*)\hat{D}(\boldsymbol{\xi})=e^{i\alpha}\hat{D}(\boldsymbol{e})$ where $\alpha$ is an irrelevant phase and $\boldsymbol{e}\equiv\sqrt{2\pi}(M^\perp)^T\boldsymbol{c}$ for some $\boldsymbol{c}\in\mathbb{Z}^{2N}$. 
To see the net result of the shift error and the counter displacement, we can again assume $M^\perp$ is in the canonical basis, where the first two rows of $M^\perp$ generates the logical operators. Hence, after the attempted correction, we will have 
\begin{equation}
\label{eq:pauli_channel}
\begin{cases} 
\bar{I} & \textrm{if } \boldsymbol{c} \in (2\mathbb{Z}, 2\mathbb{Z}, \mathbb{Z}, \mathbb{Z}\cdots,\mathbb{Z}, \mathbb{Z} ) \\ 
\bar{X} & \textrm{if } \boldsymbol{c} \in (2\mathbb{Z} + 1, 2\mathbb{Z}, \mathbb{Z}, \mathbb{Z}\cdots,\mathbb{Z}, \mathbb{Z} )  \\ 
\bar{Z} & \textrm{if } \boldsymbol{c} \in (2\mathbb{Z}, 2\mathbb{Z}+1, \mathbb{Z}, \mathbb{Z}\cdots,\mathbb{Z}, \mathbb{Z} ) \\ 
\bar{Y} & \textrm{if } \boldsymbol{c} \in (2\mathbb{Z} + 1, 2\mathbb{Z} + 1, \mathbb{Z}, \mathbb{Z}\cdots,\mathbb{Z}, \mathbb{Z} ) 
\end{cases}    
\end{equation}
on the encoded information.
Hence there will be logical error if and only if either of the first two components of $\boldsymbol{c}$ is an odd integer.

{For more general cases with non-canonical $M^{\perp}$, let $R$ denotes the unimodular matrix that canonizes the generator matrix $M$, i.e., $M'=RM$ is in canonical basis, and $M'^\perp=-\Omega(R^T)^{-1}\Omega M^\perp$ is the canonized logical operators, according to Eq.~\ref{eq:def_M_perp}. With that, Eq.~\ref{eq:pauli_channel} can be applied for $-\Omega R\Omega\boldsymbol{c}$ in a non-canonical basis.
}

{
Before proceeding, we remark that Eq.~\ref{eq:closest_point_problem} is referred to as minimum energy decoding (MED) in Ref.~\cite{conrad2022gottesman}, which is an approximation of the optimal maximum likelihood decoding (MLD) for general GKP codes. 
In particular, the MED is only optimal when $\sigma\rightarrow0$ as it searches for the most possible error, instead of the most possible coset of errors. 
Nevertheless, for certain quantum error correction codes, MED is shown to have similar performance compared to the MLD, which generally has greater time complexity \cite{bravyi2014efficient}.
We further note that in Ref.~\cite{conrad2022gottesman}, MED is only discussed as a subroutine for MLD to decode concatenated codes. 
Here, on the other hand, we present a general algorithm for decoding generic GKP codes from the lattice perspective. 
}

\subsection{Closest point search problem}
\label{sec:Closest point search problem}

The problem in Eq.~\ref{eq:closest_point_problem} is known as the closest point search problem, or the closest vector search problem, in the mathematical community, which can be stated formally as follows. For a given lattice $\Lambda\subset\mathbb{R}^n$, find {a lattice point $\boldsymbol{u}\in \Lambda$ that is closest to an input vector $\boldsymbol{t}\in\mathbb{R}^n$:}
\begin{align}
\label{eq:def_CP}
    \boldsymbol{\chi}_{\boldsymbol{t}}(\Lambda) \equiv \argmin_{\boldsymbol{u}\in\Lambda}||\boldsymbol{t}-\boldsymbol{u}|| . 
\end{align}
In the case of tie, $\boldsymbol{\chi}_{\boldsymbol{t}}(\Lambda)$ is chosen arbitrarily from the closest points.
Equivalently, if the lattice is generated by the matrix $M$, then 
\begin{align}
\label{eq:def_CP_2}
    (M^T)^{-1}\boldsymbol{\chi}_{\boldsymbol{t}}(\Lambda(M)) = \argmin_{\boldsymbol{b}\in\mathbb{Z}^n}||\boldsymbol{t}-M^T\boldsymbol{b}||,
\end{align}
where the right hand side gives the coordinates of the closest point in the basis of $M$.
We emphasize that although the closest point problem is described with respect to $M$ in Eq.~\ref{eq:def_CP}-\ref{eq:def_CP_2}, for a GKP lattice $M_\text{gkp}$, the closest point problem is to be solved for its symplectic dual $M_\text{gkp}^\perp$, instead of the lattice itself, per Eq.~\ref{eq:closest_point_problem}.
Although the closest point problem has been known to be NP-hard for decades \cite{van1981another,micciancio2001hardness}, due to its many applications in diverse areas \cite{agrell1998optimization,conway1984voronoi,damen2000lattice,blake2002lattices,micciancio2009lattice,clarkson1999frequency,schnorr1995attacking}, there has been many attempts to reduce the search time for exact solutions \cite{agrell2002closest,schnorr1995attacking,micciancio2010deterministic,dadush2014short,kannan1983improved,kannan1987minkowski,fincke1985improved}, or approximate solutions \cite{arora1997hardness,dinur2000improved,micciancio2001shortest}.
Here we discuss an \emph{exact} algorithm which is based on \cite{agrell2002closest}, and more details of the algorithm will be presented in App.~\ref{sec:More details on the closest point decoder}. {Note that we are interested in an exact algorithm (despite its exponential time cost) since we would like to use it to understand generic, unstructured, and small-sized GKP codes, as well as to benchmark efficient decoders for structured GKP codes.} 

The algorithm starts by preprocessing the generator matrix $M$ into a lower-triangular form via the transformation
\begin{align}
\label{eq:RLQ}
    M = RLQ,
\end{align}
where $R$ is a unimodular matrix and $Q$ is an orthogonal matrix. As described in Sec.~\ref{sec:The canonical basis for the GKP codes}, matrices differed by a unimodular matrix multiplied from the left generate the same lattice, and because the orthogonal transformation can be regarded as rotating the basis vectors, matrices $M$ and $L$ generates the identical lattice.
The transformation in Eq.~\ref{eq:RLQ} is also known as lattice reduction, which is a process of selecting a good basis for speeding up the closest point searching process. 
The Lenstra-Lenstra-Lov\'asz (LLL) algorithm and the Korkine-Zolotareff (KZ) algorithm are two widely used techniques for lattice reductions. They have advantages in different scenarios, which will be discussed further in Sec.~\ref{sec:Discussion and conclusion} and App.~\ref{sec:More details on the closest point decoder}.

\begin{figure}
\centering
\includegraphics[width=\linewidth]{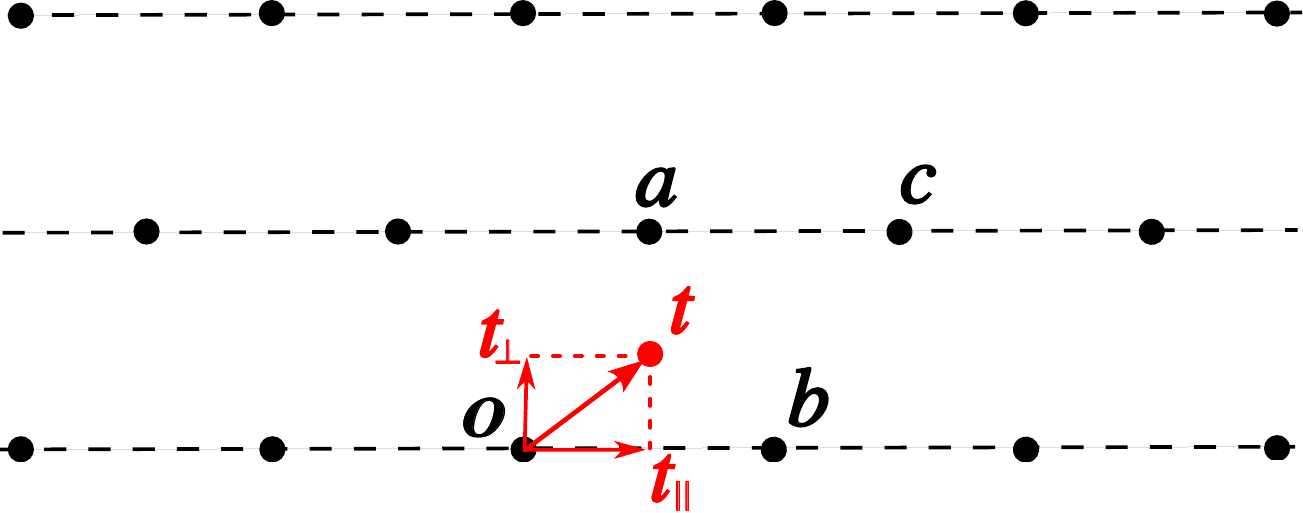}
 \caption{
A 2D lattice as a stack of 1D sublattices. The black dots represent the lattice points of a triangular lattice, and the dashlines represent the parallel 1D sublattices. The decomposition into sublattices is not unique. The red dot is the input vector $\boldsymbol{t}$, which can be decomposed as $\boldsymbol{t}=\boldsymbol{t}_\parallel+\boldsymbol{t}_\perp$ which are parallel and perpendicular to the sublattices respectively.
For illustration purpose, the point is chosen to lie \emph{slightly above} the center of the equilateral triangle formed by the points $\boldsymbol{o}$, $\boldsymbol{a}$ and $\boldsymbol{b}$. 
Hence the closest point is $\boldsymbol{a}$; on the other hand, the closest line to $\boldsymbol{t}$ is the line cross $\boldsymbol{o}$ and $\boldsymbol{b}$. The closest point algorithm will first search the $\boldsymbol{ob}$ line and randomly select $\boldsymbol{o}$ or $\boldsymbol{b}$ as the candidate closest point, which sets the upper bound for the distance between the closest point and $\boldsymbol{t}$. Hence we can ignore the rest of the 1D sublattices, and only need to search the $\boldsymbol{ac}$ line, which leads to the true closest point, namely $\boldsymbol{a}$. 
}
 \label{fig:closest_point_idea}
\end{figure}

The next step is to find the closest point in the new basis $L$. For that, we first notice
\begin{align*}
    (M^T)^{-1}\boldsymbol{\chi}_{\boldsymbol{t}}(\Lambda(M)) &= \argmin_{\boldsymbol{b}\in\mathbb{Z}^n}||Q\boldsymbol{t}-L^TR^T\boldsymbol{b}|| \\
    &=(R^T)^{-1}\argmin_{\boldsymbol{b}'\in\mathbb{Z}^n}||Q\boldsymbol{t}-L^T\boldsymbol{b}'|| \\
    &=(R^T)^{-1}(L^T)^{-1}\boldsymbol{\chi}_{Q\boldsymbol{t}}(\Lambda(L)) 
\end{align*}
such that
\begin{align}
\label{eq:transformation_of_closest_points}
    \boldsymbol{\chi}_{\boldsymbol{t}}(\Lambda(M)) = Q^T\boldsymbol{\chi}_{Q\boldsymbol{t}}(\Lambda(L)).
\end{align}
Hence, the problem reduces to  finding the closest point $\boldsymbol{\chi}_{\boldsymbol{t}'}(\Lambda(L))$ for $\boldsymbol{t}'\equiv Q\boldsymbol{t}$.
The basic idea of the searching algorithm is to view the $n$-dimensional lattice as a stack of $(n-1)$-dimensional sublattices, and search these sublattices recursively. For instance, a 2D lattice can be viewed as a collection of 1D lattices as shown in Fig.~\ref{fig:closest_point_idea}.
We can similarly decompose the input vector as $\boldsymbol{t}'=\boldsymbol{t}'_\parallel+\boldsymbol{t}'_\perp$ which are parallel and perpendicular to the sublattices respectively. 
The searching proceeds with the Schnorr-Euchner strategy\cite{schnorr1994lattice}, which sorts the sublattices in the ascending order of their vertical distances to $\boldsymbol{t}'$.
Let us denote the nearest sublattice as $\Lambda'$, then we can first identify $\boldsymbol{\chi}_{\boldsymbol{t}'}(\Lambda')$ which is the nearest point in $\Lambda'$ to $\boldsymbol{t}'$. 
It is important to note that $\boldsymbol{\chi}_{\boldsymbol{t}'}(\Lambda)\neq\boldsymbol{\chi}_{\boldsymbol{t}'}(\Lambda')$ because the nearest lattice point needs not lie within the nearest sublattice, as shown in Fig.~\ref{fig:closest_point_idea}.
Nevertheless, the distance between $\boldsymbol{\chi}_{\boldsymbol{t}'}(\Lambda')$ and $\boldsymbol{t}'$ provides an upper bound $\rho\equiv||\boldsymbol{\chi}_{\boldsymbol{t}'}(\Lambda')-\boldsymbol{t}'||$, and $\boldsymbol{\chi}_{\boldsymbol{t}'}(\Lambda)$ cannot lie in the sublattice with vertical distance larger than $\rho$. We only need to search a finite set of sublattices, and update the upper bound $\rho$ if the new candidate point has smaller distance. 
The searching is complete after all the sublattices with vertical distance smaller than $\rho$ have been visited.
Once the closest point $\boldsymbol{\chi}_{\boldsymbol{t}'}(\Lambda(L))$ is identified in the basis $L$, we can transform it back to the original basis in $M$ via Eq.~\ref{eq:transformation_of_closest_points}. This concludes the closest point searching algorithm.

\section{Searching for optimized GKP codes}
\label{sec:Searching for optimized GKP codes}

In previous sections, we illustrate that a GKP code can be viewed as a symplectic integral lattice, and decoding the GKP code is equivalent to finding the closest point in the lattice. 
In this section, we will use this machinery to analyze several known {concatenated} GKP codes {and to numerically} search for optimized GKP codes. 

\subsection{Analysis of known {concatenated} GKP codes}
\label{sec:known concatenated GKP codes}

\begin{table}[h!]
\centering
\begin{tabular}{|c || c|| c | } 
[[5,1,3]] &[[7,1,3]] & $d_0=3$ surface code \\
 \hline
  \mybox[0.25cm]{$I$}\mybox[0.25cm]{$X$}\mybox[0.25cm]{$Z$}\mybox[0.25cm]{$Z$}\mybox[0.25cm]{$X$} & \mybox[0.25cm]{$I$}\mybox[0.25cm]{$I$}\mybox[0.25cm]{$I$}\mybox[0.25cm]{$X$}\mybox[0.25cm]{$X$}\mybox[0.25cm]{$X$}\mybox[0.25cm]{$X$} & \mybox[0.25cm]{$X$}\mybox[0.25cm]{$X$}\mybox[0.25cm]{$I$}\mybox[0.25cm]{$X$}\mybox[0.25cm]{$X$}\mybox[0.25cm]{$I$}\mybox[0.25cm]{$I$}\mybox[0.25cm]{$I$}\mybox[0.25cm]{$I$}\\
  \mybox[0.25cm]{$X$}\mybox[0.25cm]{$I$}\mybox[0.25cm]{$X$}\mybox[0.25cm]{$Z$}\mybox[0.25cm]{$Z$} & \mybox[0.25cm]{$I$}\mybox[0.25cm]{$X$}\mybox[0.25cm]{$X$}\mybox[0.25cm]{$I$}\mybox[0.25cm]{$I$}\mybox[0.25cm]{$X$}\mybox[0.25cm]{$X$} & \mybox[0.25cm]{$I$}\mybox[0.25cm]{$X$}\mybox[0.25cm]{$X$}\mybox[0.25cm]{$I$}\mybox[0.25cm]{$I$}\mybox[0.25cm]{$I$}\mybox[0.25cm]{$I$}\mybox[0.25cm]{$I$}\mybox[0.25cm]{$I$}\\
  \mybox[0.25cm]{$Z$}\mybox[0.25cm]{$X$}\mybox[0.25cm]{$I$}\mybox[0.25cm]{$X$}\mybox[0.25cm]{$Z$} & \mybox[0.25cm]{$X$}\mybox[0.25cm]{$I$}\mybox[0.25cm]{$X$}\mybox[0.25cm]{$I$}\mybox[0.25cm]{$X$}\mybox[0.25cm]{$I$}\mybox[0.25cm]{$X$} & \mybox[0.25cm]{$I$}\mybox[0.25cm]{$I$}\mybox[0.25cm]{$I$}\mybox[0.25cm]{$I$}\mybox[0.25cm]{$I$}\mybox[0.25cm]{$I$}\mybox[0.25cm]{$X$}\mybox[0.25cm]{$X$}\mybox[0.25cm]{$I$}\\
  \mybox[0.25cm]{$Z$}\mybox[0.25cm]{$Z$}\mybox[0.25cm]{$X$}\mybox[0.25cm]{$I$}\mybox[0.25cm]{$X$} & \mybox[0.25cm]{$I$}\mybox[0.25cm]{$I$}\mybox[0.25cm]{$I$}\mybox[0.25cm]{$Z$}\mybox[0.25cm]{$Z$}\mybox[0.25cm]{$Z$}\mybox[0.25cm]{$Z$} & \mybox[0.25cm]{$I$}\mybox[0.25cm]{$I$}\mybox[0.25cm]{$I$}\mybox[0.25cm]{$I$}\mybox[0.25cm]{$X$}\mybox[0.25cm]{$X$}\mybox[0.25cm]{$I$}\mybox[0.25cm]{$X$}\mybox[0.25cm]{$X$}\\
                                                                                                  & \mybox[0.25cm]{$I$}\mybox[0.25cm]{$Z$}\mybox[0.25cm]{$Z$}\mybox[0.25cm]{$I$}\mybox[0.25cm]{$I$}\mybox[0.25cm]{$Z$}\mybox[0.25cm]{$Z$} & \mybox[0.25cm]{$Z$}\mybox[0.25cm]{$I$}\mybox[0.25cm]{$I$}\mybox[0.25cm]{$Z$}\mybox[0.25cm]{$I$}\mybox[0.25cm]{$I$}\mybox[0.25cm]{$I$}\mybox[0.25cm]{$I$}\mybox[0.25cm]{$I$}\\
                                                                                                  & \mybox[0.25cm]{$Z$}\mybox[0.25cm]{$I$}\mybox[0.25cm]{$Z$}\mybox[0.25cm]{$I$}\mybox[0.25cm]{$Z$}\mybox[0.25cm]{$I$}\mybox[0.25cm]{$Z$} & \mybox[0.25cm]{$I$}\mybox[0.25cm]{$Z$}\mybox[0.25cm]{$Z$}\mybox[0.25cm]{$I$}\mybox[0.25cm]{$Z$}\mybox[0.25cm]{$Z$}\mybox[0.25cm]{$I$}\mybox[0.25cm]{$I$}\mybox[0.25cm]{$I$}\\
                                                                                                  &                                                                                                                                       & \mybox[0.25cm]{$I$}\mybox[0.25cm]{$I$}\mybox[0.25cm]{$I$}\mybox[0.25cm]{$Z$}\mybox[0.25cm]{$Z$}\mybox[0.25cm]{$I$}\mybox[0.25cm]{$Z$}\mybox[0.25cm]{$Z$}\mybox[0.25cm]{$I$} \\
                                                                                                  &                                                                                                                                       & \mybox[0.25cm]{$I$}\mybox[0.25cm]{$I$}\mybox[0.25cm]{$I$}\mybox[0.25cm]{$I$}\mybox[0.25cm]{$I$}\mybox[0.25cm]{$Z$}\mybox[0.25cm]{$I$}\mybox[0.25cm]{$I$}\mybox[0.25cm]{$Z$} \\
\hline       
\end{tabular}
\caption{The stabilizers for the [[5,1,3]], [[7,1,3]] and $d_0=3$ surface codes.
}
\label{tab:stabilizers_513_713}
\end{table}

{Here, we analyze known concatenated GKP codes.} As a warm-up, we start with the [[5,1,3]] and [[7,1,3]] qubit codes, whose stabilizers are shown in Tab.~\ref{tab:stabilizers_513_713}. We form {concatenated} GKP codes by concatenating them with the hexagonal GKP code generated by
\begin{align}
    M_\text{hex}=3^{-\frac{1}{4}}\begin{bmatrix}
    2 & 0\\
    1 & \sqrt{3}
    \end{bmatrix}.
\end{align}
The resulting codes have five and seven modes respectively, and upon solving Eq.~\ref{eq:def_distances}, we find that they are balanced GKP codes with distances equal to $d=3^{1/4}\sqrt{2\pi}\approx3.2989$.
These {concatenated} GKP codes are balanced because their stabilizer groups are invariant under the cyclic transformation of the Pauli operators
\begin{align}
\label{eq:cyclic_transformation}
    X\rightarrow Y,\quad Y\rightarrow Z,\quad Z\rightarrow X,
\end{align}
which is evident from Tab.~\ref{tab:stabilizers_513_713}, and the hexagonal GKP code is itself balanced with distance $3^{-1/4}\sqrt{2\pi}$. 
In fact, for the {concatenated} GKP code with a balanced base GKP code, its distance is given by \cite{royer2022encoding}
\begin{align}
\label{eq:distance_stabilizer_GKP}
    d = \sqrt{d_0}d_\text{base},
\end{align}
where $d_0$ and $d_\text{base}$ are the distances of the qubit stabilizer code and the base GKP code respectively.
In a similar spirit, we can form another balanced GKP code by concatenating the [[5,1,3]] code with the $D_4$ code, and the resulting code has ten modes with distance $3^{1/2}\sqrt{2\pi}\approx4.3416$. 
In contrast, the concatenation of the $d_0=3$ surface code with the hexagonal GKP code is not a balanced code because its stabilizer group is not invariant under the cyclic transformation in Eq.~\ref{eq:cyclic_transformation}. Nevertheless, Eq.~\ref{eq:distance_stabilizer_GKP} still holds and the surface-hexagonal GKP code has distance $3^{1/4}\sqrt{2\pi}$. 
As a comparison, we plot the distances of these four {concatenated GKP} codes in Fig.~\ref{fig:optimized_GKP_codes}(a).

{In addition to using the code distance,} we can {also} quantify the error correction capability of a GKP code by calculating its fidelity, subject to independent and identically distributed Gaussian shift errors, as we assume in Eq.~\ref{eq:def_shift_errors}.
More specifically, we use the Monte Carlo method by sampling $10^6$ random shifts from the  Gaussian distribution $\mathcal{N}(0, \sigma^2)$, followed by using the closest point decoder {and Eq.~\ref{eq:pauli_channel}} to determine the probability that the logical information is preserved. 
The number of samples is determined such that statistical fluctuations are negligible.
In Fig.~\ref{fig:optimized_GKP_codes}(b), we show the fidelities of the four {concatenated} GKP codes discussed above with noise strength {$\sigma\approx0.5143$}. 
We notice that the fidelity of the $d_0=3$ surface-hexagonal code is similar to that of [[5,1,3]]-hexagonal or [[7,1,3]]-hexagonal codes, which are worse than the [[5,1,3]]-$D_4$ code. 
It is possible to improve the fidelity of the surface-GKP code by considering larger distances such as $d_0=5$, but the runtime of finding the closest point increases exponentially with $d_0^2$ {(i.e., the number of modes in the surface-GKP code) since we are using a general-purpose closest point decoder here}. 
This poses serious challenge to decode a single syndrome for surface-GKP codes of large distances, not mentioning that one has to repeat for $10^6$ samples in order to estimate its fidelity.
In Sec.~\ref{sec:Polynomial time closest point decoder for surface-GKP code}, we will devise {an exact and} polynomial time closest point decoder that is tailored to decode surface-GKP codes much more efficiently. With that, we will benchmark the fidelity of surface-GKP codes with larger distances and different noise strengths.

\subsection{Generalizations of the tesseract {and $D_{4}$} codes}
\label{sec:generalized_codes_intro}

\begin{figure}
\centering
\includegraphics[width=\linewidth]{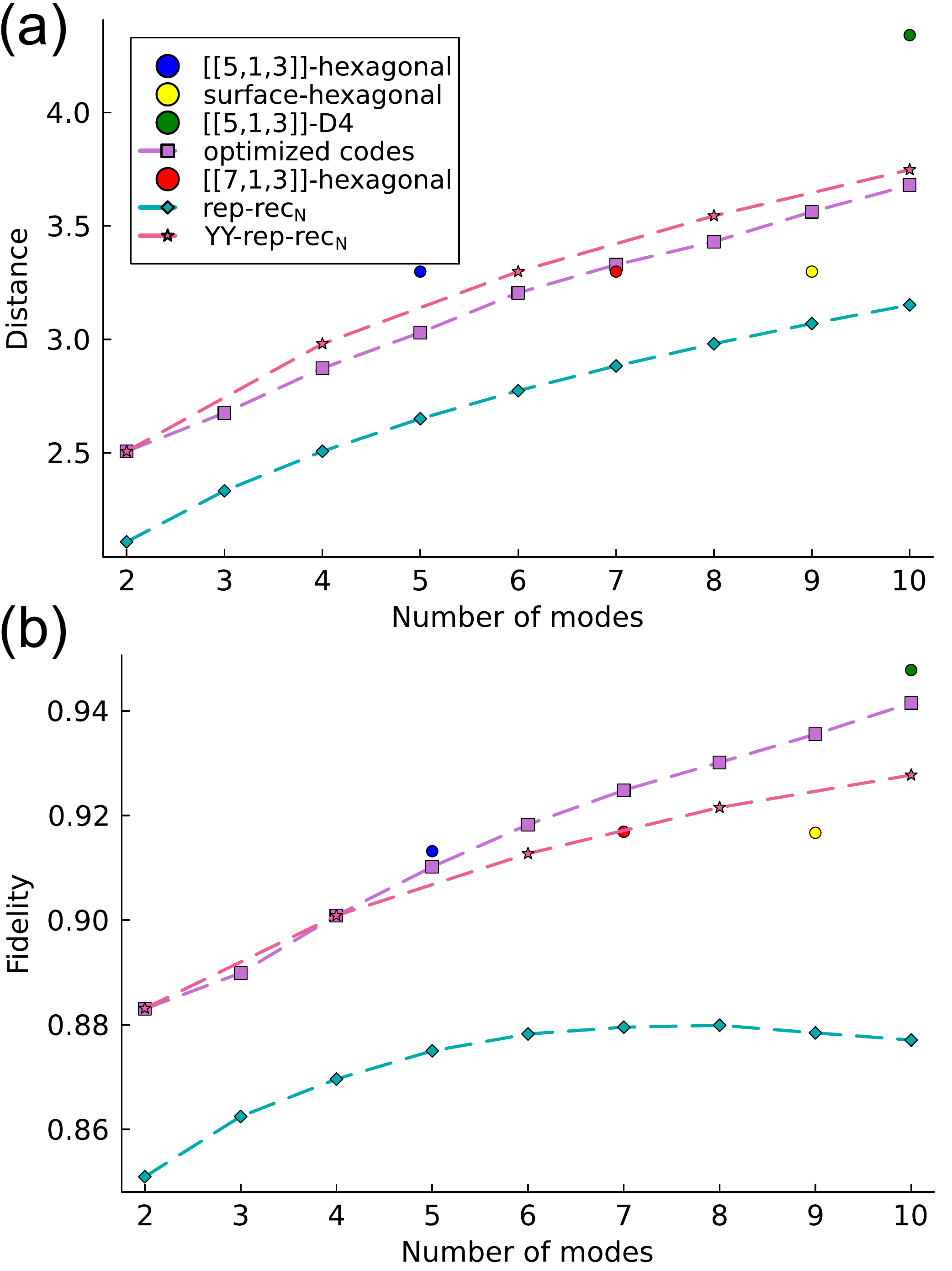}
 \caption{
 (a) The distances for the GKP codes discussed in Sec.~\ref{sec:Searching for optimized GKP codes}, as a function of number of modes. The circles indicate the {concatenated GKP} codes introduced in Sec.~\ref{sec:known concatenated GKP codes}, green diamonds and red stars are for $\text{rep-rec}_N$ and $\text{YY-rep-rec}_N$ respectively, and the {purple} squares are the numerically optimized codes. The dash lines are guides of eyes for the corresponding families of GKP codes. 
 (b) The fidelites for the same set of GKP codes (indicated with the same legends). Each data point is obtained by sampling $10^6$ random shift errors from the Gaussian distribution $\mathcal{N}(0, \sigma^2)$ with {$\sigma\approx0.5143$}. 
We emphasize that the numerically optimized codes are not \emph{optimal}, see the discussion in the main text.
}
 \label{fig:optimized_GKP_codes}
\end{figure}

In Sec.~\ref{sec:Lattice examples}, we demonstrated {two lattices}, the tesseract lattice and the $D_{2N}$ lattice, and show that both of them can be used to encode a logical qubit. For the case with two modes $(N=2)$, the $D_4$ lattice outperforms the tesseract lattice because $d^{D_4}=2^{1/4}d^\text{tess}$.
Despite that, the distance for the $D_{2N}$ code with $N>2$ is the same as the $D_4$ code because $d_Y^{D_{2N}}$ is fixed to be $\sqrt{2\pi}$ {independent of $N$}. 
Interestingly, one could generalize the tesseract lattice to higher dimensions with larger code distances than the $D_{2N}$ code.
In particular, since the tesseract lattice corresponds to a $2$-qubit repetition code, we consider the concatenation of the $N$-qubit XX repetition code with the rectangular GKP code in Eq.~\ref{eq:2D_rec} with $\eta=N^{1/4}$.
We shall denote the resulting code as rep-rec$_{N}$ which corresponds to a $2N$-dimensional lattice.
The distance of the $\text{rep-rec}_N$ code can be understood in the following way. Since the XX stabilizers cannot detect the logical $\bar{X}$ errors, the distance $d^{\text{rep-rec}_N}_X=N^\frac{1}{4} \sqrt{\pi}$, which is {the X distance of} the rectangular code when $\eta=N^{1/4}$. 
For the $\bar{Z}$ operator, it corresponds to a concatenation of $N$ copies of $\frac{1}{\sqrt{2}}(0, N^{-1/4})^T$, the {shift vector corresponding to} the $\bar{Z}$ operator of the rectangular code. 
Because the distance $d^{\text{rep-rec}_N}_Z$ is the length of the $\bar{Z}$ operator multiplied by $\sqrt{2\pi}$, we arrive at the distances for the $\text{rep-rec}_N$ code
\begin{align}
    \label{eq:def_distances_rep_rec_N}
    d^{\text{rep-rec}_N}_X = d^{\text{rep-rec}_N}_Z = N^\frac{1}{4} \sqrt{\pi}, \quad d^{\text{rep-rec}_N}_Y =N^\frac{1}{4} \sqrt{2\pi}.
\end{align}
Here we used the fact that $d^{\text{rep-rec}_N}_Y=\sqrt{(d^{\text{rep-rec}_N}_X)^2+(d^{\text{rep-rec}_N}_Z)^2}$ because the vectors for the logical $\bar{X}$ and $\bar{Z}$ operators are orthogonal to each other. 
From Eq.~\ref{eq:def_distances_rep_rec_N}, we have $d^{\text{rep-rec}_N} = N^\frac{1}{4} \sqrt{\pi}$ which is indeed larger than $\sqrt{2\pi}$, the code distance of the $D_{2N}$ code, for $N>4$.

The rep-rec$_{N}$ code is not a balanced GKP code because $d^{\text{rep-rec}_N}_{X,Z}$ are always smaller than $d^{\text{rep-rec}_N}_Y$. In order to balance the distances of different logical operators, and increase the code distance further, we consider concatenating two copies of the $\text{rep-rec}_N$ codes with the YY stabilizer. 
We shall denote the resulting code as $\text{YY-rep-rec}_N$, where $N$ is an even number.
Because the YY stabilizer detects both the logical $\bar{X}$ and $\bar{Z}$ errors, but not the logical $\bar{Y}$ error, the distances for the logical $\bar{X}$ and $\bar{Z}$ are enhanced such that
\begin{align}
    d^{\text{YY-rep-rec}_N}_X = d^{\text{YY-rep-rec}_N}_Y = d^{\text{YY-rep-rec}_N}_Z =N^\frac{1}{4} \sqrt{2\pi}.
\end{align}
The distance for the $\text{rep-rec}_N$ and the $\text{YY-rep-rec}_N$ codes are confirmed via explicitly solving Eq.~\ref{eq:def_distances}, and the results are shown in Fig.~\ref{fig:optimized_GKP_codes}(a). The green and red solid lines indicate their respective scalings with respect to the number of modes. 
We note that $\text{YY-rep-rec}_N$ code always has even number of modes and its distance is always larger than the $\text{rep-rec}_N$ code with the same number of modes, as expected.

We calculate the fidelities of the two codes with the same Monte Carlo method described in Sec.~\ref{sec:known concatenated GKP codes}, and the results are shown in Fig.~\ref{fig:optimized_GKP_codes}(b).
For {$\sigma\approx0.5143$}, the fidelity of the six-mode $\text{YY-rep-rec}_N$ code is comparable with [[7,1,3]]-hexagonal code, and could improve beyond that of the $d_0=3$ surface-hexagonal code with increasing number of modes; on the other hand, the fidelity of the $\text{rep-rec}_N$ code saturates at around $0.82$. 
We emphasize that the $\text{rep-rec}_N$ and $\text{YY-rep-rec}_N$ codes are different from the biased GKP repetition code introduced in Ref.~\cite{stafford2022biased} which exhibits a threshold of $\sigma^*\approx0.599$. 
An $N$-mode biased GKP repetition code is constructed by concatenating $N$ one-mode rectangular GKP codes with an $N$-qubit repetition code, with the aspect ratio of the inner GKP code optimized for a given set of $N$ and $\sigma$. 
Although it is similar to our construction, the aspect ratio for the inner GKP codes of both $\text{rep-rec}_N$ and $\text{YY-rep-rec}_N$ codes are fixed to be $\eta^2=N^\frac{1}{2}$. 
We choose to fix the aspect ratio for the $\text{rep-rec}_N$ code because it is a natural generalization for the two-mode tesseract code, which has aspect ratio $\sqrt{2}$. Hence from the lattice perspective, the lattices for these two GKP codes can be viewed as higher dimensional generalizations of the tesseract lattice.

In Sec.~\ref{sec:generalized_codes}, we will demonstrate that {the $\text{rep-rec}_N$ and $\text{YY-rep-rec}_N$ codes} can be decoded in runtime that is linear to the number of modes. We also perform more detailed fidelity analysis for larger instances of these two families of codes. Our result show that these two families of codes do not exhibit thresholds, and increasing the number of modes do not necessarily improve the fidelity for the range of noise studied, which is in contrast to the surface-GKP code or the biased GKP repetition code in Ref.~\cite{stafford2022biased}. This is evident for the $\text{rep-rec}_N$ code, as shown in Fig.~\ref{fig:optimized_GKP_codes}(b), and indicates that the threshold of the $\text{rep-rec}_N$ code, if any, is tied to the optimization of the biasing for the inner GKP code.

\subsection{Numerical search for optimized GKP codes}
\label{sec:Search for optimized GKP codes}

We have studied several families of GKP codes, which can all be understood as a concatenation of a certain qubit stabilizer code with a base GKP code.
The list of {concatenated GKP} codes can grow further by including, for example, Shor's nine qubit code \cite{shor1995scheme} or Bacon-Shor code \cite{bacon2006operator}, which can be similarly analyzed from a lattice perspective.
Besides the {concatenated} GKP codes, however, viewing GKP codes as lattices allows us to numerically search for optimized GKP codes with good metrics, such as code distance.

Recall from Sec.~\ref{sec:Transformation between GKP codes} that an arbitrary GKP code can be understood as a code that results from applying a Gaussian unitary operator to the square lattice GKP code. The resultant generator matrix reads
\begin{align}
\label{eq:lattice_transformation_2}
    M = M_\text{sq}(\boldsymbol{d})S^T,
\end{align}
where $M_\text{sq}(\boldsymbol{d})$ is defined in Eq.~\ref{eq:def_M_sq}, and $S$ is the $(2N)\times(2N)$ symplectic matrix for the Gaussian unitary. Here we focus on GKP codes encoding a single qubit {in $N$ modes (i.e., $d_1=2$ and $d_2=...=d_N=1$)}, and aim to optimize the symplectic matrix such that the resultant code has as large {code} distance as possible.
For this purpose, we consider the Bloch-Messiah decomposition for a general symplectic matrix $S\in\mathbb{R}^{2N\times2N}$ \cite{braunstein2005squeezing}
\begin{align}
\label{eq:BM}
    S = O_1ZO_2,
\end{align}
where {$O_1$ and $O_2$} are orthogonal symplectic matrices, and $Z=\text{diag}(e^{-r_1},e^{r_1}, ..., e^{-r_N},e^{r_N})$ with real parameters $(r_1,...,r_N)$. 
Here the diagonal matrix $Z$ represents a set of one-mode  squeezing operations, and {$O_1$ and $O_2$} correspond to the unitaries that preserve the total excitations in all the modes, such as beam-splitting.
This can be seen by noticing that the total number of excitation $\hat{n}\equiv\sum_{j=1}^N\hat{a}^\dagger_j\hat{a}_j=(\hat{\boldsymbol{x}}\cdot\hat{\boldsymbol{x}}-N)/2$, and orthogonal matrices preserve the Euclidean length in $\mathbb{R}^{2N}$. 
Further, {in the \texttt{qqpp} ordering,} a $2N\times2N$ orthogonal symplectic matrix can be written as a matrix exponential \cite{fulton2013representation}
\begin{align}
\label{eq:O_qqpp}
    {O^{\texttt{qqpp}}} = \exp \begin{bmatrix}
    X & Y \\
    -Y & X
    \end{bmatrix},
\end{align}
where $Y^T=Y$ is an $N\times N$ real symmetric matrix and $X=-X^T$ is an $N\times N$ real anti-symmetric matrix. 

Upon combining Eq.~\ref{eq:lattice_transformation_2} and \ref{eq:BM}, we see that a generic GKP code has generator matrix of the form $M = M_\text{sq}O_2^TZO_1^T$. But since $O_1$ is an orthogonal matrix which only rotates the basis vectors, the GKP code is equivalent to the lattice generated by 
\begin{align}
    M = M_\text{sq}O_2^TZ, 
\end{align}
{assuming the underlying noise model is isotropic (i.e., invariant under rotation).} Hence for a general $N$-mode GKP code, it can be parameterized by $N^2+N$ real parameters, $N$ for the squeezing parameters and the rest for the orthogonal symplectic matrices. 
We will optimize over this set of parameters, for a given number of mode, to find {good} GKP codes.

In Fig.~\ref{fig:optimized_GKP_codes}(a), we show the distances for the numerically optimized GKP codes ({purple} squares) as a function of number of modes $N$. 
Each data point is obtained by initializing {$10^4$} random symplectic matrices, and performing gradient descent with respect to the negative distance, followed by selecting the code with the largest distance. 
We further apply the same Monte-Carlo method, as described in Sec.~\ref{sec:known concatenated GKP codes}, to calculate the fidelity of the numerically optimized codes, as shown in Fig.~\ref{fig:optimized_GKP_codes}(b).
We remark that one could use the fidelity as the cost function for searching the optimized code. However, this requires one to perform Monte-Carlo sampling at each iteration step of the optimization. Since finding the closest point for a general lattice incurs significant time overhead, particularly for large lattices, this approach is inefficient in practice.
We believe distance is a reasonable indicator for the error correction capability of a GKP code, particularly for low noise regime.

For the case of two modes, the optimizer finds the $\text{rep-rec}_N$ code with $N=2$, which is equivalent to the $D_4$ code. The $D_4$ lattice has been shown to be the best quantizer in the context of classical error correction because it supports the densest sphere packing in four dimensions \cite{conway2013sphere}. Hence we believe that the optimizer has found the optimal GKP code for the case of two modes. 
However, we emphasize that the found optimized codes need not be optimal for $N>2$. Particularly, it is clear that {certain optimized code with $N\geq3$ has distance shorter than that of the $\text{YY-rep-rec}_N$ code}. This may be attributed to the fact that only {$10^4$} random ansatzs have been used in the search, and we expect that the distance of the optimized code will get closer or even beyond that of the $\text{YY-rep-rec}_N$ code if more random initial points are used.
Surprisingly, the found optimized codes in fact have comparable or better fidelities compared to the $\text{YY-rep-rec}_N$ code. 
More interestingly, for the case with nine modes, the optimizer finds a GKP code that has better distance and fidelity than the $d_0=3$ surface-hexagonal GKP code. 
{
Similarly, the optimized three-mode and seven-mode codes outperform the $\text{rep-rec}_N$ code and [[7,1,3]]-hexagonal code respectively, in both the distance and fidelity metrics. 
We have shown the generator matrices for the optimized codes with $N=3$, $N=7$ and $N=9$ in App.~\ref{app: Generators_3_9}.
}
Unfortunately, we have not been able to understand the structure of these numerically optimized codes, which will be left for future works.

{
For optimized codes with even number of modes, we notice that their distances are generally smaller than those for the $\text{YY-rep-rec}_N$ codes. On the other hand, we remark that, the fidelity of the optimized codes are generally better than that of the $\text{YY-rep-rec}_N$ codes for larger number of modes. This is evident in Fig.~\ref{fig:optimized_GKP_codes}(b) where we compare their fidelity at $\sigma\approx0.5143$. In particular, we notice that despite the optimized code with $N=4$ has smaller distance compared to the $\text{YY-rep-rec}_N$ code with the same number of modes, they have almost the same fidelity. We leave the detailed study of this family of optimized codes to future works. 
}

In summary, we have shown that general GKP codes can be viewed as parameterized lattices, and in principle one could find good GKP codes via numerically optimizing their distances. In fact we found {three code instances with $N=3$, $N=7$, and $N=9$ which outperform the known concatenated GKP codes with the same number of modes}, in both the code distance and fidelity metrics.   
We have also illustrated two generalizations of the tesseract codes, namely the $\text{rep-rec}_N$ and the $\text{YY-rep-rec}_N$ code, which exhibit good code distances and fidelities. 
For the rest of the paper, we will switch gear and focus on efficient closest point decoders for these two codes as well as the surface-GKP code.

\section{Efficient closest point decoder for structured GKP codes}
\label{sec:Efficient closest point decoder for structured GKP codes}

In this section we describe several techniques to decode lattices with well defined structures \cite{van2016cryptographic, conway2013sphere}. 
For convenience, we first rephrase the closest point problem in Eq.~\ref{eq:def_CP} for a generic set of points: 
for any discrete set of points $\Sigma\subset \mathbb{R}^n$, find the closest point
\begin{align}
\label{eq:def_chi_t_S}
    \boldsymbol{\chi}_{\boldsymbol{t}}(\Sigma) = \argmin_{\boldsymbol{r}\in\Sigma}||\boldsymbol{t}-\boldsymbol{r}|| 
\end{align}
for a given target ${\boldsymbol{t}}\in\mathbb{R}^n$. In the case of tie, $\boldsymbol{\chi}_{\boldsymbol{t}}(\Sigma)$ is chosen arbitrarily from the closest points. 

\subsection{Decoding a discrete set of points}

We note that $\Sigma$ in Eq.~\ref{eq:def_chi_t_S} needs not be a lattice, and it needs not have regular patterns. Nevertheless, we can still decompose the set $\Sigma$ into several subsets, or apply a shift to all the points in the set. Particularly, for any discrete set $\Sigma\subset \mathbb{R}^n$, if $\Sigma$ can be decomposed into the union of $k$ discrete sets as $\Sigma=\cup_{i=1}^k \Sigma_i$, then we we have
\begin{align}
\label{eq:decode_set}
    \boldsymbol{\chi}_{\boldsymbol{t}}(\cup_{i=1}^k \Sigma_i) = \boldsymbol{\chi}_{\boldsymbol{t}}(\{\boldsymbol{\chi}_{\boldsymbol{t}}(\Sigma_i): i=1,...,k\}),
\end{align}
which suggests that we can find the closest points for each subset $\Sigma_i$, followed by comparing their distances to ${\boldsymbol{t}}$ and select the one with shortest distance. This indeed works because the closest point in $\Sigma$ must lie in some subsets $\Sigma_i$, which  is by definition as close or closer than the closest point from all the other subsets. 

Further, for any discrete set of points $\Sigma\subset \mathbb{R}^n$, we can obtain a new set of points by shifting all the points by a vector ${\boldsymbol{r}}$, denoted as ${\boldsymbol{r}}+\Sigma$. The closest point in the set of shifted points can be obtained as
\begin{align}
\label{eq:decode_shifted_lattice}
    \boldsymbol{\chi}_{\boldsymbol{t}}({\boldsymbol{r}}+\Sigma) = \boldsymbol{\chi}_{\boldsymbol{t-r}}(\Sigma) + {\boldsymbol{r}}.
\end{align}
To prove this, let ${\boldsymbol{g}}=\boldsymbol{\chi}_{\boldsymbol{t}}(\Sigma)$, if all the points are shifted by $\boldsymbol{r}$, then the closest point is also shifted by the same amount, i.e., $\boldsymbol{\chi}_{\boldsymbol{t}+\boldsymbol{r}}({\boldsymbol{r}}+\Sigma)=\boldsymbol{g+r}=\boldsymbol{\chi}_{\boldsymbol{t}}(\Sigma)+{\boldsymbol{r}}$. Redefine ${\boldsymbol{t}}' = {\boldsymbol{t}+\boldsymbol{r}}$, we have $\boldsymbol{\chi}_{\boldsymbol{t}'}({\boldsymbol{r}}+\Sigma)=\boldsymbol{\chi}_{\boldsymbol{t'-r}}(\Sigma)+{\boldsymbol{r}}$ as desired.

\subsection{Decoding direct sums of lattices}

Suppose we have a generator matrix $M$, which is a direct sum of several square matrices $M=\oplus_{i=1}^kM_i$, then the corresponding lattice is also a direct sum of several sublattices, 
\begin{align*}
    \Lambda(M)=\Lambda(\oplus_{i=1}^kM_i)=\oplus_{i=1}^k\Lambda(M_i)=\oplus_{i=1}^k\Lambda_i.    
\end{align*}
Such direct sum of lattices can be decoded by simply decoding each orthogonal projection of ${\boldsymbol{t}}$ onto the space spanned by each component lattice, followed by combining the results. Formally, we have
\begin{align}
\label{eq:decode_direct_sum}
    \boldsymbol{\chi}_{\boldsymbol{t}}(\oplus_{i=1}^k\Lambda_i) = \oplus_{i=1}^k\boldsymbol{\chi}_{\pi_i({\boldsymbol{t}})}(\Lambda_i),
\end{align}
where $\pi_i$ denotes the orthogonal projection onto the space spanned by $\Lambda_i$. In practice, we select the corresponding components ${\boldsymbol{t}}_i$ in ${\boldsymbol{t}}$ and decode it with $M_i$, followed by assembling the result together to arrive at $\boldsymbol{\chi}_{\boldsymbol{t}}(\Lambda)$.

\subsection{Decoding union of cosets}
\label{sec:Decoding union of cosets}
Consider a lattice $\Lambda$, and a set of vectors ${\boldsymbol{r}}_i$ $(i=0,...,l-1)$, we can construct a union of cosets of $\Lambda$ as
\begin{align}
\label{eq:union_cosets}
    \Sigma \equiv \bigcup_{i=0}^{l-1} ({\boldsymbol{r}}_i+\Lambda).
\end{align}
Here we fix ${\boldsymbol{r}}_0={\bf 0}$ such that $\Lambda\subset\Sigma$, and other coset vectors $\boldsymbol{r}_i$ are real valued vectors in $\mathbb{R}^n$. For a lattice point $\boldsymbol{u}\in\Lambda$, $\Sigma$ contains $\boldsymbol{u}$ together with its translations by all the coset vectors.
Upon combining Eq.~\ref{eq:decode_set}-\ref{eq:decode_shifted_lattice}, the union $\Sigma$ can be decoded as
\begin{eqnarray}
    \begin{aligned}
    \label{eq:decode_union_cosets}
    &\boldsymbol{\chi}_{\boldsymbol{t}}\left(\bigcup_{i=0}^{l-1} ({\boldsymbol{r}}_i+\Lambda)\right) \\
    =& \boldsymbol{\chi}_{\boldsymbol{t}}(\{{\boldsymbol{r}}_i+\boldsymbol{\chi}_{{\boldsymbol{t}}-{\boldsymbol{r}}_i}({\Lambda}): i=0,1,...,l-1\}).
    \end{aligned}
\end{eqnarray}
We note that despite $\Sigma$ needs not be a lattice, the Euclidean dual of a lattice can be treated as a union of cosets with respect to the original lattice,exactly as shown in Eq.~\ref{eq:union_cosets}. Hence, if a lattice can be decoded efficiently, its Euclidean dual can also be decoded efficiently with Eq.~\ref{eq:decode_union_cosets}, provided only a handful of cosets to be decoded.

As will be shown in App.~\ref{sec:The concatenated GKP code as a glue lattice}, {concatenated GKP} codes, such as surface-GKP code, can be viewed as a union of cosets, where the group elements in the stabilizer group play the role of coset vectors $\boldsymbol{r}_i$ in Eq.~\ref{eq:union_cosets}. 
However, because the size of its stabilizer group generally grows exponentially with the number of modes, and naive application of Eq.~\ref{eq:decode_union_cosets} will require exponentially many cosets to be decoded.
In Sec.~\ref{sec:Polynomial time closest point decoder for surface-GKP code}, we overcome this difficulty by combining Eq.~\ref{eq:decode_union_cosets} with an MWPM algorithm, which yields a polynomial time decoder for the surface-GKP code.

\subsection{Decoding glue lattices}
\label{ref:Decoding glue lattices}
A glue lattice can be regarded as a union of cosets for a direct sum of lattices
\begin{align}
    \label{eq:glue_lattice}
    \Lambda = \bigcup_{i=0}^{l-1} ({\boldsymbol{r}}_i+\oplus_{j=1}^m\Lambda_j).
\end{align}
Since ${\boldsymbol{r}}_0={\bf 0}$, the glue lattice has a sublattice which is a direct sum of $m$ lattices. In this context, the vectors ${\boldsymbol{r}}_i$ are called gluing vectors. Combining Eq.~\ref{eq:decode_direct_sum}, \ref{eq:decode_union_cosets}, we have the closest point for the glue lattice as
\begin{eqnarray}
    \begin{aligned}
    \label{eq:decode_glue_lattice}
    &\boldsymbol{\chi}_{\boldsymbol{t}}\left(\bigcup_{i=0}^{l-1} ({\boldsymbol{r}}_i+\oplus_{j=1}^m\Lambda_j)\right)\\
    =& \boldsymbol{\chi}_{\boldsymbol{t}}\left(\{{\boldsymbol{r}}_i+\boldsymbol{\chi}_{{\boldsymbol{t}}-{\boldsymbol{r}}_i}(\oplus_{j=1}^m\Lambda_j): i=0,1,...,l-1\}\right)\\
    =&\boldsymbol{\chi}_{\boldsymbol{t}}\left(\{{\boldsymbol{r}}_i+\oplus_{j=1}^m\boldsymbol{\chi}_{\pi_j({\boldsymbol{t}}-{\boldsymbol{r}}_i)}(\Lambda_j): i=0,1,...,l-1\}\right).
    \end{aligned}
\end{eqnarray}
Glue lattices encompass concatenated GKP codes, including those obtained through Construction A \cite{conrad2022gottesman}. In Sec.~\ref{sec:YY-rep-rec_N}, we will show that the $\text{YY-rep-rec}_N$ code can be viewed as a glue lattice, and Eq.~\ref{eq:decode_glue_lattice} plays a key role in decoding the code in linear time.

\section{Linear time decoder for $D_n$ lattices and their Euclidean duals}
\label{sec:Linear time decoder for Dn}

In this section, we will present the linear time decoders for $D_n$ lattices and their Euclidean duals, denoted as $D_n^*$. These two lattices will serve as building blocks for the more complex GKP lattices as shown in Sec.~\ref{sec:generalized_codes}.
As illustrated by Conway and Sloane in Ref.~\cite{conway1982fast}, the decoding of $D_n$ and $D_n^*$ turn out to be very straightforward, after we understand a bit deeper the simplest case, namely decoding the $Z_n$ lattices.

\subsection{Linear time decoder for $Z_n$ lattices}
\label{sec:decoder_Z^N}

The $n$-dimensional $Z_n$ lattice is an integer lattice with generator $M_{Z_n}=I_n$. The closest point in the $Z_n$ lattice for an arbitrary point $\boldsymbol{t}\in\mathbb{R}^n$ is given by
\begin{align}
\label{eq:def_chi_t_Zn}
    % f(\boldsymbol{x}) = (\lfloor x_1\rceil, ..., \lfloor x_n\rceil).
    \boldsymbol{\chi}_{\boldsymbol{t}}(Z_n) = (\lfloor t_1\rceil, ..., \lfloor t_n\rceil).
\end{align}
Here $\lfloor t\rceil$ denotes the closest integer to $t\in\mathbb{R}$, and in case of a tie, the integer with the smallest absolute value is chosen. This algorithm is presented as ClosestPointZn in Alg.~\ref{alg: ClosestPointZn}.

\begin{algorithm}
\caption{ClosestPointZn($\boldsymbol{t}$)}
\label{alg: ClosestPointZn}
{\bf Input: } The error syndrome $\boldsymbol{t}\in\mathbb{R}^{n}$; \\ 
{\bf Output: } The optimal integer $\boldsymbol{b}\in\mathbb{Z}^{n}$; \\ 
%  $\boldsymbol{b} \leftarrow f(\boldsymbol{s})$
$\boldsymbol{b} \leftarrow(\lfloor t_1\rceil, ..., \lfloor t_n\rceil)$
\end{algorithm}

For decoding the $D_n$ lattice, finding the nearest point for $Z_n$ is not sufficient, and we have to find the \emph{second} nearest point for $\boldsymbol{t}$. 
For that, we introduce the function $w(t)$, the second nearest integer to a real number $t$, 
\begin{align}
\label{def:w(t)}
    w(t) = 
    \begin{cases}
      \lfloor t\rceil+1 & \text{if } x\geq \lfloor t\rceil\\
      \lfloor t\rceil-1 & \text{if } x< \lfloor t\rceil
    \end{cases}.
\end{align}
In Ref.~\cite{conway1982fast}, $w(t)$ is called rounding $x$ \emph{the wrong way}, and is the key to find the {second} closest point in the $Z_n$ lattice for a given point $\boldsymbol{t}$. The idea is to find the component of $\boldsymbol{t}$, say $t_k$, which is the furthest from its closest integer, and round it the wrong way. Mathematically, let $\boldsymbol{\chi}'_{\boldsymbol{t}}(Z_n)$ denotes the second nearest point for the given $\boldsymbol{t}$, then we have
\begin{align}
\label{eq:def_chi_t_Zn_2}
    \boldsymbol{\chi}'_{\boldsymbol{t}}(Z_n) = (\lfloor t_1\rceil, ..., \lfloor t_{k-1}\rceil, w(x_k), \lfloor t_{k+1}\rceil, ..., \lfloor t_n\rceil), 
\end{align}
where % $k$ is determined such that 
\begin{align}
    k \equiv \argmax_{1\leq k\leq n} |t_k - \lfloor t_k\rceil|.
\end{align}
Alternatively, since $t_k$ is furthest from its nearest integer $\lfloor t_k\rceil$, then among all the components of $\boldsymbol{t}$, $t_k$ must be the closest to the second nearest integer $w(t_k)$. Hence we can also write $k \equiv \argmin_{1\leq k\leq n} |w(t_k)|$. $\boldsymbol{\chi}'_{\boldsymbol{t}}(Z_n)$ is indeed the second closest point to $\boldsymbol{t}$ as its norm is larger than $\boldsymbol{\chi}_{\boldsymbol{t}}(Z_n)$, and if we were to round the other component $t_{i\neq k}$ in the wrong way (and round $t_k$ the correct way), the resulting $\boldsymbol{\chi}''_{\boldsymbol{t}}(Z_n)$ will have a larger norm than $\boldsymbol{\chi}'_{\boldsymbol{t}}(Z_n)$ by the definition of $k$ above. 
We will now illustrate why the function $\boldsymbol{\chi}'_{\boldsymbol{t}}(Z_n)$ can help us to find the closest point in the $D_n$ lattice.

\subsection{Linear time decoder for $D_n$ lattices}
\label{sec:decoderDN}

In order to find the closest point in the $D_n$ lattice for a given $\boldsymbol{t}$, recall that $D_n$ is a sublattice of $Z_n$ where the sum of the components of any lattice point is always even. In order to identify $\boldsymbol{\chi}_{\boldsymbol{t}}(D_n)$, we first find the closest and second closest points, namely $\boldsymbol{\chi}_{\boldsymbol{t}}(Z_n)$ and $\boldsymbol{\chi}'_{\boldsymbol{t}}(Z_n)$, in the $Z_n$ lattice. We note that since the two only differ by one component, i.e., 
$$||\boldsymbol{\chi}_{\boldsymbol{t}}(Z_n)-\boldsymbol{\chi}'_{\boldsymbol{t}}(Z_n)||=1,$$
one and only one of them lies in the $D_n$ lattice. Hence, the closest point in $D_n$ lattice is whichever of $\boldsymbol{\chi}_{\boldsymbol{t}}(Z_n)$ and $\boldsymbol{\chi}'_{\boldsymbol{t}}(Z_n)$ that has the even sum of the components. Since both  $\boldsymbol{\chi}_{\boldsymbol{t}}(Z_n)$ and $ \boldsymbol{\chi}'_{\boldsymbol{t}}(Z_n)$ can be found with linear runtime, $D_n$ can be decoded with runtime linear to $n$. The decoder is presented as ClosestPointDn in Alg.~\ref{alg: ClosestPointDn}. In App.~\ref{app:decoder_scaled_Z^N_D^N}, we generalize this decoder to $D_n$ lattices with different lattice spacings in different directions.

\begin{algorithm}
\caption{ClosestPointDn($\boldsymbol{t}$)}
\label{alg: ClosestPointDn}
{\bf Input: } The error syndrome $\boldsymbol{t}\in\mathbb{R}^{n}$; \\ 
{\bf Output: } The optimal integer $\boldsymbol{b}\in\mathbb{Z}^{n}$; \\ 
 $\boldsymbol{b}_1 \leftarrow \boldsymbol{\chi}_{\boldsymbol{t}}(Z_n)$; ~ *Defined in Eq.~\ref{eq:def_chi_t_Zn}*\\
 $\boldsymbol{b}_2 \leftarrow \boldsymbol{\chi}'_{\boldsymbol{t}}(Z_n)$; ~ *Defined in Eq.~\ref{eq:def_chi_t_Zn_2}*\\
 \If {\text{sum($\boldsymbol{b}_1$) is even}}{
    $\boldsymbol{b} \leftarrow \boldsymbol{b}_1$
    \Else{
    $\boldsymbol{b} \leftarrow \boldsymbol{b}_2$
    }
}
\end{algorithm}

\subsection{Linear time decoder for $D_n^*$ lattices}
An Euclidean dual lattice $\Lambda^*$ is defined to be the set of all the points that have integer inner product with all the points in the original $\Lambda$. In other words,
$$\Lambda^* \equiv \left\{\boldsymbol{u} ~|~ \boldsymbol{u}^T\boldsymbol{v}\in\mathbb{Z}, ~ \forall \boldsymbol{v}\in\Lambda\right\}.$$
The Euclidean dual lattice can be generated by $M^*=(M^T)^{-1}$ \cite{royer2022encoding}, which can be seen by noticing $(M^T{\bf b})^T((M^*)^T{\bf a})={\bf b}^T(M(M^*)^T){\bf a}\in Z$ for arbitrary integer vectors ${\bf a}$ and ${\bf b}$. 
Here we present the linear decoder for the $D_n^*$ lattices, the Euclidean dual of the $D_n$ lattices. 
It turns out that the $D_n^*$ lattice is the union of two cosets of $Z_n$\cite{conway1982fast, conway2013sphere},
\begin{align}
\label{eq:D_n_dual}
    D_n^* = \bigcup_{i=0}^1({\boldsymbol{r}}_i+Z_n),
\end{align}
where the $n$-components vectors ${\boldsymbol{r}}_i$ are defined as
\begin{align}
\label{eq:D_n_dual_cosets}
    {\boldsymbol{r}}_0 = (0,...,0)^T,\quad
    {\boldsymbol{r}}_1 = \left(\frac{1}{2},...,\frac{1}{2}\right)^T.
\end{align}
This can be seen by noticing that $\boldsymbol{r}_1^T\boldsymbol{u}\in\mathbb{Z}$ for all $\boldsymbol{u}\in D_n$, and $\text{det}(M_{D_n^*})=\frac{1}{2}=\text{det}(M_{D_n})^{-1}$.
Hence with Eq.~\ref{eq:decode_union_cosets}, we have
\begin{align}
    \boldsymbol{\chi}_{\boldsymbol{t}}(D_n^*) = \boldsymbol{\chi}_{\boldsymbol{t}}(\{{\boldsymbol{r}}_i+\boldsymbol{\chi}_{{\boldsymbol{t}}-{\boldsymbol{r}}_i}({Z_n}): i=0,1\}),
\end{align}
which can be found in runtime proportional to $n$. In particular, this suggests to decode $Z_n$ twice with ${\boldsymbol{t}}$ and ${\boldsymbol{t}}-{\boldsymbol{r}}_1$ respectively, followed by picking the one with closest distance to ${\boldsymbol{t}}$. The decoder for $D_n^*$ is presented in Alg.~\ref{alg: ClosestPointDnDual}.

\begin{algorithm}
\caption{ClosestPointDnDual($\boldsymbol{t}$)}
\label{alg: ClosestPointDnDual}
{\bf Input: } The error syndrome $\boldsymbol{t}\in\mathbb{R}^{2N}$; \\ 
{\bf Output: } The optimal integer $\boldsymbol{b}\in\mathbb{Z}^{2N}$; \\ 
$\boldsymbol{b}_1 \leftarrow $ClosestPointZn($\boldsymbol{t}$);\\
$\boldsymbol{b}_2 \leftarrow $ClosestPointZn($\boldsymbol{t}-\boldsymbol{r}_1$);\\
\eIf{\text{sum($\boldsymbol{b}_1-\boldsymbol{t}$) $\leq$ sum($\boldsymbol{b}_2-\boldsymbol{t}$)}}{
   $\boldsymbol{b} \leftarrow \boldsymbol{b}_1$;
   }{
   $\boldsymbol{b} \leftarrow \boldsymbol{b}_2$;
  }
\end{algorithm}

\section{Linear time decoders for the $\text{rep-rec}_N$ and $\text{YY-rep-rec}_N$ codes}
\label{sec:generalized_codes}

In this section, we present the linear time decoder for the $\text{rep-rec}_N$ and $\text{YY-rep-rec}_N$ codes introduced in Sec.~\ref{sec:generalized_codes_intro}, which are based on the strategies in Sec.~\ref{sec:Efficient closest point decoder for structured GKP codes}-\ref{sec:Linear time decoder for Dn}. We will use these decoders to benchmark the fidelities of these two codes for different number of modes and noise strengths.

\subsection{Linear time decoder for the $\text{rep-rec}_N$ code}

\begin{figure}
\centering
\includegraphics[width=\linewidth]{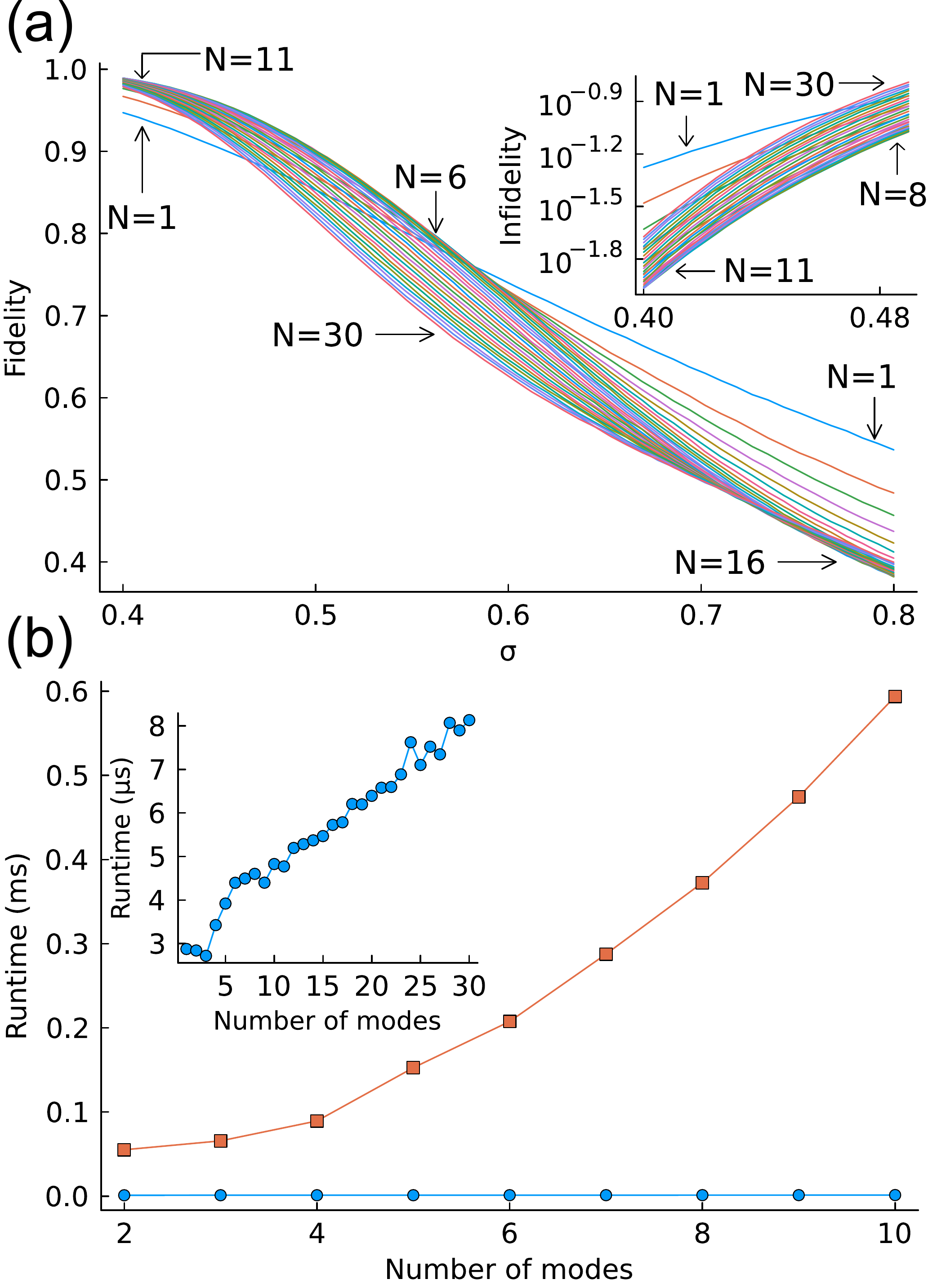}
 \caption{
 The numerical results for the $\text{rep-rec}_N$ code.
(a) The fidelity of the $\text{rep-rec}_N$ code as a function of the number of modes and noise strength $\sigma$. Each line corresponds to a given number of modes, which varies from  $N=1$ to $N=30$. {For $\sigma=0.4, 0.5714, 0.8$, we have indicated the number of modes that support minimum and maximum fidelities. The top-right inset shows the infidelities between $\sigma=0.4$ and $0.4898$ for different number of modes in the log scale. The number of modes that support minimum and maximum infidelities are indicated for $\sigma=0.4$ and $0.4898$ respectively.}
(b) The comparison of runtimes for the exponential time closest point decoder (square) and linear time closest point decoder (circle) for increasing number of modes. 
{For each data point, we average over all the samples for all the values of $\sigma$ considered. }
The inset shows the runtimes of the linear time decoder for different number of modes, up to $N=30$.
}
 \label{fig:rep-rec_N}
\end{figure}

Recall that the $\text{rep-rec}_N$ code is the concatenation of $N$-mode repetition code with the rectangular GKP code, defined in Eq.~\ref{eq:2D_rec}, with $\eta=N^{1/4}$.
The generator matrix can be written as
\begin{align}
\label{eq:def-rep-rec_N}
    M_{\text{rep-rec$_{N}$}} = M_\text{rep}^\text{(sq)}(S^T_\text{rec})^{\oplus N},
\end{align}
where $M_\text{rep}^\text{(sq)}$ is the repetition code concatenated with the square GKP code, and 
\begin{align}
\label{eq:rec_gkp}
    S_\text{rec} =
    \begin{bmatrix}
    N^{1/4} & 0\\
    0 & N^{-1/4}
    \end{bmatrix}.
\end{align}
For example, the generator matrix of the three-mode code reads
\begin{align}
    M_{\text{rep-rec}_3} = 
    \frac{1}{\sqrt{2}}
    \begin{bmatrix}
    1 & 0 & 1 & 0 & 0 & 0\\
    0 & 2 & 0 & 0 & 0 & 0\\
    0 & 0 & 1 & 0 & 1 & 0\\
    0 & 0 & 0 & 2 & 0 & 0\\
    0 & 0 & 0 & 0 & 2 & 0\\
    0 & 0 & 0 & 0 & 0 & 2
    \end{bmatrix}
    \begin{bmatrix}
    3 & 0 & 0 & 0 & 0 & 0\\
    0 & \frac{1}{3} & 0 & 0 & 0& 0\\
    0 & 0 & 3 & 0 & 0& 0\\
    0 & 0 & 0 & \frac{1}{3} & 0& 0\\
    0 & 0 & 0 & 0 & 3& 0\\
    0 & 0 & 0 & 0 & 0& \frac{1}{3}\\
    \end{bmatrix}^\frac{1}{4}.
\end{align}
We see that the $M_{\text{rep-rec}_3}$ takes a form of direct sum of two block matrices. To make this more explicit, we switch to the \texttt{qqpp} convention, and the generator takes a block diagonal form
\begin{align*}
    M^\text{(\texttt{qqpp})}_{\text{rep-rec}_3} = \frac{1}{\sqrt{2}}
    \begin{bmatrix}
    1 & 1 & 0 & 0 & 0 & 0\\
    0 & 1 & 1 & 0 & 0 & 0\\
    0 & 0 & 2 & 0 & 0 & 0\\
    0 & 0 & 0 & 2 & 0 & 0\\
    0 & 0 & 0 & 0 & 2 & 0\\
    0 & 0 & 0 & 0 & 0 & 2
    \end{bmatrix}
    \begin{bmatrix}
    3 & 0 & 0 & 0 & 0 & 0\\
    0 & 3 & 0 & 0 & 0 & 0\\
    0 & 0 & 3 & 0 & 0 & 0\\
    0 & 0 & 0 & \frac{1}{3} & 0 & 0\\
    0 & 0 & 0 & 0 & \frac{1}{3} & 0\\
    0 & 0 & 0 & 0 & 0 & \frac{1}{3}
    \end{bmatrix}^\frac{1}{4}.
\end{align*}
We can notice that for the $\hat{p}_i$ subspace, the generator matrix is an identity matrix multiplied by a factor of $\frac{\sqrt{2}}{3^{1/4}}$, hence it generates a scaled $Z_N$ lattice. Similarly for the $\hat{q}_i$ subspace, the sum of the components of the basis vectors are all even numbers multiplied by a factor of $\frac{3^{1/4}}{\sqrt{2}}$, which generates a scaled $D_n$ lattice.
More generally, with the \texttt{qqpp} ordering, the generator for the $\text{rep-rec}_N$ code reads
\begin{align}
\label{eq:generator_rep-rec}
    M^\text{(\texttt{qqpp})}_{\text{rep-rec}_N} = \left(\frac{N^{1/4}}{\sqrt{2}}M_\text{$D_N$}\right)\oplus\left(\frac{\sqrt{2}}{N^{1/4}}M_\text{$Z_N$}\right),
\end{align}
which is a direct sum of scaled $D_N$ and $Z_N$ lattice.

To decode the $\text{rep-rec}_N$ code, we will need to consider its logical operators which turns out to be the direct sum of the Euclidean duals of the $Z_N$ and $D_N$ lattices.
To see that, recall that in the \texttt{qqpp} convention, the symplectic form is defined in Eq.~\ref{eq:Omega_CSS}
such that we have
\begin{eqnarray}
\begin{aligned}
\label{eq:M_rep_rec_perp}
    M_{\text{rep-rec}_N}^{\text{(\texttt{qqpp})}\perp} &= \Omega^\text{(\texttt{qqpp})} ((M^\text{(\texttt{qqpp})}_{\text{rep-rec}_N})^{T})^{-1}(\Omega^\text{(\texttt{qqpp})})^{-1}\\
    &= 
    \begin{bmatrix}
    \frac{N^{1/4}}{\sqrt{2}}(M_\text{$Z_N$}^{T})^{-1} & 0_N\\
    0_N & \frac{\sqrt{2}}{N^{1/4}}(M_\text{$D_N$}^T)^{-1}
    \end{bmatrix} \\
    &=  \left(\frac{N^{1/4}}{\sqrt{2}}M_{Z_N^*}\right) \oplus \left(\frac{\sqrt{2}}{N^{1/4}}M_{D_N^*}\right) \\
    &=\left(\frac{N^{1/4}}{\sqrt{2}}M_{Z_N}\right) \oplus \left(\frac{\sqrt{2}}{N^{1/4}}M_{D_N^*}\right).
\end{aligned}
\end{eqnarray}
Here $Z_N^*=Z_N$ and $D_N^*$ denote the Euclidean dual lattices for $Z_N$ and $D_N$ respectively, and both lattices can be decoded in linear time as demonstrated in Sec.~\ref{sec:Efficient closest point decoder for structured GKP codes}. 
With Eq.~\ref{eq:decode_direct_sum}, the closest point for $\Lambda(M_{\text{rep-rec}_N}^{\text{(\texttt{qqpp})}\perp})$ is simply the assembly of those from the $Z_N$ and $D^*_N$ lattices, and hence the $\text{rep-rec}_N$ code can be decoded in runtime proportional to $2N$. We present the closest point decoder for the $\text{rep-rec}_N$ code in Alg.~\ref{alg: ClosestPointRepRecN}.

\begin{algorithm}
\caption{DecodeRepRecN($\boldsymbol{t}$)}
\label{alg: ClosestPointRepRecN}
{\bf Input: } The error syndrome $\boldsymbol{t}\in\mathbb{R}^{2N}$; \\ 
{\bf Output: } The optimal integer $\boldsymbol{b}\in\mathbb{Z}^{2N}$; \\ 
 $\boldsymbol{t}_q \leftarrow \boldsymbol{t}[1:2:end]$; \\
 $\boldsymbol{t}_p \leftarrow \boldsymbol{t}[2:2:end]$; \\
 $\boldsymbol{b}_q \leftarrow $ClosestPointZn($\boldsymbol{t}_q$); \\
 $\boldsymbol{b}_p \leftarrow $ClosestPointDnDual($\boldsymbol{t}_p$); \\
 $\boldsymbol{b}[1:2:end]\leftarrow\boldsymbol{b}_q$;\\
 $\boldsymbol{b}[2:2:end]\leftarrow\boldsymbol{b}_p$
\end{algorithm}

In order to characterize the error correction capability of the code, we calculate the fidelity of the $\text{rep-rec}_N$ code with the same Monte Carlo method and Gaussian noise distribution $\mathcal{N}(0, \sigma^2)$, as described in Sec.~\ref{sec:known concatenated GKP codes}.
In Fig.~\ref{fig:rep-rec_N}(a), we show the fidelity of the $\text{rep-rec}_N$ code as a function of the noise strength $\sigma$ and the number of modes, up to $N=30$. One immediately notices a band-like feature, which indicates that increasing the number of modes needs not improve the fidelity for the $\text{rep-rec}_N$ code. In particular, for the low noise regime with $\sigma=0.4$, we see that upon increasing the number of modes, the fidelity reaches maximum at $N=11$, beyond which the fidelity starts to decrease. This is confirmed in the top-right inset, where we show the infidelity between $\sigma=0.4$ and $0.4898$. Upon increasing the noise strength, we see that the code with $N=6$ has the highest fidelity at $\sigma=0.5714$, and the code with $N=30$, the largest number of modes studied here, has the lowest fidelity. 
This is similar to what we found in Fig.~\ref{fig:optimized_GKP_codes} that the fidelity of the $\text{rep-rec}_N$ code attains the maximum value of $0.82$ for $N=7$ at {$\sigma\approx0.5143$}.
For Fig.~\ref{fig:rep-rec_N}(a), at the high noise regime with $\sigma=0.8$, we find that one-mode $\text{rep-rec}_N$ code outperforms other codes as expected, i.e., the noise rate is high enough and thus increasing the number of modes only degrades the fidelity. 
Our result indicates that the $\text{rep-rec}_N$ code does not exhibit noise threshold below which increasing the number of modes will consistently improve its error correction capability.
This is in sharp contrast to the biased GKP repetition code introduced in Ref.~\cite{stafford2022biased}, and as discussed in Sec.~\ref{sec:generalized_codes_intro}, the difference can be attributed to the fact that we have fixed the aspect ratio of the inner rectangular GKP code to be $\eta^2=N^\frac{1}{2}$.
For biased GKP repetition code with generic aspect ratio, its generator matrix in the \texttt{qqpp} ordering can be obtained by substituting $\eta$ for $N^{1/4}$ in Eq.~\ref{eq:generator_rep-rec}, hence it can also be decoded with our closest point decoder. 

In Fig.~\ref{fig:rep-rec_N}(b), we compare the runtime of the linear-time decoder to the exponential-time closest point decoder, presented in Sec.~\ref{sec:Closest point search problem}, where for each data point, we average over all the samples for all the values of $\sigma$ considered.
The time overhead for the exponential-time decoder is increasing rather rapidly compared to the linear time decoder, as expected, and the difference is around five orders of magnitude for the case of ten modes. In the inset, we confirm that the runtime of the linear time decoder increases linearly with the number of modes.

There is an important remark before we proceed. 
Recall from Eq.~\ref{eq:closest_point_problem} that for both exponential-time and linear-time decoders, we are finding the closest lattice point in the symplectic dual lattice  for a given $\boldsymbol{\eta}(\boldsymbol{s})$. where $\boldsymbol{s}$ is a syndrome measurement result. From its definition in Eq.~\ref{eq:def_eta_s}, $\boldsymbol{\eta}(\boldsymbol{s})$ can be obtained from the syndrome $\boldsymbol{s}$ in linear time if $M^\perp$ is a sparse matrix with small number of nonzero entries in each column. This is indeed the case for the $\text{rep-rec}_N$ code as shown in Eq.~\ref{eq:M_rep_rec_perp}, hence it guarantees that the calculation of $\boldsymbol{\eta}(\boldsymbol{s})$ would not incur additional overhead for decoding the $\text{rep-rec}_N$ code. The similar conclusion holds for the multimode GKP codes discussed below, including the $\text{YY-rep-rec}_N$ code and surface-GKP codes.

\subsection{Linear time decoder for the $\text{YY-rep-rec}_N$ code}
\label{sec:YY-rep-rec_N}

Recall that for the $\text{rep-rec}_N$ code is not a balanced code as it has $d^{\text{rep-rec}_N}_X=d^{\text{rep-rec}_N}_Z=\sqrt{\pi}N^{1/4}$, which are always smaller than $d^{\text{rep-rec}_N}_Y=\sqrt{2\pi}N^{1/4}$. 
In Sec.~\ref{sec:generalized_codes_intro}, we introduced  the $\text{YY-rep-rec}_N$ code for balancing the code distances which concatenates two copies of the $\text{rep-rec}_N$ codes with the YY stabilizer. 
To see how to decode the $\text{YY-rep-rec}_N$ code efficiently, we start with its generator matrix
\begin{align}
    M_{\text{YY-rep-rec}_N} = M_\text{conc}^{\text{(sq)}}(S^T_\text{rec})^{\oplus2N}
\end{align}
where $S_\text{rec}$ is given in Eq.~\ref{eq:rec_gkp}.
Here $M_\text{conc}^{\text{(sq)}}$ is the $XX$ repetition codes concatenated with the YY stabilizer, and as examples, the stabilizers for the cases of $N=2,3$ are shown in Tab.~\ref{tab:stabilizer_YY_rep}. 
\begin{table}[h!]
\centering
\begin{tabular}{c|c c || c | cc } 
$N=2$ & & & $N=3$ & & \\
 \hline
 & \mybox[0.25cm]{$X$}\mybox[0.25cm]{$X$} & \mybox[0.25cm]{$I$}\mybox[0.25cm]{$I$} & & \mybox[0.25cm]{$X$}\mybox[0.25cm]{$X$}\mybox[0.25cm]{$I$} & \mybox[0.25cm]{$I$}\mybox[0.25cm]{$I$}\mybox[0.25cm]{$I$} \\
 & \mybox[0.25cm]{$I$}\mybox[0.25cm]{$I$} & \mybox[0.25cm]{$X$}\mybox[0.25cm]{$X$} & & \mybox[0.25cm]{$I$}\mybox[0.25cm]{$X$}\mybox[0.25cm]{$X$} & \mybox[0.25cm]{$I$}\mybox[0.25cm]{$I$}\mybox[0.25cm]{$I$} \\
 & \mybox[0.25cm]{$Y$}\mybox[0.25cm]{$Z$} & \mybox[0.25cm]{$Y$}\mybox[0.25cm]{$Z$} & & \mybox[0.25cm]{$I$}\mybox[0.25cm]{$I$}\mybox[0.25cm]{$I$} & \mybox[0.25cm]{$X$}\mybox[0.25cm]{$X$}\mybox[0.25cm]{$I$} \\
 &  &  & & \mybox[0.25cm]{$I$}\mybox[0.25cm]{$I$}\mybox[0.25cm]{$I$} & \mybox[0.25cm]{$I$}\mybox[0.25cm]{$X$}\mybox[0.25cm]{$X$} \\
 &  &  & & \mybox[0.25cm]{$Y$}\mybox[0.25cm]{$Z$}\mybox[0.25cm]{$Z$} & \mybox[0.25cm]{$Y$}\mybox[0.25cm]{$Z$}\mybox[0.25cm]{$Z$} \\
 \hline
  $\bar{Z}$ & \mybox[0.25cm]{$Z$}\mybox[0.25cm]{$Z$} & \mybox[0.25cm]{$Z$}\mybox[0.25cm]{$Z$} & $\bar{Z}$ & \mybox[0.25cm]{$Z$}\mybox[0.25cm]{$Z$}\mybox[0.25cm]{$Z$} & \mybox[0.25cm]{$Z$}\mybox[0.25cm]{$Z$}\mybox[0.25cm]{$Z$}\\
 $\bar{X}$ & \mybox[0.25cm]{$I$}\mybox[0.25cm]{$X$} & \mybox[0.25cm]{$Z$}\mybox[0.25cm]{$Z$} & $\bar{X}$ & \mybox[0.25cm]{$I$}\mybox[0.25cm]{$I$}\mybox[0.25cm]{$X$} & \mybox[0.25cm]{$Z$}\mybox[0.25cm]{$Z$}\mybox[0.25cm]{$Z$}
\end{tabular}
\caption{The stabilizers for the $XX$ repetition code concatenated with the YY stabilizer for $N=2,3$. For both cases, the last stabilizer is the YY stabilizer, which is the tensor product of the logical $\bar{Y}$ operators in the two blocks. We have also shown the logical operators $\bar{X}$ and $\bar{Z}$.}
\label{tab:stabilizer_YY_rep}
\end{table}

We now show that the stablizer part of the $\text{YY-rep-rec}_N$ code is a glue lattice, i.e., 
\begin{align}
\label{eq:Lambda-YY-rep-rec_N}
    \Lambda(M_\text{conc}^{\text{(sq)}}) = \Lambda_1 \cup \left(\frac{1}{\sqrt{2}}\boldsymbol{g}_{YY}+\Lambda_1\right),
\end{align}
where $\boldsymbol{g}_{YY}$ is the binary vector corresponding to the YY stabilizer and 
\begin{align}
    \label{eq:Lambda_1}
    \Lambda_1=\Lambda(M_\text{rep}^\text{(sq)}\oplus M_\text{rep}^\text{(sq)}) = \Lambda(M_\text{rep}^\text{(sq)})\oplus\Lambda( M_\text{rep}^\text{(sq)}).
\end{align}
Here $\Lambda( M_\text{rep}^\text{(sq)})$ is the stabilizer part of the $\text{rep-rec}_N$ code, and $M_\text{rep}^\text{(sq)}$ is defined in Eq.~\ref{eq:def-rep-rec_N}. 
As discussed in Sec.~\ref{sec:The concatenated GKP code}, $M_\text{conc}^{\text{(sq)}}$ is constructed by replacing $(2N-1)$ rows in $\sqrt{2}I_{4N}$ by a set of vectors $\left\{\frac{1}{\sqrt{2}}\boldsymbol{g}_j^T\right\}$ where each $\boldsymbol{g}_j$ corresponds to a stabilizer generator. 
Because the $(2N-2)$ stabilizers from the $XX$ repetition codes can be separated into two disjoint blocks, as evident from Tab.~\ref{tab:stabilizer_YY_rep}, the replacement with these vectors yield a direct sum of two $\Lambda(M_\text{rep}^\text{(sq)})$ lattices.
To see that Eq.~\ref{eq:Lambda-YY-rep-rec_N} holds,
suppose $\Lambda_1$ is generated by a set of basis vectors 
\begin{align*}
    \left\{{\boldsymbol{r}}_j, ~ j=1,...,4N\right\},
\end{align*}
then $\Lambda(M_\text{conc}^\text{(sq)})$ is span by the same set of basis vectors except the last one,
\begin{align*}
    \left\{{\boldsymbol{r}}_1, ..., {\boldsymbol{r}}_{4N-1}, \frac{1}{\sqrt{2}}\boldsymbol{g}_{YY}\right\},
\end{align*}
which is evident from the construction of $M_\text{conc}^\text{(sq)}$. 
For a given $\boldsymbol{x}\in\Lambda(M_\text{conc}^\text{(sq)})$, let 
\begin{align*}
    \boldsymbol{x}=\sum_{i=1}^{4N-1}a_i{\boldsymbol{r}}_i+\frac{b}{\sqrt{2}}\boldsymbol{g}_{YY}
\end{align*}
for some integers $a_i$ and $b$. As one can show, $\sqrt{2}\boldsymbol{g}_{YY}\in\Lambda(\sqrt{2}I_{4N})$ because $\boldsymbol{g}_{YY}$ is a binary vector and $\Lambda(\sqrt{2}I_{4N})$ is a $4N$ dimensional square lattice with lattice spacing $\sqrt{2}$. 
Because $\Lambda(\sqrt{2}I_{4N})$ is a sublattice of $\Lambda_1$, if $b$ is an even integer, we have $\boldsymbol{x}\in\Lambda_1$, otherwise, if $b$ is odd, then 
$$
\boldsymbol{x} -\frac{1}{\sqrt{2}}\boldsymbol{g}_{YY} \in \Lambda_1.
$$
This concludes that $\boldsymbol{x}$ is an element of the glue lattice shown in Eq.~\ref{eq:Lambda-YY-rep-rec_N}, and hence $\Lambda(M_\text{conc}^\text{(sq)})$ is a sublattice of the latter. Similarly, we can show that the glue lattice in Eq.~\ref{eq:Lambda-YY-rep-rec_N} is a sublattice of $\Lambda(M_\text{conc}^\text{(sq)})$ and hence the two represent the identical lattice.

The fact that $\Lambda(M_\text{conc}^\text{(sq)})$ is a glue lattice is important for decoding the $\text{YY-rep-rec}_N$ code, whose symplectic dual lattice is generated by
\begin{eqnarray}
\label{eq:M_YY-rep-rec_N_perp}
    \begin{aligned}
    M_{\text{YY-rep-rec}_N}^\perp &= \Omega(M_{\text{YY-rep-rec}_N}^T)^{-1}\Omega^{-1}\\
    &=\Omega(M_{\text{conc}}^{\text{(sq)},T})^{-1}(S^{-1}_\text{rec})^{\oplus2N}\Omega^{-1}\\
    &=(M_{\text{conc}}^{\text{(sq)}})^\perp \Omega(S^{-1}_\text{rec})^{\oplus2N}\Omega^{-1}.
    \end{aligned}
\end{eqnarray}
% where we have used Eq.~\ref{eq:identity_1}. 
%
Here $(M_{\text{conc}}^{\text{(sq)}})^\perp$ generates the symplectic dual lattice for the XX repetition code concatenated with the YY stabilizer.
Since the code encodes a single qubit, i.e., $\det(M_{\text{conc}}^{\text{(sq)}})=2$, we have $\det((M_{\text{conc}}^{\text{(sq)}})^\perp)=\frac{1}{2}=\frac{1}{4}\det(M_{\text{conc}}^{\text{(sq)}})$. Hence the symplectic dual lattice can be viewed as a union of four cosets of the original lattice, each of which corresponds to a logical operator. Since the original lattice $\Lambda(M_\text{conc}^{\text{(sq)}})$ is a union of two cosets, per Eq.~ \ref{eq:Lambda-YY-rep-rec_N}, we have
\begin{eqnarray}
% \label{eq:YY+rec_perp}
\label{eq:Lambda_perp_YY_rec}
    \begin{aligned}
    \Lambda((M_{\text{conc}}^{\text{(sq)}})^\perp) &= \bigcup_{i=0}^7(\frac{1}{\sqrt{2}}\boldsymbol{g}_i+\Lambda(M_{\text{conc}}^{\text{(sq)}})).
    \end{aligned}
\end{eqnarray}
which is a union of eight cosets. Here $\boldsymbol{g}_0=\boldsymbol{0}$, and
\begin{align*}
    \boldsymbol{g}_{1}=\boldsymbol{g}_{\bar{X}},\quad \boldsymbol{g}_{2}= \boldsymbol{g}_{\bar{X}}+\boldsymbol{g}_{\bar{Z}},\quad \boldsymbol{g}_{3}=\boldsymbol{g}_{\bar{Z}},
\end{align*}
%
% $\boldsymbol{g}_{1,2,3}=\boldsymbol{g}_{\bar{X}}, \boldsymbol{g}_{\bar{X}}+\boldsymbol{g}_{\bar{Z}}, \boldsymbol{g}_{\bar{Z}}$ 
%
are the binary vectors for the logical $\bar{X}$, $\bar{Y}$ and $\bar{Z}$ operators. Together with
\begin{align*}
    \boldsymbol{g}_{4}=\boldsymbol{g}_{YY}+\boldsymbol{g}_{0}, &\quad \boldsymbol{g}_{5}=\boldsymbol{g}_{YY}+\boldsymbol{g}_{1}, \\
    \boldsymbol{g}_{6}=\boldsymbol{g}_{YY}+\boldsymbol{g}_{2}, &\quad \boldsymbol{g}_{7}=\boldsymbol{g}_{YY}+\boldsymbol{g}_{3}, 
\end{align*}
they form the normalizer group for the stabilizer part of the $\text{YY-rep-rec}_N$ code.

The structure of the symplectic dual lattice, as presented in Eq.~\ref{eq:Lambda_perp_YY_rec}, can be greatly simplified if we work in the \texttt{qqpp} ordering where (we omit the superscript \texttt{qqpp} for clarity)
\begin{align}
M_\text{rep}^\text{(sq)}\oplus M_\text{rep}^\text{(sq)} = \frac{1}{\sqrt{2}}
\begin{bmatrix}
    M_{D_N} & 0_N & 0_N & 0_N\\
    0_N & M_{D_N} & 0_N & 0_N\\
    0_N & 0_N & 2I_{N} & 0_N\\
    0_N & 0_N & 0_N & 2I_{N}
\end{bmatrix}.
\end{align}
In this basis, the lattice $\Lambda_1$ in Eq.~\ref{eq:Lambda_1} is a direct sum of two lattices $\Lambda_1 = \Lambda_1^{(q)}\oplus\Lambda_1^{(p)}$ where
\begin{align}
\label{eq:L1_q_L1_p}
     \Lambda_1^{(q)}= \frac{1}{\sqrt{2}}\Lambda(M_{D_N}^{\oplus2}) ~ \text{ and } ~ \Lambda_1^{(p)} = \sqrt{2}\Lambda(I_{2N})
\end{align}
are the sublattices for the $\hat{q}$ and $\hat{p}$ subspaces respectively.
The decoupled form of $\Lambda_1$ is useful because, from Tab.~\ref{tab:stabilizer_YY_rep}, we notice that $\boldsymbol{g}_{\bar{Z}}$ has support only at the $\hat{p}$ subspace. 
For instance, the logical operator $\bar{Z}\equiv ZZZZ$ for $N=2$, which corresponds to the binary vector $(0,0,0,0,1,1,1,1)^T$. Hence the lattice
\begin{align}
    \Lambda_2 \equiv \frac{1}{\sqrt{2}}\boldsymbol{r}_{\bar{Z}} + \Lambda_1 = \Lambda_2^{(q)}\oplus\Lambda_2^{(p)}
\end{align}
is also a direct sum of two lattices, where $\Lambda_2^{(q)}=\Lambda_1^{(q)}$ for the $\hat{q}$ subspace, and 
\begin{align}
\label{eq:L2_p}
    \Lambda_2^{(p)} = \frac{1}{\sqrt{2}}\boldsymbol{g}^{(p)}_{\bar{Z}} + \sqrt{2}\Lambda(I_{2N})
\end{align}
for the $\hat{p}$ subspace. Here $\boldsymbol{g}^{(p)}_{\bar{Z}}$ is the second half of $\boldsymbol{g}_{\bar{Z}}$ which has all entries being unity. Upon combining Eq.~\ref{eq:L1_q_L1_p}-\ref{eq:L2_p}, we have
\begin{eqnarray}
    \begin{aligned}
    \label{eq:def_Lambda3}
    \Lambda_3\equiv&\Lambda_1 \cup \left(\frac{1}{\sqrt{2}}\boldsymbol{g}_{\bar{Z}}+\Lambda_1\right) \\
    % \Lambda_3\equiv&\Lambda_1 \cup \Lambda_2 \\
    =& \left[\Lambda_1^{(q)}\oplus\Lambda_1^{(p)}\right]\cup \left[\Lambda_1^{(q)}\oplus(\frac{1}{\sqrt{2}}\boldsymbol{g}^{(p)}_{\bar{Z}} + \sqrt{2}\Lambda(I_{2N}))\right]\\
    =&\Lambda_1^{(q)}\oplus\left[\Lambda_1^{(p)}\cup(\frac{1}{\sqrt{2}}\boldsymbol{g}^{(p)}_{\bar{Z}} + \sqrt{2}\Lambda(I_{2N}))\right]\\
    =&\frac{1}{\sqrt{2}}\Lambda(M_{D_N}^{\oplus2})\oplus\sqrt{2}\Lambda(M_{D^*_{2N}}).
    \end{aligned}
\end{eqnarray}
Here we have used the fact that the $D_{2N}^*$ lattice is a union of two cosets as shown in Eq.~\ref{eq:D_n_dual}-\ref{eq:D_n_dual_cosets}.
Similar strategy can be applied to simplify the lattice
\begin{align}
\label{eq:def_Lambda_4}
    \Lambda_4\equiv&\Lambda_3 \cup \left(\frac{1}{\sqrt{2}}\boldsymbol{g}_{YY}+\Lambda_3\right).
\end{align}
We first notice that $\boldsymbol{g}_{YY} = \boldsymbol{g}'_{YY} + \boldsymbol{g}_{\bar{Z}}$ where $\boldsymbol{g}'_{YY}$ has nonzero entries only at the first and the $N+1$-th positions. For instance, we have $\boldsymbol{g}'_{YY}=(1,0,1,0,0,0,0,0)^T$ for $N=2$.
Since $\frac{1}{\sqrt{2}}\boldsymbol{g}_{\bar{Z}}\in\Lambda_3$ by definition, the glue vector in Eq.~\ref{eq:def_Lambda_4} can be replaced by $\frac{1}{\sqrt{2}}\boldsymbol{g}_{YY}'$ which has no support in the $\hat{p}$ subspace. Hence $\Lambda_4$ reduces to a direct sum of two lattices for the $\hat{q}$ and $\hat{p}$ subspaces respectively. 
In particular, the lattice for the $\hat{p}$ subspace is $\Lambda_4^{(p)}=\sqrt{2}\Lambda(M_{D^*_{2N}})$, which is the same as $\Lambda_3^{(p)}$, whereas that for the $\hat{q}$ subspace reads
\begin{align}
\label{eq:lattice_YY''}
    \Lambda_4^{(q)} = \Lambda_3^{(q)} \cup \left(\frac{1}{\sqrt{2}}\boldsymbol{g}_{YY}'^{(q)}+\Lambda_3^{(q)}\right) \equiv \frac{1}{\sqrt{2}}\Lambda_4^{(q)'},
\end{align}
where $\boldsymbol{g}_{YY}'^{(q)}$ is the first half of $\boldsymbol{g}_{YY}'$ and
\begin{align}
\label{eq:lattice_YY''_2}
    \Lambda_4^{(q)'} = \Lambda(M_{D_{N}}^{\oplus 2})\cup(\boldsymbol{g}_{YY}'^{(q)}+\Lambda(M_{D_{N}}^{\oplus 2})).
\end{align}
Since the sum of the components of $\boldsymbol{g}_{YY}'^{(q)}$ is equal to 2, an even number, and similarly for the vectors in $\Lambda(M_{D_{N}}^{\oplus 2})$, $\Lambda_4^{(q)'}$ is in fact a $2N$-dimensional sublattice of the $D_{2N}$ lattice. 
Because $\det(M_{D_{N}}^{\oplus 2})=4$ and $\Lambda_4^{(q)'}$ is a union of two cosets of $\Lambda(M_{D_{N}}^{\oplus 2})$, the volume of the fundamental parallelotope of $\Lambda_4^{(q)'}$ is equal to 2, the same as that for the $D_{2N}$ lattice. We conclude that $\Lambda_4^{(q)'}=\Lambda(D_{2N})$ and hence 
\begin{align}
\label{eq:Lambda_4_2}
    \Lambda_4 = \frac{1}{\sqrt{2}}\Lambda(M_{D_{2N}})\oplus\sqrt{2}\Lambda(M_{D^*_{2N}}).
\end{align}
Upon combining Eq.~\ref{eq:def_Lambda3}, \ref{eq:def_Lambda_4} and \ref{eq:Lambda_4_2}, we have 
\begin{align}
    \Lambda((M_{\text{conc}}^{\text{(sq)}})^\perp) = \bigcup_{i=0}^{1} (\frac{1}{\sqrt{2}}{\boldsymbol{g}}_{\bar{X}}+\Lambda_4), 
\end{align}
which is a glue lattice with only one glue vector that is proportional to the binary vector for the logical $\bar{X}$ operator. 

The simple structure $\Lambda((M_{\text{conc}}^{\text{(sq)}})^\perp)$ enables a linear time decoder for the code. Recall the generator $M_{\text{YY-rep-rec}_N}^\perp$ defined in Eq.~\ref{eq:M_YY-rep-rec_N_perp}, in the the \texttt{qqpp} ordering, we have (we omit the superscript \texttt{qqpp} again)
\begin{align*}
    S'_\text{rec} \equiv &\Omega(S^{-1}_\text{rec})^{\oplus2N}\Omega^{-1} \\
    =& 
    \begin{bmatrix}
    0_{2N} & I_{2N}\\
    -I_{2N} & 0_{2N}
    \end{bmatrix}\begin{bmatrix}
    N^{-1/4}I_{2N} & 0_{2N}\\
    0_{2N} & N^{1/4}I_{2N}
    \end{bmatrix}
    \begin{bmatrix}
    0_{2N} & -I_{2N}\\
    I_{2N} & 0_{2N}
    \end{bmatrix}\\
    =&\begin{bmatrix}
    N^{1/4}I_{2N} & 0_{2N}\\
    0_{2N} & N^{-1/4}I_{2N}
    \end{bmatrix}.
\end{align*}
With that, the symplectic dual lattice for the $\text{YY-rep-rec}_N$ code reads
\begin{align}
    \Lambda(M_{\text{YY-rep-rec}_N}^\perp) = \bigcup_{i=0}^{1} \left(\frac{1}{\sqrt{2}}{\tilde{\boldsymbol{g}}}_{\bar{X}}+\Lambda_5\right)
\end{align}
where $\tilde{\boldsymbol{g}}_{\bar{X}} = S'_\text{rec}\boldsymbol{g}_{\bar{X}}$, and $\Lambda_5$ is a direct sum of two lattices generated by 
\begin{align}
    \left(\frac{N^{1/4}}{\sqrt{2}}M_{D_{2N}}\right) \oplus \left(\frac{\sqrt{2}}{N^{1/4}}M_{D_{2N}^*}\right).
\end{align}
Because $\Lambda(M_{\text{YY-rep-rec}_N}^\perp)$ is a glue lattice, we can use Eq.~\ref{eq:decode_glue_lattice} to find its closest point efficiently. In particular, since both $D_{2N}$ and $D_{2N}^*$ lattices can be decoded with runtime proportional to $2N$, and $\Lambda(M_{\text{YY-rep-rec}_N}^\perp)$ consists of only two cosets, we conclude that the $\text{YY-rep-rec}_N$ can be decoded with runtime proportional to $4N$. We present its closest point decoder in Alg.~\ref{alg: ClosestPointYYRepRecN}.

\begin{figure}
\centering
\includegraphics[width=\linewidth]{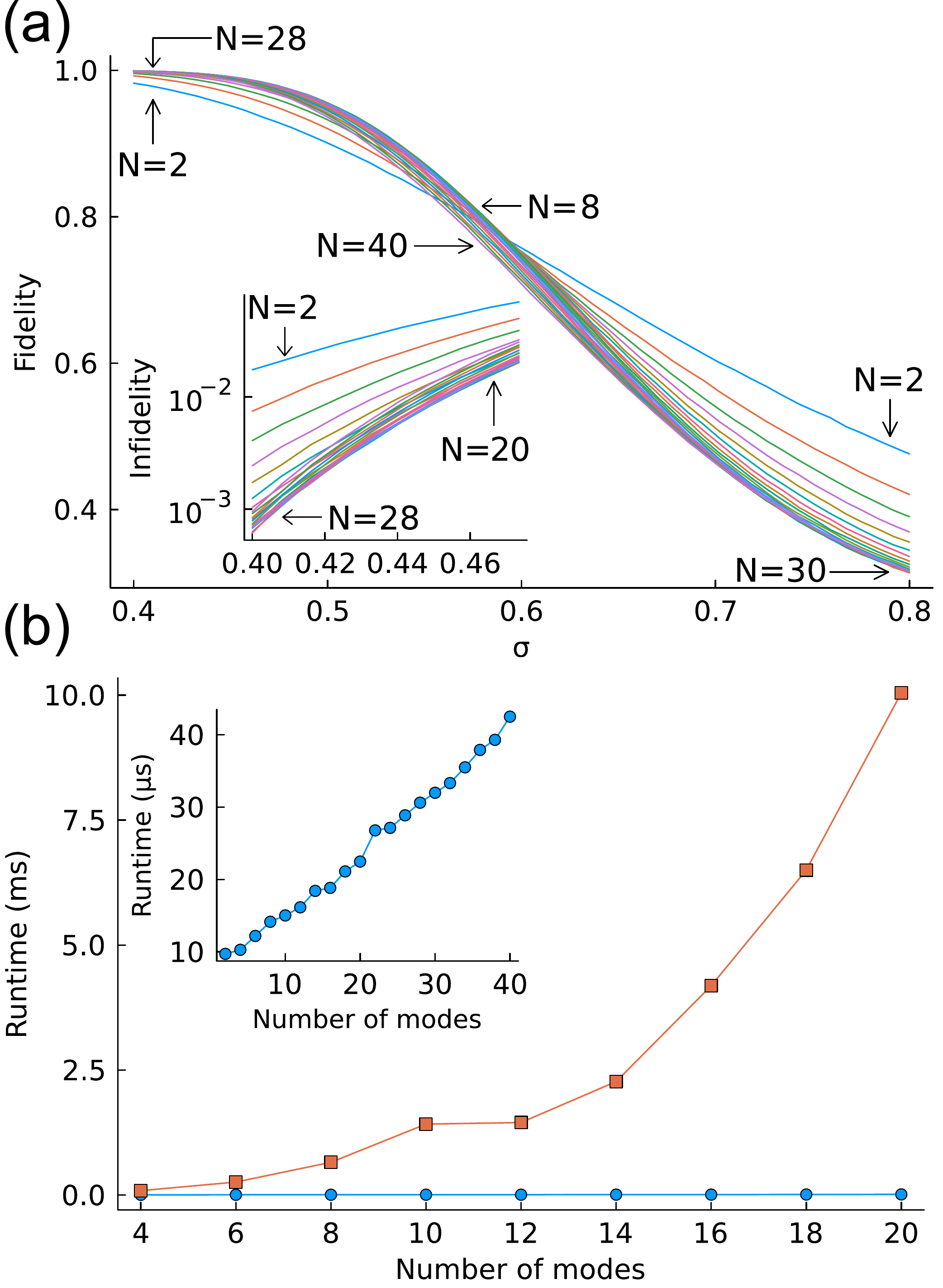}
 \caption{
(a) The fidelity of the $\text{YY-rep-rec}_N$ code as a function of the number of modes and noise strength $\sigma$. Each line corresponds to a given number of modes, which are even integers from $N=2$ to $N=40$. 
{For $\sigma=0.4, 0.5714, 0.8$, we have indicated the number of modes that support minimum and maximum fidelities.
The lower-left inset shows the infidelities between $\sigma=0.4$ and $0.4734$ for different number of modes in the log scale. The number of modes that support minimum and maximum infidelities are indicated for $\sigma=0.4$ and $0.4734$ respectively.}
(b) The comparison of runtimes for the exponential time decoder (square) and linear time decoder (circle) for increasing number of modes. 
{For each data point, we average over all the samples for all the values of $\sigma$ considered. }
The inset shows the runtimes of the linear time decoder for different number of modes, up to $N=40$.
}
 \label{fig:YY-rep-rec_N}
\end{figure}

We calculate the fidelity of the $\text{YY-rep-rec}_N$ code, up to $N=40$, with the same Gaussian noise model as in Fig.~\ref{fig:optimized_GKP_codes}(b) and Fig.~\ref{fig:rep-rec_N}(a), and the results are shown in Fig.~\ref{fig:YY-rep-rec_N}(a). 
The fidelity curve exhibits a similar band-like feature, similar to that of the $\text{rep-rec}_N$ code shown in Fig.~\ref{fig:YY-rep-rec_N}(a).
We have indicated the number of modes that support minimum and maximum fidelities from the low to high noise regimes, which show that increasing the number of modes needs not consistently improve the fidelity of the $\text{YY-rep-rec}_N$ code. For instance, as shown in the lower-left inset, the infidelity of the code at $\sigma=0.4$ and $0.4734$ reach minimum for $N=28$ and $20$ respectively, which outperform the code with $N=40$. 
As a result, we conclude that the $\text{YY-rep-rec}_N$ code, similar to the  $\text{rep-rec}_N$ code, does not exhibit a noise threshold. 
We compare the runtime between the exponential-time, general-purpose closest point decoder and the linear time decoder tailored to the $\text{YY-rep-rec}_N$ code, as shown in Fig.~\ref{fig:YY-rep-rec_N}(b). 
Each data point corresponds to the runtime averaged over all the samples for all the values of $\sigma$ considered.
As expected, the time overhead of the former is increasing rapidly, and the runtime difference is around four orders of magnitude for the case of twenty modes. In the inset, we confirm that the runtime of the linear time decoder increases linearly with the number of modes.

\begin{algorithm}
\caption{DecodeYYRepRecN($\boldsymbol{t}$)}
\label{alg: ClosestPointYYRepRecN}
{\bf Input: } The error syndrome $\boldsymbol{t}\in\mathbb{R}^{2N}$; \\ 
{\bf Output: } The optimal integer $\boldsymbol{b}\in\mathbb{Z}^{2N}$; \\ 
 $\boldsymbol{t}_q \leftarrow \boldsymbol{t}[1:2:end]$; \\
 $\boldsymbol{t}_p \leftarrow \boldsymbol{t}[2:2:end]$; \\
 $\boldsymbol{b}_{1,q} \leftarrow $ClosestPointDn($\boldsymbol{t}_q$); \\
 $\boldsymbol{b}_{1,p} \leftarrow $ClosestPointDnDual($\boldsymbol{t}_p$); \\
 $\boldsymbol{b}_1[1:2:end]\leftarrow\boldsymbol{b}_{1,q}$;\\
 $\boldsymbol{b}_1[2:2:end]\leftarrow\boldsymbol{b}_{1,p}$;\\
 $\tilde{\boldsymbol{g}}_{\bar{X},q}\leftarrow \tilde{\boldsymbol{g}}_{\bar{X}}[1:2:end]$;\\
 $\tilde{\boldsymbol{g}}_{\bar{X},p}\leftarrow \tilde{\boldsymbol{g}}_{\bar{X}}[2:2:end]$;\\
 $\boldsymbol{b}_{2,q} \leftarrow $ClosestPointDn($\boldsymbol{t}_q-\tilde{\boldsymbol{g}}_{\bar{X},q}$); \\
 $\boldsymbol{b}_{2,p} \leftarrow $ClosestPointDnDual($\boldsymbol{t}_p-\tilde{\boldsymbol{g}}_{\bar{X},p}$); \\
 $\boldsymbol{b}_2[1:2:end]\leftarrow\boldsymbol{b}_{2,q}$;\\
 $\boldsymbol{b}_2[2:2:end]\leftarrow\boldsymbol{b}_{2,p}$;\\
\eIf{\text{sum($\boldsymbol{b}_1-\boldsymbol{t}$) $\leq$ sum($\boldsymbol{b}_2-\boldsymbol{t}$)}}{
   $\boldsymbol{b} \leftarrow \boldsymbol{b}_1$;
   }{
   $\boldsymbol{b} \leftarrow \boldsymbol{b}_2$;
  }
\end{algorithm}

\section{Polynomial time closest point decoder for surface-GKP code}
\label{sec:Polynomial time closest point decoder for surface-GKP code}

In Sec.~\ref{sec:YY-rep-rec_N}, we show that decoding the $\text{YY-rep-rec}_N$ code is equivalent to finding the closest point for a glue lattice, as presented in Eq.~\ref{eq:Lambda_perp_YY_rec}. 
This is certainly not unique to the  $\text{YY-rep-rec}_N$ code, and in App.~\ref{sec:The concatenated GKP code as a glue lattice}, we generalize the argument to show that a general {concatenated GKP} code can also be viewed as a glue lattice.
Here we focus on the surface-GKP codes which encode a single logical qubit by concatenating a $[[N, 1, d_0]]$ surface code with $N$ square GKP codes. We note $N=d_0^2$ for surface-GKP codes.
As shown in App.~\ref{sec:The concatenated GKP code as a glue lattice}, the symplectic dual lattice of the surface-GKP code can be written as 
\begin{align}
\label{eq:conc_sq_dual_as_union_cosets_main}
    \Lambda(M_\text{surf}^\perp) = \bigcup_{j=0}^{2^{N+1}-1} \big[\frac{1}{\sqrt{2}}{\boldsymbol{g}}_j + \Lambda(\sqrt{2}I_{2N})\big].
\end{align}
where the vectors $\left\{{\boldsymbol{g}}_j\right\}$ correspond to the elements in the normalizer group. For notation simplicity, we assume $\boldsymbol{g}_0=\boldsymbol{0}$ and {$\boldsymbol{g}_{1},\cdots, \boldsymbol{g}_{N-1}$} generate the full stabilizer group, and {$\boldsymbol{g}_{N}, \boldsymbol{g}_{N+1}$} are the logical $\bar{X}$ and $\bar{Z}$ operators.
The problem of decoding the surface-GKP code is equivalent to finding the closest point in $\Lambda(M_\text{surf}^\perp)$ for a given syndrome. 
However, because the lattice is a union of $2^{N+1}$ cosets, direct application of either Eq.~\ref{eq:decode_union_cosets} or \ref{eq:decode_glue_lattice} will incur an exponential runtime for the decoding. 
We now show that the search of $2^{N+1}$ cosets can be efficiently performed with an MWPM algorithm, and the surface-GKP code can be decoded in polynomial time. 

As discussed in Sec.~\ref{sec:Closest point search problem}, we aim to find the closest point for a given $\boldsymbol{t}$, i.e., 
\begin{align}
    \boldsymbol{\chi}_{\boldsymbol{t}}(\sqrt{2\pi}(M_\text{surf}^\perp)) = \argmin_{\boldsymbol{u}\in\sqrt{2\pi}\Lambda(M_\text{surf}^\perp)}||\boldsymbol{t}-\boldsymbol{u}||.
\end{align}
% where $\Lambda(M_\text{surf}^\perp)$ is the dual lattice for the surface-GKP code. 

\begin{widetext}

\begin{figure}[h!]
\centering
\includegraphics[width=\linewidth]{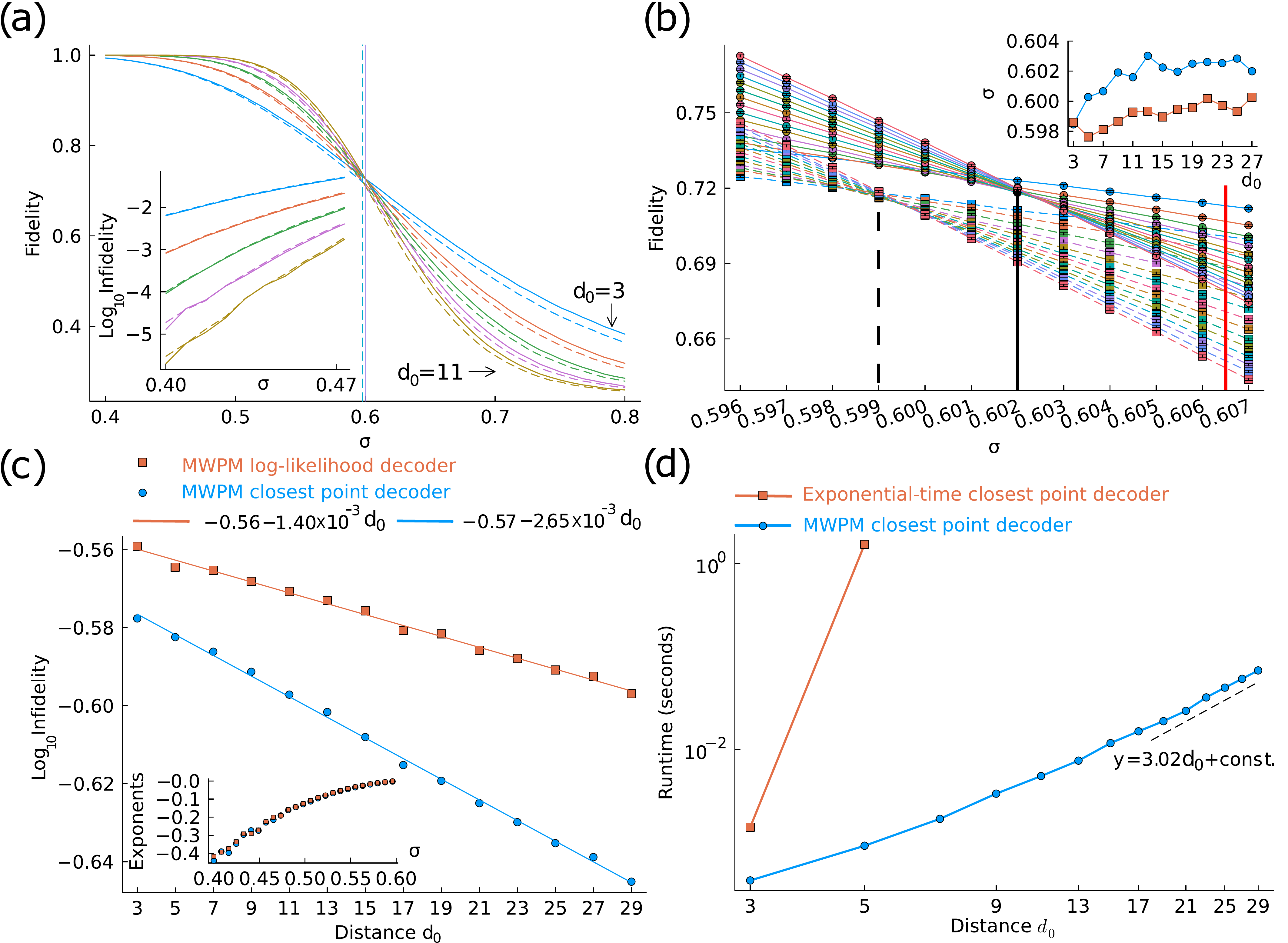}
 \caption{
Numerical results for surface-GKP codes. 
(a) The fidelity of the surface-GKP codes as a function of the noise strength $\sigma$ and distance $d_0$, which are odd integers from $d_0=3$ to $d_0=11$, as indicated. 
The solid and dash lines correspond to the MWPM closest point decoder and the MWPM log-likelihood decoder respectively, and the vertical lines indicate their thresholds. 
The bottom-left inset shows the infidelities near $\sigma=0.4$ for different $d_0$.
(b) The fidelity of the surface-GKP codes near the threshold. The noise strength $\sigma$ is scanned from $0.596$ to $0.607$ with resolution $0.001$, and the distances are odd integers from $d_0=3$ to $d_0=29$. Each data point is obtained from $10^7$ Monte-Carlo samples, which is 10 times as large as that for the data points in other subplots.
{The standard error of the data points are of the order $10^{-4}$, and the error bars of the fidelity are all smaller than the markers of the data points. 
The black solid and dash vertical lines correspond to the thresholds for the MWPM closest point and log-likelihood decoders, which read $0.602$ and $0.599$ respectively. The red solid line corresponds to $1/\sqrt{e}=0.6065\cdots$ which is an important quantity from the quantum information theory point of view because it is the value of $\sigma$ at which the known lower bound to the quantum capacity of a Gaussian random displacement channel vanishes. See the main text for more discussions.
Inset shows the crossings as a function of $d_0$, which shows that the crossing from the MWPM closest point is always higher than that for the log-likelihood decoder by 0.002 for large $d_0$.
}
%
% Inset shows the crossing as a function of $d_0$, from which the thresholds for the MWPM closest point and log-likelihood decoders read $0.602$ and $0.599$ respectively. See the main text for more details.
%
(c) The logarithm of the infidelity as a function of $d_0$ for $\sigma=0.5959$. The blue circles and red squares correspond to the MWPM closest point and log-likelihood decoders respectively. The data points are fitted with linear functions and the slopes for the two decoders read $-2.65\times10^{-3}$ and $-1.40\times10^{-3}$ respectively. Inset shows the slopes as a function of $\sigma$ up to $\sigma=0.60$. See the main text for more discussion. 
{
(d) Comparison of the runtimes for the MWPM closest point decoder and the exponential time closest point decoder for the surface-GKP codes. Both horizontal and vertical axes are in log scales. The red squares correspond to the exponential-time decoder whose runtime scales exponentially with $d_0^2$ (hence only two datapoints are calculated); the blue circles correspond to the runtime of the MWPM closest point decoder which scales as $d_0^{3.02}$ for large $d_0$.
}
}
 \label{fig:surface-GKP code}
\end{figure}

\end{widetext}

For simplicity, we will consider the scaled lattice
\begin{align}
    \Lambda'\equiv \sqrt{2}\Lambda(M_\text{surf}^\perp) =\bigcup_{j=0}^{2^{N+1}-1} \big[{\boldsymbol{g}}_j + \Lambda(2I_{2N})\big],
\end{align}
which is an integral lattice, and find the closest point for the scaled vector $\boldsymbol{t}'\equiv \frac{1}{\sqrt{\pi}}\boldsymbol{t}$, namely
\begin{align}
\label{eq:def_chi_t_surf}
    \boldsymbol{\chi}'\equiv \boldsymbol{\chi}_{\boldsymbol{t}'}(\Lambda') =\frac{1}{\sqrt{\pi}} \boldsymbol{\chi}_{\boldsymbol{t}}(\sqrt{2\pi}\Lambda(M_\text{surf}^\perp)).
\end{align}
Since $\boldsymbol{\chi}'$ is an integer valued vector, we first observe that each component $\chi'_i$ has to be either the closest or second closest integer of $t'_i$. In other words, $\boldsymbol{\chi}'$ is contained in the following set of $2^{2N}$ vectors
\begin{align}
    S_{\boldsymbol{\chi}'}\equiv \left\{\boldsymbol{\chi}'|\chi'_i = \lfloor t'_i\rceil~ \text{or}~ w(t'_i),~ \text{for } 1\leq i\leq 2N\right\},
\end{align}
where $w(x)$ is the second closest integer for $x\in\mathbb{R}$, as defined in Eq.~\ref{def:w(t)}. 
To see that, suppose $\chi'_i$ is neither the closest nor the second closest integer of $t'_i$, because $\Lambda(2I_{2N})$ is a sublattice of $\Lambda'$, we can use the $i$-th basis vector of $\Lambda(2I_{2N})$ to translate $\chi'_i$ to either the closest or second closest integer of $t'_i$. 
The resultant vector is guaranteed to have closer distance to $\boldsymbol{t}'$, compared to the vector before the translation, hence $\boldsymbol{\chi}'$ must be in the set $S_{\boldsymbol{\chi}'}$.

However, directly searching through $S_{\boldsymbol{\chi}'}$ is not only impractical for large $N$, but also unnecessary because not all the vectors in $S_{\boldsymbol{\chi}'}$ are in the lattice $\Lambda'$. 
For instance, although 
\begin{align}
    \boldsymbol{\chi}'' \equiv (\lfloor t'_1\rceil, ..., \lfloor t'_{2N}\rceil)
\end{align}
is the closest possible integer valued vector to $\boldsymbol{t}'$, it is however not necessarily in $\Lambda'$. Instead, we have to round certain components of $\boldsymbol{\chi}''$ in the wrong way, similar to how we find the closest point in the $D_n$ lattice, as shown in Sec.~\ref{sec:decoderDN}. 
For this purpose, we further observe that $\Lambda'$ can be written in the following way
\begin{align}
\label{eq:Lambda'_2}
    \Lambda' = \left\{\boldsymbol{v}\in\mathbb{Z}^{2N}~|~ \text{mod}(\boldsymbol{g}_i^T\Omega\boldsymbol{v}, 2)=0, 1\leq i\leq N-1\right\}.
\end{align}
To see that Eq.~\ref{eq:Lambda'_2} holds, we note that $\text{mod}(\boldsymbol{g}_i^T\Omega\boldsymbol{v}, 2)=0$ is equivalent to $\frac{1}{\sqrt{2}}\boldsymbol{g}_i^T\Omega\frac{\boldsymbol{v}}{\sqrt{2}}\in\mathbb{Z}$, which suggests that $\frac{\boldsymbol{v}}{\sqrt{2}}$ has integer valued symplectic product with all the vectors in $\Lambda(M_\text{surf})$. It follows that $\boldsymbol{v}$ is in $\sqrt{2}\Lambda(M_\text{surf}^\perp)$ or equivalently $\Lambda'$. 

With these observations, finding the closest point $\boldsymbol{\chi}'$ reduces to finding a vector in $S_{\boldsymbol{\chi}'}$, that is closest to $\boldsymbol{t}'$ while satisfying $\text{mod}(\boldsymbol{g}_i^T\Omega\boldsymbol{\chi}', 2)=0$ for $1\leq i \leq N-1$. Despite $\boldsymbol{\chi}''$ needs not satisfy the latter condition, it can be used as an ansatz and we enumerate the stabilizers with $\text{mod}(\boldsymbol{g}_i^T\Omega\boldsymbol{\chi}'', 2)\neq0$. Our goal is to round certain entries of $\boldsymbol{\chi}''$ in the wrong way such that the resultant vector $\boldsymbol{\chi}'$ is guaranteed to be in $\Lambda'$. Since $\boldsymbol{\chi}''$ is the closest integer valued vector, changing certain entries from $\lfloor t'_{i}\rceil$ to $w(t_i')$ will increase the distance to $\boldsymbol{t}'$, we would like to minimize the increased distance while ensuring $\boldsymbol{\chi}'\in\Lambda'$.

In the above description, the stabilizers of the surface-GKP code is not explicitly invoked, hence the conclusion can be applied to the surface-GKP code with different choice of stabilizer generators, or other {concatenated GKP} codes. 
However we emphasize that for the decoding strategy presented below to work, we do need to assume that a given shift error can induce at most two syndrome errors, a property shared by the surface-GKP code and some other concatenated GKP codes. {Certain stabilizer codes, such as the color code, do not have this property, hence the closest point decoder presented cannot be applied to such cases.}
Also we have restricted our attention to the case where the base GKP code is a square GKP code.

With these constraints stated, finding the closest point $\boldsymbol{\chi}'$ can be further reduced to the MWPM problem for a weighted graph
\begin{align}
\label{eq:def_G}
    G=(V, E, W).
\end{align}
Here $V=\left\{v_i\right\}$ contains a set of $N-1$ vertices, each corresponds to a stabilizer generator $\boldsymbol{g}_i$. 
For any pair of distinct vertices $v_{i}$ and $v_j$, they will share an edge $e\in E$ if $\boldsymbol{g}_{i}$ and $\boldsymbol{g}_{j}$ share one or more nonzero entries.
In particular, let $g_{jk}$ denotes the $k$-th entry of $\boldsymbol{g}_j$, and 
\begin{align}
    S_{ij} \equiv \left\{k ~|~ g_{ik}=g_{jk}=1\right\}
\end{align}
be the set of shared entries of $\boldsymbol{g}_{i}$ and $\boldsymbol{g}_{j}$. Since we would like to minimize the total weight of a matching, the weight of the edge between vertices $v_{i,j}$ is assigned to be
\begin{align}
\label{eq:def_W}
    \min_{k\in S_{ij}}[(w(t'_k) - t'_k)^2-(\lfloor t'_k\rceil - t'_k)^2],
\end{align}
which is the increased distance if we round the $k$-th element of $\boldsymbol{\chi}''$ in the wrong way.  
After the weighted graph is set up, we define a set of highlighted vertices
\begin{align}
\label{eq:def_H}
    H = \left\{v_i\in G ~|~ \text{mod}(\boldsymbol{g}_i^T\Omega\boldsymbol{\chi}'', 2)\neq0\right\}.
\end{align}
If the number of highlighted vertices is odd, we add a highlighted boundary vertex into $H$ such that the number of highlighted vertices is always even. 
We then apply the MWPM algorithm for the weighted graph $G$ to identify a set of edges, which match the highlighted vertices pairwise.
The selected edges correspond to the entries in $\boldsymbol{\chi}''$ that need to be rounded in the wrong way, in order to obtain a lattice point $\boldsymbol{\chi}'\in\Lambda'$.
By construction, $\boldsymbol{\chi}'$ is the closest point in $S_{\boldsymbol{\chi}'}$ that has even symplectic product with all the stabilizers. 
Hence, from Eq.~\ref{eq:def_chi_t_surf}, we arrive at the desired closest point, which is simply $\sqrt{\pi}\boldsymbol{\chi}'$.

The above algorithm can find the closest point efficiently, because the MWPM of a graph can be found in runtime that is polynomial to the number of vertices and edges \cite{fowler2013minimum, higgott2023sparse}. 
{It can be further sped up for the surface-GKP code, if we use the the \texttt{qqpp} convention, where the generator matrix is given by}
\begin{align}
    M_\text{surf}^{\texttt{qqpp}} = 
    \begin{bmatrix}
    M_\text{surf}^{(q)} & 0_N \\
    0_N & M_\text{surf}^{(p)}
    \end{bmatrix}.
\end{align}
Here $M_\text{surf}^{(q)}$ and $M_\text{surf}^{(p)}$ are the generators for the lattices in the $\hat{q}$ and $\hat{p}$ subspaces respectively.
In this convention, the vector $\boldsymbol{g}_i$ has support only in the $\hat{q}$ or $\hat{p}$ subspace if it corresponds to an X or Z stabilizer respectively. As a result, the weighted graph defined in Eq.~\ref{eq:def_G} is split into two disjoint subgraphs $G^{(q)}$ and $G^{(p)}$. The subgraph $G^{(q)}$ consists of $\frac{N-1}{2}$ vertices, each corresponds to a $X$-stabilizer, whereas the $\frac{N-1}{2}$ vertices in $G^{(p)}$ correspond to the Z-stabilizers. 
The set of highlighted vertices for the subgraphs are defined as
\begin{eqnarray}
\label{eq:def_H_pq}
    \begin{aligned}
    H^{(q)} &= \left\{v_i\in G^{(q)} ~|~ \text{mod}((\boldsymbol{g}^{(q)}_i)^T\boldsymbol{\chi}''^{(p)}, 2)\neq0\right\},\\
    H^{(p)} &= \left\{v_i\in G^{(p)} ~|~ \text{mod}((\boldsymbol{g}^{(p)}_i)^T\boldsymbol{\chi}''^{(q)}, 2)\neq0\right\}.
    \end{aligned}
\end{eqnarray}
Here $\boldsymbol{g}_i^{(q)}$ and $\boldsymbol{g}_i^{(p)}$ denote the vectors for the X and Z stabilizers respectively, and $\boldsymbol{\chi}''^{(p,q)}$ are the projections of $\boldsymbol{\chi}''$ onto the $\hat{q}$ and $\hat{p}$ subspaces.
If either $H^{(p),(q)}$ has odd number of highlighted vertices, we add a boundary vertex into the set such that both $H^{(p),(q)}$ have even number of highlighted vertices. 
We then apply the MWPM algorithm separately for the two disjoint subgraphs, which yield two sets of selected edges. 
Upon combining the two sets of selected edges, they correspond to the entries in $\boldsymbol{\chi}''$ that need to be rounded in the wrong way. 
We present the algorithm for decoding the square-GKP code in Alg.~\ref{alg: ClosestPointSurf}.

{Let us compare the MWPM closest point decoder to MWPM log-likelihood decoder studied in \cite{noh2020fault}. For the latter decoder, the surface-GKP code is decoded by solving exactly the same MWPM problem, but with the crucial difference that the edge weights are not given by Eq.~\ref{eq:def_W}. Instead, in the MWPM log-likelihood decoder, the weights are likelihood functions that estimate the probability of the logical errors due to the given syndromes, which depend on the noise model chosen. The decoder performs well when the shift error is small; however, such log-likelihood estimation is typically unreliable if the shift error is large and close to the decision boundaries, where the crossover between correctable and uncorrectable shift occurs.
On the other hand, the closest point decoder is designed to identify the closest point exactly regardless the size of the shift error. Further, because the closest point decoder does not assume the distribution of the error shifts (other than the fact that shorter shifts are more likely), it is not only more reliable but also can be applied to many other noise models.
}

% In Fig.~\ref{fig:surface-GKP code_1}(a),
In Fig.~\ref{fig:surface-GKP code}(a),
we provide numerical evidence that the MWPM closest point decoder outperforms the MWPM log-likelihood decoder. We plot the fidelity of the surface-GKP codes as a function of noise strength $\sigma$ and distance $d_0$ for the two decoders. For a given distance $d_0$, the solid and dash lines represent the MWPM closest point and log-likelihood decoders respectively, which shows that the fidelity from the former is always higher than the latter. We note from the bottom right of the plot that the difference in the fidelity is larger for larger noise strength, hence the noisier the hardware, there will be more benefit for using the closest point decoder.
In the left-bottom inset, we show the infidelities near the low noise regime which are almost indistinguishable for the two decoders.
More interestingly, we can notice that the threshold for the closest point decoder is slightly larger than that for the log-likelihood decoder, as indicated by the solid and dash vertical lines respectively. 
In order to more precisely quantify the difference, we perform a more careful calculation as shown in Fig.~\ref{fig:surface-GKP code}(b). Here we scan through $\sigma=0.596$ and $\sigma=0.607$ with a resolution of $0.001$, and the distances for the surface code are between $d_0=3$ and $d_0=29$. In order to better suppress the statistical fluctuations, each data point in Fig.~\ref{fig:surface-GKP code}(b) is obtained via the Monte-Carlo method with $10^7$ samples, which is 10 times as large as that for other plots in Fig.~\ref{fig:surface-GKP code}.
The fidelities for the MWPM log-likelihood and closest point decoders are indicated by the circular and square markers respectively, and the dash and solid lines are guides of eyes. 
To obtain the threshold for either decoder, we determine the crossing $\sigma^*$ where the fidelity of the distance $d_0+2$ surface-GKP code is larger than that of the distance $d_0$ surface-GKP code, for a given $d_0$. Then we investigate to what value the crossing point converges as we increase $d_{0}${, as shown in the inset. We calculate the mean and standard deviation for the crossings for $d_0>13$, and obtain $\sigma^*=0.6025\pm0.0004$ for the closest point decoder, and $\sigma^*=0.5996\pm0.0004$ for the log-likelihood decoder. Since the resolution of $\sigma$ in our simulation is $0.001$, we choose to report only three digits for the thresholds.

}

In the above threshold estimate, we made sure that our analysis reliably yields the threshold $\sigma^{*}$ values up to three significant digits with $10^{7}$ samples. On the other hand, previous works have calculated the threshold $\sigma$ mostly up to two significant digits and obtained $\sigma^{*}=0.60$ for the surface GKP code using the MWPM log-likelihood decoder \cite{vuillot2019quantum,fukui2018high}, $\sigma^{*}=0.58$ for the surface GKP code with designed noise bias \cite{hanggli2020enhanced}, and $\sigma^{*}=0.59$ for the color-GKP code \cite{zhang2021quantum}. One obvious reason why we care about three significant digits is because the difference between the MWPM closest point decoder and the MWPM log-likelihood decoder can only be resolved in the third significant digit. Another reason is because the best known lower bound to the quantum capacity of a Gaussian random displacement channel (with noise standard deviation $\sigma$) is given by \cite{holevo2001evaluating}
\begin{align}
    \max\Big{(}\log_{2}\Big{(} \frac{1}{e\sigma^{2}} \Big{)}, 0\Big{)},   
\end{align} 
and vanishes when $\sigma \ge 1/\sqrt{e} = 0.6065\cdots$. Hence, showing that a code has a threshold higher than $1/\sqrt{e} = 0.6065\cdots$ has significant implications on the Gaussian quantum capacity as it means that the code would then establish a better lower bound to the quantum capacity of a Gaussian random displacement channel than what has been known in the past two decades. 
{Although this is not the case with the surface GKP code decoded by the MWPM closest point decoder, since its threshold $\sigma^*$ is only $0.602$, we have made important progress towards this goal. As shown in Fig.~\ref{fig:surface-GKP code}, the gap between the threshold of the surface GKP code and $1/\sqrt{e}$ (the red vertical line) has been decreased by almost a half with the closest point decoder.
}

In Fig.~\ref{fig:surface-GKP code}(c),
we show the logarithm of the infidelity as a function of distance $d_0$ for $\sigma=0.5959$, which is slightly below the thresholds for both decoders. The red squares and blue circles correspond to the MWPM log-likelihood and closest point decoders respectively. We notice that not only is the infidelity for the closest point decoder smaller than that of the log-likelihood decoder, the infidelity is also decreasing much faster, as we increase the code distance. We fit the data points with linear functions, as shown by the solid lines, and find that the slope for the closest point decoder is almost twice as large as that for the log-likelihood decoder. This suggests that, for $\sigma=0.5959$, as the code distance of the surface-GKP code is increased, closest point decoder suppresses the logical error rate twice as faster as the log-likelihood decoder does. In the inset, we show the slopes as a function of $\sigma$ up to $\sigma=0.60$, which shows that their relative difference is getting smaller for smaller $\sigma$, another evidence that the two decoders perform similarly for small noise regime.

In Fig.~\ref{fig:surface-GKP code}(d), 
we further compare the runtime of the exponential-time closest point decoder (red squares) with the MWPM closest point decoder (blue circles) { in a log-log plot}. For the former, it is clear that the required runtime increases significantly from $d_0=3$ to $d_0=5$. Hence we only show two data points for the exponential-time closest point decoder. {For the MWPM closest point decoder, the run time is significantly reduced by three orders of magnitude for $d_0=5$, and scales like $d_0^{3.02}$ for large $d_0$.}

\begin{algorithm}
\caption{DecodeSurfaceGKP($\boldsymbol{t}$)}
\label{alg: ClosestPointSurf}
{\bf Input: } The error syndrome $\boldsymbol{t}\in\mathbb{R}^{2N}$ and $\left\{\boldsymbol{g}_i\right\}$ \\ 
{\bf Output: } The closest point $\boldsymbol{\chi}\in\mathbb{R}^{2N}$; \\ 
$\boldsymbol{\chi}''^{(q)} \leftarrow(\lfloor t_1\rceil, \lfloor t_3\rceil, ..., \lfloor t_{2N-1}\rceil)$\\
$\boldsymbol{\chi}''^{(p)} \leftarrow(\lfloor t_2\rceil, \lfloor t_4\rceil, ..., \lfloor t_{2N}\rceil)$\\
$G^{(q)} \leftarrow (V^{(q)}, E^{(q)}, W^{(q)})$ \quad  *Defined similarly as in Eq.~\ref{eq:def_G}-\ref{eq:def_W}*\\
$G^{(p)} \leftarrow (V^{(p)}, E^{(p)}, W^{(p)})$ \quad  *Defined similarly as in Eq.~\ref{eq:def_G}-\ref{eq:def_W}*\\
$H^{(q)} \leftarrow \left\{v_i ~|~ \text{mod}((\boldsymbol{g}^{(q)}_i)^T\boldsymbol{\chi}''^{(p)}, 2)\neq0\right\}$ ~ *Defined in Eq.~\ref{eq:def_H_pq}*\\
$H^{(p)} \leftarrow \left\{v_i ~|~ \text{mod}((\boldsymbol{g}^{(p)}_i)^T\boldsymbol{\chi}''^{(q)}, 2)\neq0\right\}$ ~ *Defined in Eq.~\ref{eq:def_H_pq}*\\
$\boldsymbol{e}^{(q)} \leftarrow \text{MWPM}(G^{(q)}, H^{(q)})$\\
$\boldsymbol{e}^{(p)} \leftarrow \text{MWPM}(G^{(p)}, H^{(p)})$\\
$\boldsymbol{e}[1:2:end]\leftarrow\boldsymbol{e}^{(q)}$;\\
 $\boldsymbol{e}[2:2:end]\leftarrow\boldsymbol{e}^{(p)}$;\\
\For{$1\leq i\leq 2N$}{
\If{e[i]==1}{
   $\chi''[i] = w(t_i)$
   }
}
$\boldsymbol{\chi}\leftarrow\sqrt{\pi} \boldsymbol{\chi}''$
\end{algorithm}

\section{Discussion and conclusion}
\label{sec:Discussion and conclusion}

In this work, we have investigated the quantum error correction with GKP codes from a lattice perspective, and there are three main results. 
We first reviewed that a general $N$-mode GKP code can be viewed as a $2N$-dimensional symplectic integral lattice, and showed that decoding the GKP code is equivalent to finding the closest point in the lattice for the given error syndrome.
Because the closest point search problem has been studied extensively in the classical error correction literature, we formulated a closest point decoder for general GKP codes.
Second, we provide a proof-of-concept demonstration that it is possible to numerically search optimized GKP code from a lattice perspective. The numerically found codes, despite not optimal, exhibit better error correction properties at low error rate, compared to the known GKP codes, such as the [[7,1,3]]-hexagonal codes, or the $d_0=3$ surface-hexagonal GKP code. 
Third, we show that despite the fact that closest point decoder incurs exponential time cost in the number of modes for general GKP codes, it is possible to devise efficient closest point decoders for structured GKP codes.
In particular, we proposed two generalizations of the tesseract codes, namely the $\text{rep-rec}_N$ and $\text{YY-rep-rec}_N$ codes, which exhibits good quantum error correction properties, and show that they can be decoded in runtime that is linear to the number of modes.
For the surface-GKP code, with the help of a MWPM algorithm, a polynomial-time closest point decoder is introduced which outperforms the previous MWPM log-likelihood decoder and yields a noise threshold of $\sigma^*=0.602$.

A few remarks are in order. 
Recall that in our numerical search for optimized GKP codes, we started with {$10^4$} initial points and perform the optimization with respect to the distance.
Because of the relatively small size of the trial ansatz, the optimized codes do not necessarily have the optimal distances. In fact, for certain number of modes, we have examples {of analytically constructed GKP codes} that outperform the {numerically} optimized code either in terms of distance or fidelity.
It is possible to find better GKP codes by optimizing on top of these known GKP codes, but for certain dimensions, we do not have known GKP codes with good error correction capabilities.
Generally, one would need to scale to much larger set of initial points for finding the optimal GKP code with large number of modes, which will incur significant time overhead.
There would be similar overhead costs if one chose to optimize with respect to fidelity, instead of distance, because Monte-Carlo sampling is required at each iteration step of the optimization.
The bottleneck can be partially mitigated if a more efficient algorithm is used to find the closest point.
In this work, we choose to adopt the algorithm in Ref.~\cite{agrell2002closest} as our closest point decoder because of its simplicity, but it is certainly not the most efficient algorithm.
To the best of our knowledge, the best deterministic closest point search algorithm was proposed by D. Micciancio and P. Voulgaris (MV) in Ref.~\cite{micciancio2010deterministic}. The core of the MV algorithm is a more efficient method to determine the Voronoi cell of the lattice, such that the complexity of the algorithm is $2^{O(n)}$. Despite the fact that it is still exponential in the dimension of the lattice (because closest point search problem is NP-hard), it improves the $n^{O(n)}$ runtime of the previously known algorithms \cite{kannan1983improved}.
Subsequently, Daniel Dadush and Nicolas Bonifas gave a randomized algorithm that provides quadratic speed up compared to the MV algorithm\cite{dadush2014short}. The algorithm is Las Vegas in the sense that it always gives the correct result but the runtime differs depending on the inputs. 
It would be {interesting} to implement these algorithms and use them to search for optimized GKP codes.

Note also that the efficiency of decoding a GKP code relies heavily on the basis chosen. In the main text and App.~\ref{sec:More details on the closest point decoder}, we discussed the KZ and LLL algorithms for finding a good basis for different applications. 
Further, for a given GKP code, a set of good basis vectors will help its experimental realizations. {In various proposals for implementing GKP codes \cite{royer2020stabilization, campagne2020quantum, sivak2022real}, the time it takes to stabilize a GKP code is proportional to the Euclidean length of the GKP stabilizer generators that are being measured. Thus especially for numerical optimized codes, it is important to look for good lattice generators that can speed up the closest point search algorithms and are practical for experimental implementations. As a related note, it could also be interesting to numerically optimize GKP codes using a well structured ansatz (e.g., geometric locality or bounded length of all stabilizer generators) such that it is guaranteed that the numerically found GKP code can be readily implemented experimentally.  }

Another interesting future direction would be benchmarking closest point decoders for other families of {concatenated GKP} codes.
For instance, recently a MWPM decoder is proposed for decoding the color code with a Möbius geometry, which  demonstrates a logical failure rate that is competitive with the optimal performance of the surface code \cite{sahay2022decoder}.
Given that we have known the closest point decoder can help to increase the fidelity and noise threshold for the surface code, it would be interesting to see if it can help in the similar manner for the color code.
However, the MWPM closest point decoder cannot be directly applied to the color code because the decoder assumes a given shift error can induce at most two syndrome errors. 
Hence finding efficient closest point decoder for the color code and other families of {concatenated GKP} codes is a challenging but urgent topic in its own right. 

Lastly we remark that we have only focused on the minimum energy decoding via the closest point problem, which is optimal only in the $\sigma\rightarrow 0$ limit \cite{conrad2022gottesman}. A truly optimal decoding strategy is the maximum likelihood decoding which is more involved than the closest point decoding. An interesting future work would be to investigate the maximum-likelihood decoders and see if the error correction performance of multimode GKP codes can be significantly improved in the large $\sigma$ regime (e.g., close to where the quantum capacity nearly vanishes).

\section{Acknowledgements}
It is a pleasure to thank Arne Grimsmo, John Preskill and Mackenzie Shaw for useful discussions. 
ML would like to thank Péter Kómár and Eric Kessler for their supports of the project.
{We also would like to thank Francesco Arzani and Timo Hillmann for the very insightful discussions on constructing lattices for concatenated GKP codes.}
We would like to acknowledge the AWS EC2 resources
which were used for part of the simulations performed in
this work.

\appendix

\section{Details of constructing lattices for concatenated GKP codes}
\label{app:Details of constructing lattices for concatenated GKP codes}

In this appendix, we provide more details for constructing lattices from the concatenated GKP codes. Specifically, we will focus on concatenating a $[[N, k]]$ stabilizer code {with $N$ single mode square GKP codes to encode $k$ qubits}.

As explained in Sec.~\ref{sec:The concatenated GKP code}, we start by constructing a separable lattice generated by $N$ copies of the square code
\begin{align}
    M^\text{(sq)} = \sqrt{2}I_{2N}
\end{align}
We will replace $N-k$ rows in $M^\text{(sq)}$ by the set of vectors $\left\{\frac{1}{\sqrt{2}}\boldsymbol{g}^T_j, j=1,...,N-k\right\}$, where $\boldsymbol{g}_j$ are the binary vectors for the generators of the stabilizer group. 
To make sure the resultant matrix, denoted as $M_\text{conc}^{\text{(sq)}}$ is full rank, for each $\boldsymbol{g}_j$, let the $l$-th element be the first nonzero element in $\boldsymbol{g}_j$, we replace the $l$-th row of $M^\text{(sq)}$ by $\boldsymbol{g}_j^T/\sqrt{2}$ if it has not been replaced before, otherwise we look for the next nonzero element in $\boldsymbol{g}_j$ until an appropriate replacement is done. 
We repeat the process for all the stabilizer generators $\boldsymbol{g}_j$ and the desired $M_\text{conc}^{\text{(sq)}}$ is arrived.
This algorithm is shown in Alg.~\ref{alg:concatenate_stabilizer_GKP}.

{
We used Alg.~\ref{alg:concatenate_stabilizer_GKP} for the multimode GKP codes discussed in the main text, but it turns out for certain GKP codes, the algorithm will not give a full rank matrix. A more general approach works with the standard form of the stabilizer code. For that, we can view the set of binary vectors $\left\{\boldsymbol{g}_j^T\right\}$ as a $(N-k)\times(2N)$ matrix $G$ with components $g_{jl}$. As shown in Section 10.5.7 of Ref.~\cite{nielsen2002quantum}, via the Gaussian elimination and relabelling the qubits if needed, one could bring the matrix $G$ into the standard form 
\begin{equation*}
% g_{jl}^\text{standard} = 
\begin{array}{@{} c @{}}
    \begin{array}{@{} r @{}}
      r~\{\hspace{\nulldelimiterspace} \\
      N-k-r~\{\hspace{\nulldelimiterspace} \\
    \end{array}
    \left [
      \begin{array}{ *{6}{c} }
        % \hspace{0.15cm}I\hspace{0.15cm} & \hspace{1.15cm}A_1\hspace{0.15cm} & \hspace{0.15cm}A_2\hspace{0.15cm} & B & 0 & C\\
        % \undermat{r}{\hspace{0.15cm}0\hspace{0.15cm}} & \undermat{N-k-r}{0} & \undermat{\hspace{-0.1cm}k}{0} & \undermat{\hspace{-0.1cm}r}{D} & \undermat{\hspace{-0.1cm}N-k-r}{I} & \undermat{\hspace{-0.1cm}k}{E}
        {\hspace{0.23cm}I} \hspace{0.cm}& {\hspace{0.5cm}A_1} \hspace{0.3cm}& {\hspace{0.23cm}A_2} \hspace{0.3cm}& {\hspace{0.13cm}B} \hspace{0.1cm}& {\hspace{0.53cm}0} \hspace{0.4cm}& {\hspace{0.2cm}C}\\
        \undermat{r}{\hspace{0.23cm}0} \hspace{0.cm}& \undermat{N-k-r}{\hspace{0.5cm}0} \hspace{0.3cm}& \undermat{k}{\hspace{0.23cm}0} \hspace{0.3cm}& \undermat{r}{\hspace{0.13cm}D} \hspace{0.1cm}& \undermat{N-k-r}{\hspace{0.53cm}I} \hspace{0.4cm}& \undermat{k}{\hspace{0.2cm}E}
      \end{array}
    \right ]  
    \vspace{0.5cm}
    \mathstrut
  \end{array}.
\end{equation*}
Here $r$ is the rank of the left $(N-k)\times N$ submatrix of $G$, $I$ is the identity matrix, and $A_1$, $A_2$, $B,C,D,E$ are all integer valued matrices. 
With that, the generator matrix for the concatenated GKP code, in the \texttt{qqpp} ordering, can be constructed as
\begin{align*}
&M^\text{(sq)}_\text{conc} = \\ 
&\frac{1}{\sqrt{2}}
\begin{array}{@{} c @{}}
    \left [
      \begin{array}{ *{6}{c} }
        {\hspace{0.23cm}I} \hspace{0.cm}& {\hspace{0.5cm}A_1} \hspace{0.3cm}& {\hspace{0.23cm}A_2} \hspace{0.3cm}& {\hspace{0.13cm}B} \hspace{0.1cm}& {\hspace{0.53cm}0} \hspace{0.4cm}& {\hspace{0.2cm}C}\\
        {\hspace{0.23cm}0} \hspace{0.cm}& {\hspace{0.5cm}0} \hspace{0.3cm}& {\hspace{0.23cm}0} \hspace{0.3cm}& {\hspace{0.13cm}D} \hspace{0.1cm}& {\hspace{0.53cm}I} \hspace{0.4cm}& {\hspace{0.2cm}E}\\
        {\hspace{0.23cm}0} \hspace{0.cm}& {\hspace{0.5cm}0} \hspace{0.3cm}& {\hspace{0.23cm}2I} \hspace{0.3cm}& {\hspace{0.13cm}0} \hspace{0.1cm}& {\hspace{0.53cm}0} \hspace{0.4cm}& {\hspace{0.2cm}0}\\
        {\hspace{0.23cm}0} \hspace{0.cm}& {\hspace{0.5cm}0} \hspace{0.3cm}& {\hspace{0.23cm}0} \hspace{0.3cm}& {\hspace{0.13cm}2I} \hspace{0.1cm}& {\hspace{0.53cm}0} \hspace{0.4cm}& {\hspace{0.2cm}0}\\
        {\hspace{0.23cm}0} \hspace{0.cm}& {\hspace{0.5cm}2I} \hspace{0.3cm}& {\hspace{0.23cm}0} \hspace{0.3cm}& {\hspace{0.13cm}0} \hspace{0.1cm}& {\hspace{0.53cm}0} \hspace{0.4cm}& {\hspace{0.2cm}0}\\
        \undermat{r}{\hspace{0.23cm}0} \hspace{0.cm}& \undermat{N-k-r}{\hspace{0.5cm}0} \hspace{0.3cm}& \undermat{k}{\hspace{0.23cm}0} \hspace{0.3cm}& \undermat{r}{\hspace{0.13cm}0} \hspace{0.1cm}& \undermat{N-k-r}{\hspace{0.53cm}0} \hspace{0.4cm}& \undermat{k}{\hspace{0.2cm}2I}
      \end{array}
    \right ]
    \begin{array}{@{} r @{}}
      \}r\hspace{1.36cm}\\
      \}N-k-r \\
      \}k\hspace{1.36cm}\\
      \}r\hspace{1.36cm} \\
      \}N-k-r\\
      \}k\hspace{1.36cm}
    \end{array}    
    \vspace{0.5cm}
    \mathstrut
  \end{array}.
\end{align*}
To see that $M^\text{(sq)}_\text{conc}$ is a valid GKP lattice, we first note that since $\left\{\boldsymbol{g}_j\right\}$ corresponds to stabilizers that commute with each other, we have $\text{mod}(\boldsymbol{g}_j^T\Omega^\texttt{qqpp}\boldsymbol{g}_l,2)=0$ for all $j, l=1,..,N-k$. Hence the Gram matrix $M^\text{(sq)}_\text{conc}\Omega^\texttt{qqpp}(M^\text{(sq)}_\text{conc})^T$ is indeed integer valued. 
Further, the matrix has determinant $2^k$, which can be seen by swapping the two columns labeled by $N-k-r$. 
Thus, we conclude that $M^\text{(sq)}_\text{conc}$ is a GKP code that encodes $k$ qubits. 

It is important to note that, swapping the columns of $M^\text{(sq)}_\text{conc}$ is equivalent to multiplying a non-sympletic orthogonal matrix from the right of $M^\text{(sq)}_\text{conc}$, which in general leads to a non-symplectic integral matrix $M'$. Despite $M'$ has the same determinant as $M^\text{(sq)}_\text{conc}$, it cannot be regarded as a GKP code. This can also be seen from the fact that swapping the columns of $G$ generally spoils the commutation relations of the stabilizers. 
}

\begin{algorithm}
\caption{ConcatenatedGKP}\label{alg:concatenate_stabilizer_GKP}
\KwData{The set of binary vectors $\left\{\boldsymbol{g}_j, j=1,...,N-k\right\}$}
\KwResult{The matrix $M_\text{conc}^{\text{(sq)}}$}
 $M_\text{conc}^{\text{(sq)}}\leftarrow \sqrt{2}I_{2N}$;\\
 \For{$1\leq j \leq N-k$}{
 \For{$1\leq l \leq 2N$}{
    \If{$g_{jl}\neq 0$}{
   \If{the $l$-th row of $M_\text{conc}^{\text{(sq)}}$ has not been replaced}{
   $M_\text{conc}^{\text{(sq)}}[l,:]\leftarrow\frac{1}{\sqrt{2}}\boldsymbol{g}_j$;\\
   {\bf break};
   }
   }
 }
 }
\end{algorithm}

\section{Algorithm for canonizing a GKP lattice}
\label{app:canonize_A}

In this section, we give an algorithm to construct a unimodular matrix $R$, which is integer valued with $|\det(R)|=1$, such that, for the given anti-symmetric matrix A, 
\begin{align}
\label{eq:canonical_form_2}
    RAR^T = \text{diag}(\boldsymbol{d})\otimes\omega = \text{diag}(\boldsymbol{d})\otimes
    \begin{bmatrix}
    0 & 1\\
    -1 & 0
    \end{bmatrix}
\end{align}
where $\boldsymbol{d}=(d_1,...,d_n)$ are non-negative integers, and we will assume $n$ is an even integer in this section. 
% The algorithms consists of the following steps.
The algorithms consists of the following subroutines.

The first subroutine, which we call PutFirstRowToZero, find a unimodular matrix $R_1$ for a given $n\times n$ anti-symmetric matrix $A$, such that the first row and column of $A^{(1)}\equiv R_1AR_1^T$ each has only one nonzero entry, say $A^{(1)}_{12}$ and $A^{(1)}_{21}=-A^{(1)}_{12}$. They are the greatest common divisor (GCD) of the original row, i.e. $A^{(1)}_{12}=\text{GCD}(A_{11},A_{12},...,A_{1n})$. We shall illustrate the details of this subroutine below. 

The second subroutine, which we call Tridiagonalize, recursively applies the subroutine PutFirstRowToZero for the $(n-1)\times(n-1)$ dimensional submatrix of $A$, such that the resulting matrix is tridiagonal. The output of Tridiagonalize is a matrix $R_2$, which is the product of outputs from PutFirstRowToZero. 
Since the product of unimodular matrices is also unimodular, we arrive at a unimodular matrix $R_2$ such that $A^{(2)}\equiv R_2AR_2^T$ is tridiagonal 
\begin{align}
\label{eq:A'}
    A^{(2)} = \begin{bmatrix}0 & A^{(2)}_{12} & 0 & 0 & ...\\ -A^{(2)}_{12} & 0 & A^{(2)}_{23} & 0 & ...\\ 0 & -A^{(2)}_{23} & 0 & A^{(2)}_{34} & ... \\ 0 & 0 & -A^{(2)}_{34} & 0 & ... \\ ... 
    \end{bmatrix}.
\end{align}

The third subroutine, which we call CanonizeTridiagonal, ensures that the element $A^{(2)}_{12}$ divides all the entries in the matrix $A^{(2)}$, i.e., $A'_{23}, A'_{34}, ...$, as shown in Eq.~\ref{eq:A'}. If it is already the case, then we can proceed to the next subroutine, otherwise we use the following unimodular matrix 
\begin{align}
\label{eq:R_3'}
    R'_3=\begin{bmatrix}1 & 1 & 1 & 1 & ...\\  0 & 1 & 0 & 0 & ... \\ 0& 0 & 1 & 0 & ... \\ 0 & 0 & 0 & 1 & ...\\ ... \\ \end{bmatrix},
\end{align}
to arrive at
\begin{align*}
    A^{(3)}&=R'_3A^{(2)}R_3^{'T} \\
    &= \begin{bmatrix}0 & A^{(2)}_{12}-A^{(2)}_{23} & A^{(2)}_{23}-A^{(2)}_{34} & ...\\ -(A^{(2)}_{12}-A^{(2)}_{23}) & 0 & A^{(2)}_{23} & ...\\ -(A^{(2)}_{23}-A^{(2)}_{34}) & -A^{(2)}_{23} & 0 & ... \\ ... \end{bmatrix}.
\end{align*}
We can again apply the subroutine PutFirstRowToZero such that the first row of $A^{(3)}$ has only nonzero entry $A^{(3)}_{12}$. After that, by construction, $A^{(3)}_{12}$ is the GCD of the original row, we have
\begin{align}
\label{eq:A''_12_as_GCD}
A^{(3)}_{12}&=\text{GCD}(A^{(2)}_{12}-A^{(2)}_{23},A^{(2)}_{23}-A^{(2)}_{34},A^{(2)}_{34}-A^{(2)}_{45},...)\nonumber\\
&=\text{GCD}(A^{(2)}_{12},A^{(2)}_{23},A^{(2)}_{34},...)
\end{align}
Note that $A^{(3)}$ needs not retain the tridiagonal form after the application of PutFirstRowToZero, but we can always apply the subroutine Tridiagonalize again to restore it. Since all the entries in $A^{(3)}$ are integer-valued linear combinations of $A^{(2)}_{12}, A^{(2)}_{23}, ...$, we have that $A^{(3)}_{12}$ divides all the elements in the resultant matrix, by Eq.~\ref{eq:A''_12_as_GCD}. We shall collect all the unimodular matrices involved in this process as $R_3$ such that
\begin{align*}
    A^{(3)}\equiv R_3A^{(2)}R_3^T = \begin{bmatrix}0 & A^{(3)}_{12} & 0 & 0 & ...\\ -A^{(3)}_{12} & 0 & A^{(3)}_{23} & 0 & ...\\ 0 & -A^{(3)}_{23} & 0 & A^{(3)}_{34} & ... \\ 0 & 0 & -A^{(3)}_{34} & 0 & ... \\ ... 
    \end{bmatrix},
\end{align*}
which shares the  same form as $A^{(2)}$ with $A^{(3)}_{12}$ divides all the entries in the matrix.

The fourth subroutine, which we call BlockTridiagonalize, put the tridiagonal matrix $A^{(3)}$ into a block diagonal form, and hence arrive at the desired form as shown in Eq.~\ref{eq:canonical_form_2}. For that, we can construct the following unimodular matrix
\begin{align}
\label{eq:R_4'}
    R_4'=\begin{bmatrix}1 & 0 & 0 & 0 & ...\\  0 & 1 & 0 & 0 & ... \\ A^{(3)}_{23}/A^{(3)}_{12} & 0 & 1 & 0 & ... \\ 0 & 0 & 0 & 1 & ...\\ ... \\ \end{bmatrix},
\end{align}
such that 
\begin{align*}
    RA^{(3)}R^T &= \begin{bmatrix}0 & A^{(3)}_{12} & 0 & 0 & ...\\ -A^{(3)}_{12} & 0 & 0 & 0 & ...\\ 0 & 0 & 0 & A^{(3)}_{34} & ... \\ 0 & 0 & -A^{(3)}_{34} & 0 & ... \\ ... \end{bmatrix} \\
    &= A^{(3)}_{12}\omega \oplus A^{(4)'}.
\end{align*}
We note that since $A^{(3)}_{12}$ divides all the matrix elements in $A^{(3)}$, the matrix $R_4'$ is unimodular. 
Since $A^{(4)'}$ is an $(n-2)\times(n-2)$ anti-symmetric tridiagonal matrix, the recursive applications of the subroutines CanonizeTridiagonal and BlockTridiagonalize will arrive at the canonical form of $A$, our initial given matrix. 
We shall collect all the unimodular matrices involved in this process as $R_4$.
An optional subroutine could be devised to perform additional row and column swapping such that $|d_1|\geq|d_2|\geq...\geq|d_n|$ for the canonical form shown in Eq.~\ref{eq:canonical_form_2}.

Here we provide more details for the subroutine PutFirstRowToZero. Consider the following antisymmetric matrix $A$, 
\begin{align}
    A = \begin{bmatrix}0 & A_{12} & A_{13} & A_{14} & ...\\ -A_{12} & 0 & A_{23} & A_{24} & ...\\ -A_{13} & -A_{23} & 0 & A_{34} & ... \\ -A_{14} & -A_{24} & -A_{34} & 0 & ... \\ ... \end{bmatrix}.
\end{align}
For any integer pairs $A_{12}, A_{13}$, by the Bezout's identity, we can use the extended Euclidean algorithm to find another integer pairs $(x_3, x_{13})$ such that
\begin{align}
    A_{12}x_3+A_{13}x_{13}=\text{GCD}(A_{12},A_{13})\equiv g_3
\end{align}
Then we can construct an unimodular matrix 
\begin{align}
    R=\begin{bmatrix}1 & 0 & 0 & 0 & ...\\  0 & x_3 & x_{13} & 0 & ... \\ 0 & -A_{13}/g_3 & A_{12}/g_3 & 0 & ... \\ 0 & 0 & 0 & 1 & ...\\ ... \\ \end{bmatrix},
\end{align}
such that
\begin{align*}
    &RAR^T = \\
    &\begin{bmatrix}0 & g_3 & 0 & ...\\ -g_3 & 0 & A_{23} & ...\\ 0 & -A_{23} & 0 & ... \\ -A_{14} & -(A_{24}x_3+A_{34}x_{13}) & \frac{A_{13}A_{24}-A_{12}A_{34}}{g_3} & ... \\ ... \end{bmatrix}.
\end{align*}
The procedure can be proceed for the pair $(g_3, A_{14})$, and we denote the corresponding Bezout coefficients as $(x_4, x_{14})$ such that $g_3x_4+A_{14}x_{14}=\text{GCD}(g_3,A_{14})\equiv g_4$. 
One can confirm that with the following product of unimodular matrices
\begin{align}
\label{eq:R_PutFirstRowToZero}
    R=...\times\begin{bmatrix}1 & 0 & 0 & 0 & ...\\  0 & x_4 & 0 & x_{14} & ... \\ 0 & 0 & 1 & 0 & ... \\ 0 & -\frac{A_{14}}{g_4} & 0 & \frac{g_3}{g_4} & ...\\ ... \\ \end{bmatrix}
    &\times\begin{bmatrix}1 & 0 & 0 & 0 & ...\\  0 & x_3 & x_{13} & 0 & ... \\ 0 & -\frac{A_{13}}{g_3} & \frac{A_{12}}{g_3} & 0 & ... \\ 0 & 0 & 0 & 1 & ...\\ ... \\ \end{bmatrix},
\end{align}
we have
\begin{align}
    RAR^T = \begin{bmatrix}0 & g & 0 & 0 & ...\\ -g & 0 & A'_{23} & A'_{24} & ...\\ 0 & -A'_{23} & 0 & A'_{34} & ... \\ 0 & -A'_{24} & -A'_{34} & 0 & ... \\ ... \end{bmatrix}.
\end{align}
The resulting matrix has the desired property that the first row and column has only one nonzero entry $A'_{12}$ and $A'_{21}=-A'_{12}$, which is the greatest common divisor of the original row.

Below we present the algorithm CanonizeGKPLattice in Alg.~\ref{alg:CanonizeGKPLattice}, which canonize a given GKP code, with the help of the above subroutines.

\begin{algorithm}[h!]
\caption{GCD($a,b$)}\label{alg:GCD}
{\bf Input: } Two integers $a,b$; \\ 
{\bf Output: } The greatest common divisor $g$, and two integers $x,y$ satisfying $ax+by=g$ \\ 
\eIf{a=0}{
   $g,x,y\leftarrow b, 0, 1$\\
   }{
   $g,x_1,y_1\leftarrow \text{GCD}(|b|\%|a|, |a|)$ \\
    $x \leftarrow y_1-\lfloor\frac{|b|}{|a|}\rfloor x_1$\\
    $y\leftarrow x_1$\\
    \If{$g<0$}{
    $x,y\leftarrow -x, -y$\\
    }
    $x,y\leftarrow \text{sign}(a)x, \text{sign}(b)y$\\
  }
\end{algorithm}
\begin{algorithm}[h!]
\caption{PutFirstRowToZero(A)}\label{alg:PutFirstRowToZero}
{\bf Input: } An tridiagonal antisymmetric matrix $A$ \\ 
{\bf Output: } A unimodular matrix $R_1$ \\ 
$g_3, x_3, x_{13}\leftarrow \text{GCD}(A_{12}, A_{13})$\\
% $g_4, x_4, x_{14}\leftarrow \text{GCD}(A_{14}, g_3)$\\
\For{$3\leq i\leq n-1$}{
$g_{i+1}, x_{i+1}, x_{1, i+1}\leftarrow \text{GCD}(A_{1, i+1}, g_i)$\\
}
$R_1\leftarrow \text{Matrix defined in Eq.~\ref{eq:R_PutFirstRowToZero}}$
\end{algorithm}
\begin{algorithm}[h!]
\caption{Tridiagonalize(A)}\label{alg:Tridiagonalize}
{\bf Input: } An tridiagonal antisymmetric matrix $A$ \\ 
{\bf Output: } A unimodular matrix $R_2$ \\ 
$R_2\leftarrow I_{n}$\\
\For{$1\leq i\leq n-1$}{
$R_1\leftarrow \text{PutFirstRowToZero}(A[i:end, i:end])$\\
$R_2\leftarrow (I_{i-1}\oplus R_1)R_2$
}
\end{algorithm}
\begin{algorithm}[h!]
\caption{CanonizeTridiagonal(A)}\label{alg:CanonizeTridiagonal}
{\bf Input: } An tridiagonal antisymmetric matrix $A$ \\ 
{\bf Output: } A unimodular matrix $R_3$ \\ 
$R_3'\leftarrow \text{Matrix defined in Eq.~\ref{eq:R_3'}}$\\
$A^{(3)}\leftarrow R'_3AR_3^{'T}$\\
$R_2\leftarrow \text{Tridiagonalize}(A^{(3)})$\\
$R_3\leftarrow R_2R_3'$
\end{algorithm}
\begin{algorithm}[h!]
\caption{BlockTridiagonalize(A)}\label{alg:BlockTridiagonalize}
{\bf Input: } An tridiagonal antisymmetric matrix $A$ \\ 
{\bf Output: } A unimodular matrix $R_4$ \\ 
$R_4'\leftarrow \text{Matrix defined in Eq.~\ref{eq:R_4'}}$\\
$A^{(4)'}\leftarrow R'_4AR_4^{'T}$\\
$R_4''\leftarrow \text{Tridiagonalize}(A^{(4)'}[3:end, 3:end])$\\
$R_4\leftarrow (I_2\oplus R_4'')R_4'$
\end{algorithm}

\begin{algorithm}
\caption{CanonizeGKPLattice(M)}\label{alg:CanonizeGKPLattice}
{\bf Input: } A GKP lattice generator $M$ \\ 
{\bf Output: } The canonical basis $M'$ for the GKP code \\ 
$A \leftarrow M\Omega M^T$ \\
$R_2\leftarrow $ Tridiagonalize($A$) \\
$R_4\leftarrow $ BlockTridiagonalize($R_2AR_2^T$) \\
$R \leftarrow R_4R_2$\\
$M' \leftarrow R  M$
\end{algorithm}

\section{More details on the closest point decoder}
\label{sec:More details on the closest point decoder}

In this section, we provide more details for the closest point decoder. Given an arbitrary point $\boldsymbol{t}\in\mathbb{R}^n$, and the generator matrix $M$ for a $n$-dimensional lattice $\Lambda$, we describe an algorithm to compute the point $\boldsymbol{\chi}_{\boldsymbol{t}}(\Lambda(M))\in\Lambda$ that is closest to $\boldsymbol{t}$ \cite{agrell2002closest}.
In Sec.~\ref{sec:Closest point search problem}, the algorithm is described in two parts, the preprocessing part and the decoding part. We first describe the LLL and KZ reductions, for preprocessing the generator matrix. 

For a matrix $M$, both LLL and KZ reduction produce a lower triangular matrix $L$, as given in Eq.~\ref{eq:RLQ},
\begin{align}
\label{eq:def_L}
    L = \begin{bmatrix}
    \boldsymbol{v}_1^T\\
    \boldsymbol{v}_2^T\\
    \vdots\\
    \boldsymbol{v}_n^T
    \end{bmatrix} = 
    \begin{bmatrix}
    v_{11} & 0 & ... & 0\\
    v_{21} & v_{22} & ... & 0\\
    \vdots & \vdots & \ddots & \vdots\\
    v_{n1} & v_{n2} & ... & v_{nn}
    \end{bmatrix},
\end{align}
where $\boldsymbol{v}_k^T$ is the $k$-th row of $M$, and $v_{kj}$ denotes its $j$-th entry.
For later purpose, we will set the diagonal components $v_{kk}$ to be all positive by multiplying $-1$ to the $k$-th row if needed.
The matrix $L$ is defined recursively to be LLL-reduced if $n=1$ or if the following conditions hold
\begin{eqnarray}
\label{eq:def_LLL_basis}
    \begin{aligned}
    ||\boldsymbol{v}_1||&\leq\frac{2}{\sqrt{3}}||\boldsymbol{v}_2||,\\
    |v_{k1}|&\leq\frac{|v_{11}|}{2} \text{ for } k=2,...,n,
    \end{aligned}
\end{eqnarray}
and the submatrix
\begin{align}
\label{eq:def_submatrix}
    \begin{bmatrix}
    v_{21}& ... & 0\\
    \vdots & \ddots & \vdots\\
    v_{n2} & ... & v_{nn}
    \end{bmatrix}
\end{align}
is also LLL-reduced. Similarly, the KZ-reduced basis is also defined recursively if $n=1$ or if the following conditions hold
\begin{eqnarray}
\label{eq:def_KZ_basis}
    \begin{aligned}
    \boldsymbol{v}_1 &\text{ is the shortest nonzero vector in $\Lambda$},\\
    |v_{k1}|&\leq\frac{|v_{11}|}{2} \text{ for } k=2,...,n,
    \end{aligned}
\end{eqnarray}
and the submatrix in Eq.~\ref{eq:def_submatrix} is KZ-reduced. 
Clearly the two base only differ by the first conditions in Eq.~\ref{eq:def_LLL_basis} and \ref{eq:def_KZ_basis}. If $L$ is KZ-reduced, then it is also LLL-reduced but the reverse is not necessarily true. %
However KZ reduction typically requires runtime that is exponential to the matrix size whereas LLL reduction operates in polynomial time.
Depending on the problem at hand, sometimes it is advantageous to use one reduction over the other, which will be discussed at the end of this section.

As we describe in Sec.~\ref{sec:Closest point search problem}, an $n$-dimensional lattice can always be decomposed into layers of $n-1$ dimensional sublattices. 
Mathematically this corresponds to decompose the generator matrix as
\begin{align}
    L = \begin{bmatrix}
    L'\\
    \boldsymbol{v}_n^T
    \end{bmatrix}, 
\end{align}
where $L'$ is an $(n-1)\times n$ matrix which is the generator matrix for the sublattice. The $n\times1$ vector can be further decomposed into $\boldsymbol{v}_n=\boldsymbol{v}_\parallel+\boldsymbol{v}_\perp$ where 
\begin{align*}
    \boldsymbol{v}_\parallel =(v_{n1}, ..., v_{n,n-1}, 0)^T, \quad 
    \boldsymbol{v}_\perp =(0, ..., 0, v_{nn})^T,
\end{align*}
are parallel and perpendicular to the sublattices respectively. 
In this setup, the sublattices can be labeled by $u_n\in\mathbb{Z}$, and the distance between two adjacent layers is simply $v_{nn}$. Since the decoding algorithm will be described as a recursive procedure, the subscript in $u_n$ help keep tracking the dimension of the lattice.

For a given $\boldsymbol{t}\in\mathbb{R}^n$, we can similarly decompose it as  $\boldsymbol{t}=\boldsymbol{t}_\parallel+\boldsymbol{t}_\perp$, and from which we can identify the index
\begin{align}
    u_n^*\equiv \left\lfloor\frac{\boldsymbol{t}^T\boldsymbol{v}_\perp}{||\boldsymbol{v}_\perp||^2}\right\rceil = \left\lfloor\frac{t_n}{v_{nn}}\right\rceil ,
\end{align}
for the sublattice that is nearest to $\boldsymbol{t}$. Here $\lfloor\cdot\rceil$ denotes the nearest integer. 
For a sublattice labeled by $u_n$, its vertical distance with $\boldsymbol{t}$ is given by
\begin{align}
    y_n = \left|u_n-\frac{t_n}{v_{nn}}\right|||\boldsymbol{t}_\perp||.
\end{align}
The nearest lattice point $\boldsymbol{\chi}_{\boldsymbol{t}}(\Lambda)$ needs not lie in the nearest sublattice labeled by $u^*_n$, but it has to lie within a set of nearest sublattices labeled by
\begin{align}
\label{eq:nearest_sublattices}
    \left\{u_n^*,u_n^*-1,u_n^*+1,u_n^*-2,u_n^*+2,...\right\},
\end{align}
which include the nearest sublattice, the next nearest sublattice, and further. 
In Eq.~\ref{eq:nearest_sublattices}, the nearest sublattices are ordered according to their vertical distances to $\boldsymbol{t}$, according to the Schnorr-Euchner strategy \cite{schnorr1994lattice}.
We note that the number of sublattices in Eq.~\ref{eq:nearest_sublattices} can be bounded if an upper bound for $\rho$ is known, because $\boldsymbol{\chi}_{\boldsymbol{t}}(\Lambda)$ cannot lie in sublattices with distance $y_n$ that is larger than the bound. 
Hence we could start the search of nearest point from the nearest sublattice, and once a candidate lattice point for $\boldsymbol{\chi}_{\boldsymbol{t}}(\Lambda)$ is identified, its distance to $\boldsymbol{x}$ will serve as the bound $\rho$ until the next candidate point with shorter distance is identified. 

We have reduced the problem of finding the closest lattice point in an $n$-dimensional lattice to finding it in a set of $(n-1)$-dimenisonal lattices.
This dimensional reduction can proceed further, and since the generator matrix is lower triangular, what we have described above also apply to all $k$-dimensional lattices with $1\leq k\leq n-1$.
Suppose we are searching a $k$-dimensional sublattice with the set of nearest sublattices labeled by $\left\{u_k^*,u_k^*-1,u_k^*+1,...\right\}$. There are three possibilities.

1. $k=1$. Since the sublattice of a 1D lattice is a point, we have arrive at a candidate closest point namely $u_1^*$. If the distance between the lattice point is smaller than the bound $\rho$, then we update the bound and the candidate closest point; otherwise we discard the point founded. After that, we set $k=2$.

2. $n-1\geq k>1$. We search the closest point in each subspace via dimension reduction, and keep updating the best candidate closest point and the upper bound $\rho$. This is done until none of the subspace in the set $\left\{u_k^*,u_k^*-1,u_k^*+1,...\right\}$ has vertical distance to $\boldsymbol{t}$ less than $\rho$. Then we set $k$ to $k+1$.

3. $k=n$. 
This suggests that we have searched all the subspaces in Eq.~\ref{eq:nearest_sublattices} that could possibly contain the closest point. Hence we output the best candidate lattice point found.

The above closest point algorithm can be significantly sped up if the $(n-1)$-dimensional subspaces in Eq.~\ref{eq:nearest_sublattices} are as separate as possible, which minimizes the number of subspaces to be searched within the bound $\rho$. 
In the extreme case, if the spacing between the $(n-1)$-dimensional subspaces is much larger than that of all the lower dimensional lattices, then the closest point will be very likely contained in the nearest plane. In this case, the dimensionality of the problem is effectively reduced by one. 
Simiarly, the spacing between the points in the 1D sublattice should be as small as possible. If the 1D sublattice is so dense that all the higher dimensional lattices have much larger spacings, then we only need to search the closest sublattices, which again reduce the dimensionality of the problem by one. 
The KZ reduction yields an optimal basis that complies to the above two observations \cite{agrell2002closest}. 
From Eq.~\ref{eq:def_KZ_basis}, we see that KZ reduction produces the smallest possible value for $v_{11}$ in Eq.~\ref{eq:def_L}, and hence the 1D sublattice is densely packed.
Since the reduction is applied recursively, $v_{22}$ and other diagonal elements are minimized subsequently. Because changing the basis does not change the volume of the Voronoi cell $\det(L)\equiv\prod_{i=1}^nv_{ii}$, the order of minimization naturally produces $v_{nn}$ that is maximized, hence we conclude that the spacing between the $(n-1)$-dimensional subspaces are maximized \cite{agrell2002closest}.
Unfortunately, the runtime of KZ reduction scales exponentially with the dimensionality of the lattice. On the other hand, LLL reduction, which operates in  polynomiall time in $n$, only produces an approximately optimal basis because the first condition in Eq.~\ref{eq:def_LLL_basis} is not optimal. 
Because of the trade-offs between the runtime and basis quality, one should choose different reduction methods for different problems. 
For the purpose of characterizing a GKP code, if we would like to compute its distance, we will use the LLL algorithm because the decoding algorithm will only be ran a handful of times; if the fidelity of the GKP code is the desired quantity, because it typically involves decoding a few million error syndromes or more, we will use the KZ reduction to preprocess the lattice once for subsequent decodings. For further comparison of the two reductions, readers are referred to the detailed review in \cite{wubben2011lattice}, and the benchmarking results in  \cite{agrell2002closest}. A detailed implementation of the algorithm presented here can also be found in \cite{agrell2002closest}.

\section{Linear time decoder for scaled $Z_n$ and $D_n$ lattices}
\label{app:decoder_scaled_Z^N_D^N}

In Sec.~\ref{sec:decoder_Z^N}-\ref{sec:decoderDN}, we reviewed that it is possible to construct decoders with runtime proportional to the dimensionalities of the $Z_n$ and $D_n$ lattices. Here we generalize the algorithms for these two type of lattices with differen lattice constants. We start with the $Z_n$ lattice with lattice constants given by $\lambda_i$ ($1\leq i\leq n$) for different axes. In other word, the generator matrix of the lattice is a diagonal matrix $\text{diag}(\lambda_1, ..., \lambda_n)$. 
Let $y_k\equiv \lfloor x_1/\lambda_1\rceil$, then the closest point can be found as follows
\begin{align}
    f(\boldsymbol{x}, {\boldsymbol \lambda}) = (\lfloor y_1\rceil\lambda_1, ..., \lfloor y_n\rceil\lambda_n).
\end{align}
In order to find the second closest point in the scaled $Z_n$ lattice, from the discussion in Sec.~\ref{sec:decoder_Z^N}, we would like to find the $k$-th component, and substitute $\lfloor x_k/\lambda_k\rceil\lambda_k$ with $w(x_k/\lambda_k)\lambda_k$. The resulting vector is denoted as
\begin{eqnarray}
    \begin{aligned}
    g(\boldsymbol{x}, {\boldsymbol\lambda}) = (&\lfloor y_1\rceil\lambda_1, ..., \lfloor y_{k-1}\rceil\lambda_{k-1},  \\
    &w(y_k)\lambda_k, \lfloor y_{k+1}\rceil\lambda_{k+1}, ..., \lfloor y_n\rceil\lambda_n), 
    \end{aligned}
\end{eqnarray}
which clearly has larger norm compared to $f(\boldsymbol{x}, {\boldsymbol \lambda})$ by the definition of $w(x)$ in Eq.~\ref{def:w(t)}. To make sure $g(\boldsymbol{x}, {\boldsymbol\lambda})$ has the second smallest distance from $\boldsymbol{x}$, the change in the distance (squared) has to be as small as possible, hence we have
\begin{align}
    k = \argmin_{1\leq k\leq n} |(x_k - \lfloor y_k\rceil\lambda_k)^2 - (x_k - w(y_k)\lambda_k)^2|.
\end{align}
Since all of the above operations can be done in time proportional to the dimension of the lattice, $f(\boldsymbol{x}, {\boldsymbol \lambda})$ and $g(\boldsymbol{x}, {\boldsymbol \lambda})$ can be found efficiently for the scaled $Z_n$ lattice.

If we color the scaled $Z_n$ lattice in a checkerboard fashion, we have the scaled $D_n$ lattice. For example, the scaled $D_4$ lattice has the following generator matrix\cite{royer2022encoding}
\begin{align}
\begin{bmatrix}
1 & 1 & 0 & 0\\
0 & 1 & 1 & 0\\
0 & 0 & 1 & 1\\
0 & 0 & 0 &  2
\end{bmatrix}
\begin{bmatrix}
\lambda_1 & 0 & 0 & 0\\
0 & \lambda_2 & 0 & 0\\
0 & 0 & \lambda_3 & 0\\
0 & 0 & 0 &  \lambda_4
\end{bmatrix}
= 
\begin{bmatrix}
\lambda_1 & \lambda_2 & 0 & 0\\
0 & \lambda_2 & \lambda_3 & 0\\
0 & 0 & \lambda_3 & \lambda_4\\
0 & 0 & 0 &  2\lambda_4
\end{bmatrix}.
\end{align}
Following the same idea in Sec.~\ref{sec:decoderDN}, in order to find the closest point in the scaled $D_n$ lattice, we first find the closest and the second closest points in the $Z_n$ lattice. In the units of the lattice constants $\boldsymbol\lambda$, they are given by the following integer-valued vectors
\begin{align}
    (\lfloor y_1\rceil, ..., \lfloor y_n\rceil)
\end{align}
and
\begin{align}
    (\lfloor y_1\rceil, ..., \lfloor y_{k-1}\rceil, w(y_k), \lfloor y_{k+1}\rceil, ..., \lfloor y_n\rceil)
\end{align}
respectively. Again, since the norm of their difference is 1, we simply need to determine which one of the above vectors has an even sum of components, then the corresponding lattice point is the closest point in the scaled $D_n$ lattice.

\section{The {concatenated GKP} code as a glue lattice}
\label{sec:The concatenated GKP code as a glue lattice}

In this section, we show that the {concatenated GKP} code introduced in Sec.~\ref{sec:The concatenated GKP code} can be viewed as a glue lattice. The argument generalizes the one presented in Sec.~\ref{sec:YY-rep-rec_N} in the context of the $\text{YY-rep-rec}_N$ code.

We start with concatenating a $[[N, k]]$ stabilizer code with a one-mode square GKP code that encode a single qubit. The resultant code encodes $k$ logical qubit in $N$ mode, and we will show that the code can be viewed as a union of cosets.  

As explained in Sec.~\ref{sec:The concatenated GKP code}, the $2N\times2N$ dimensional matrix $M_\text{conc}^\text{(sq)}$ is obtained by replacing $N-k$ rows in $\sqrt{2}I_{2N}$ by the vectors in Eq.~\ref{eq:basis_construction_A_lattice}. 
The lattice generated by $M_\text{conc}^\text{(sq)}$ is a $2N$ dimensional lattice
\begin{align}
\label{eq:conc_sq_as_union_cosets}
    \Lambda(M_\text{conc}^\text{(sq)}) = \bigcup_{j=0}^{2^{N-k}-1} \big[\frac{1}{\sqrt{2}}{\boldsymbol{g}}_j + \Lambda(\sqrt{2}I_{2N})\big].
\end{align}
where the vectors ${\boldsymbol{g}}_j$ correspond to the elements in the stabilizer group. For notation simplicity, we assume $\boldsymbol{g}_0=\boldsymbol{0}$ and $\boldsymbol{g}_{1,...,N-k}$ generate the full stabilizer group.
To see that Eq.~\ref{eq:conc_sq_as_union_cosets} holds, we will generalize the argument presented in Sec.~\ref{sec:YY-rep-rec_N}.
Suppose $\Lambda(\sqrt{2}I_{2N})$ is generated by a set of basis vectors 
\begin{align*}
    \left\{{\boldsymbol{r}}_j, ~ j=1,...,2N\right\},
\end{align*}
then $\Lambda(M_\text{conc}^\text{(sq)})$ is span by the same set of basis vectors except $N-k$ of them
\begin{align*}
    \left\{{\boldsymbol{r}}_1, ..., {\boldsymbol{r}}_{N+k}, \frac{1}{\sqrt{2}}\boldsymbol{g}_1, ..., \frac{1}{\sqrt{2}}\boldsymbol{g}_{N-k}\right\},
\end{align*}
which is evident from the construction of $M_\text{conc}^\text{(sq)}$. 
For a given $\boldsymbol{x}\in\Lambda(M_\text{conc}^\text{(sq)})$, let 
\begin{align*}
    \boldsymbol{x}=\sum_{i=1}^{N+k}a_i{\boldsymbol{r}}_i+\frac{1}{\sqrt{2}}\sum_{j=1}^{N-k}b_{j}\boldsymbol{g}_j
\end{align*}
for some integers $a_i$ and $b_j$. 
As one can show, $\sqrt{2}\boldsymbol{g}_j\in\Lambda(\sqrt{2}I_{2N})$ for all $j=1,...,N-k$ because $\boldsymbol{g}_j$ is a binary vector and $\Lambda(\sqrt{2}I_{2N})$ is a $2N$ dimensional square lattice with lattice spacing $\sqrt{2}$. 
We can define $b_j=b_j'+b_j''$ where $b_j'$ is the nearest even integer to $b_j$ such that the difference $b_j''$ is non-negative. Since $b_j'$ is even, $b_j'\boldsymbol{g}_j/\sqrt{2}$ is in the lattice $\Lambda(\sqrt{2}I_{2N})$ such that we have
\begin{align}
    \boldsymbol{x} -\tilde{\boldsymbol{g}} \in \Lambda(\sqrt{2}I_{2N}),
\end{align}
where $\tilde{\boldsymbol{g}} \equiv \frac{1}{\sqrt{2}}\sum_{j=1}^{N-k}b''_{j}\boldsymbol{g}_j$ is the binary vector of an element in the stabilizer group. We conclude that $\boldsymbol{x}$ is an element of the union of cosets shown in Eq.~\ref{eq:conc_sq_as_union_cosets}, and hence $\Lambda(M_\text{conc}^\text{(sq)})$ is a sublattice of the latter
\begin{align*}
    \Lambda(M_\text{conc}^\text{(sq)}) \subset \bigcup_{j=0}^{2^{N-k}-1} \big[\frac{1}{\sqrt{2}}\tilde{\boldsymbol{g}}_j + \Lambda(\sqrt{2}I_{2N})\big].
\end{align*}
Similarly, we can show that the union of cosets is a sublattice of $\Lambda(M_\text{conc}^\text{(sq)})$ and hence the two represent the identical lattice.

In order to decode the {concatenated GKP} code, we consider its symplectic dual lattice $\Lambda((M_\text{conc}^\text{(sq)})^\perp)$.
Since the {concatenated GKP} code encodes $k$ logical qubits, $\Lambda((M_\text{conc}^\text{(sq)})^\perp)$ can be viewed as a union of $2^{2k}$ cosets, each of which corresponds to a logical operator. Combining with Eq.~\ref{eq:conc_sq_as_union_cosets}, we have 
\begin{align}
\label{eq:conc_sq_dual_as_union_cosets}
    \Lambda((M_\text{conc}^\text{(sq)})^\perp) = \bigcup_{j=0}^{2^{N+k}-1} \big[\frac{1}{\sqrt{2}}\tilde{\boldsymbol{g}}_j' + \Lambda(\sqrt{2}I_{2N})\big].
\end{align}
Here $\tilde{\boldsymbol{g}}_j'$ correspond to the elements in the normalizer group, that contains both the stabilizers and the logical operators, and we assume $\boldsymbol{g}'_0=\boldsymbol{0}$ and $\boldsymbol{g}'_{1,...,N+k}$ generate the full normalizer group.
In the main text, we focus on decoding the surface-GKP code, and Eq.~\ref{eq:conc_sq_dual_as_union_cosets_main} is a special case of Eq.~\ref{eq:conc_sq_dual_as_union_cosets} with $k=1$.

Further, we can consider concatenating the $[[N, k]]$ stabilizer code with a general, non-square one-mode GKP code. Per Eq.~\ref{eq:concatenated_GKP}, the generator for the {concatenated GKP} code can be written as
\begin{align}
    M = M^{\text{(sq)}}_\text{conc}S^T = M^{\text{(sq)}}_\text{conc}(S_\text{base}^T)^{\oplus N},
\end{align}
where the symplectic matrix $S$ takes a block diagonal form. Hence, for each $\boldsymbol{u}\in\Lambda(M)$, there is a $\boldsymbol{v}\in\Lambda((M_\text{conc}^\text{(sq)}))$ such that $\boldsymbol{u}=S\boldsymbol{v}$. Thus, from Eq.~\ref{eq:conc_sq_as_union_cosets}, we have
\begin{align*}
    \Lambda(M) &= \bigcup_{j=0}^{2^{N-k}-1} \big[\frac{1}{\sqrt{2}}{S\boldsymbol{g}}_j + \Lambda(\sqrt{2}S^T)\big]\\
    &= \bigcup_{j=0}^{2^{N-k}-1} \big[\frac{1}{\sqrt{2}}{S\boldsymbol{g}}_j + \Lambda(\sqrt{2}(S_\text{base}^T)^{\oplus N})\big]\\
    &= \bigcup_{j=0}^{2^{N-k}-1} \big[\frac{1}{\sqrt{2}}{S\boldsymbol{g}}_j + \oplus_{i=1}^N\Lambda(\sqrt{2}S_\text{base}^T)\big].
\end{align*}
Hence this shows that the $[[N, k]]$ stabilizer code with a general one-mode GKP code corresponds to a glue lattice.

\section{Generator matrices for numerically optimized GKP codes with $N=3$, $N=7$ and $N=9$}
\label{app: Generators_3_9}

{
In Sec.~\ref{sec:Search for optimized GKP codes}, we exhibit three instances of numerically optimized GKP codes, with three modes, seven modes and nine modes, that have better QEC properties than the known GKP codes with the same number of modes. %
Here we provide their generator matrices for the interested readers. 

Recall from Sec.~\ref{sec:Search for optimized GKP codes} that the generator matrix of a generic GKP code can be written as
\begin{align}
\label{eq:MsqTO2TZ}
    M = M_\text{sq}O_2^TZ,
\end{align}
up to a rotation of basis vectors. Here $ M_\text{sq}=\textrm{diag}(\sqrt{2}, 1, \cdots, 1 ) \otimes I_2$ is the $N$-mode square lattice GKP that encodes a single qubit. Hence the generator matrix is fully determined by the diagonal matrix 
% $Z=\text{diag}(e^{r_1},e^{-r_1}, ..., e^{r_N},e^{-r_N})$
$Z=\text{diag}(r_1, r_1^{-1}, ..., r_N, r_N^{-1})$ 
with positive parameters 
$$
{\bf r}\equiv(r_1,...,r_N),
$$
and the sympletic orthogonal matrix $O_2$. Since $O_2$ takes a block form in the  \texttt{qqpp} ordering, as given in Eq.~\ref{eq:O_qqpp}, we can write
\begin{align}
    M = M_\text{sq}T^{-1}(O_2^{{\texttt{qqpp}}})^TTZ,
\end{align}
where 
\begin{align}
\label{eq:O_2_qqpp}
    O_2^{\texttt{qqpp}} = \exp \begin{bmatrix}
    X & Y \\
    -Y & X
    \end{bmatrix},
\end{align}
and $T$ is the basis transformation from the \texttt{qqpp} ordering to the \texttt{qpqp} ordering. 
Recall that $Y^T=Y$ is an $N\times N$ real symmetric matrix and $X=-X^T$ is an $N\times N$ real anti-symmetric matrix. 
For $N=3$, the basis transformation matrix is given by 
\begin{align}
\label{eq:T_3}
    T = \begin{bmatrix}
     1 & 0 & 0 & 0 & 0 & 0\\
     0 & 0 & 1 & 0 & 0 & 0\\
     0 & 0 & 0 & 0 & 1 & 0\\
     0 & 1 & 0 & 0 & 0 & 0\\
     0 & 0 & 0 & 1 & 0 & 0\\
     0 & 0 & 0 & 0 & 0 & 1
    \end{bmatrix},
\end{align}
and similar matrices can be constructed for $N=7$ and $N=9$. Below, we will report the matrices $X$, $Y$ and the vector ${\bf r}$ for the optimized codes with $N=3$, $N=7$ and $N=9$.
For other optimized modes with different $N$, similar quantities can be found in our package LatticeAlgorithms.jl.

For $N=3$, these quantities read
\begin{align}
\label{eq:XY_3}
    X &= \left[
        \begin{array}{ccc}
        0.000 & -0.129 & 0.303 \\
        0.129 & 0.000 & -0.804 \\
        -0.303 & 0.804 & 0.000 \\
        \end{array}
    \right],\nonumber\\
    Y &= \left[
        \begin{array}{ccc}
        1.126 & -0.674 & -1.101 \\
        -0.674 & 0.556 & 0.252 \\
        -1.101 & 0.252 & 0.235 \\
        \end{array}
    \right],\\
    {\bf r} &= \begin{bmatrix}
     0.329 & 3.018 & 0.326
    \end{bmatrix}.\nonumber
\end{align}
% For clarity, we have reported only the upper triangular parts of the matrices and kept only three significant digits. 
%
As one can check, upon combining Eq.~\ref{eq:MsqTO2TZ}-\ref{eq:XY_3}, we arrive at a three-mode GKP code with distance $2.670$, which is better than $3^{1/4}\sqrt{\pi}\approx2.33$, the three-mode repetition code.
}
\begin{widetext}
{
For $N=7$, the quantities read
\begin{align*}
    X &= \left[
        \begin{array}{ccccccc}
        0.000 & 0.080 & 0.237 & -0.221 & 0.599 & -0.597 & -0.220 \\
        -0.080 & 0.000 & -0.250 & 0.261 & 0.652 & -0.556 & -0.074 \\
        -0.237 & 0.250 & 0.000 & -0.884 & 0.039 & -0.121 & 0.335 \\
        0.221 & -0.261 & 0.884 & 0.000 & 0.796 & -0.431 & -0.199 \\
        -0.599 & -0.652 & -0.039 & -0.796 & 0.000 & -0.506 & -0.239 \\
        0.597 & 0.556 & 0.121 & 0.431 & 0.506 & 0.000 & -0.105 \\
        0.220 & 0.074 & -0.335 & 0.199 & 0.239 & 0.105 & 0.000 \\
        \end{array}
    \right],\\
    Y &= \left[
        \begin{array}{ccccccc}
        -0.356 & -0.519 & -0.077 & -0.616 & -0.256 & 0.374 & 0.826 \\
        -0.519 & -0.661 & -0.073 & 0.527 & 0.300 & 0.239 & 0.629 \\
        -0.077 & -0.073 & -1.209 & 0.381 & -0.201 & 0.014 & -0.223 \\
        -0.616 & 0.527 & 0.381 & -0.621 & -0.670 & 1.313 & -0.129 \\
        -0.256 & 0.300 & -0.201 & -0.670 & -0.992 & -0.253 & -0.246 \\
        0.374 & 0.239 & 0.014 & 1.313 & -0.253 & -0.114 & 0.184 \\
        0.826 & 0.629 & -0.223 & -0.129 & -0.246 & 0.184 & -0.882 \\
        \end{array}
    \right],\\ 
    {\bf r} &= \begin{bmatrix}
      0.362 & 2.650 &2.613 &2.471&0.382&0.366&0.341
    \end{bmatrix}.
\end{align*}
One can check that the corresponding GKP code has distance $3.326$, which is better than $3^{1/4}\sqrt{2\pi}\approx3.299$, the distance for the [[7,1,3]]-hexagonal GKP code.

For $N=9$, the quantities read
\begin{align*}
    X &= \left[
        \begin{array}{ccccccccc}
        0.000 & 0.334 & -0.761 & 0.958 & -0.031 & 0.347 & 0.322 & 0.098 & 0.064 \\
        -0.334 & 0.000 & -0.072 & 0.755 & 0.162 & 0.830 & 0.534 & 0.207 & 0.557 \\
        0.761 & 0.072 & 0.000 & 0.094 & 0.810 & 0.121 & -0.415 & 0.422 & 0.155 \\
        -0.958 & -0.755 & -0.094 & 0.000 & 0.079 & -0.879 & 0.213 & -0.098 & 0.553 \\
        0.031 & -0.162 & -0.810 & -0.079 & 0.000 & 0.379 & -0.444 & -0.081 & 0.180 \\
        -0.347 & -0.830 & -0.121 & 0.879 & -0.379 & 0.000 & -0.356 & 0.022 & 0.194 \\
        -0.322 & -0.534 & 0.415 & -0.213 & 0.444 & 0.356 & 0.000 & -0.001 & -0.538 \\
        -0.098 & -0.207 & -0.422 & 0.098 & 0.081 & -0.022 & 0.001 & 0.000 & 0.167 \\
        -0.064 & -0.557 & -0.155 & -0.553 & -0.180 & -0.194 & 0.538 & -0.167 & 0.000 \\
        \end{array}
    \right],\\
    Y &= \left[
        \begin{array}{ccccccccc}
        0.377 & 0.298 & -0.200 & 0.225 & -0.411 & 0.010 & 0.072 & 0.438 & -0.261 \\
        0.298 & 0.510 & -0.112 & 0.706 & 0.324 & 0.366 & -0.349 & -0.207 & -0.948 \\
        -0.200 & -0.112 & -0.181 & 0.233 & 0.265 & -0.051 & 0.783 & -0.436 & 0.379 \\
        0.225 & 0.706 & 0.233 & -0.325 & 0.112 & -0.157 & 0.171 & 0.343 & -0.012 \\
        -0.411 & 0.324 & 0.265 & 0.112 & 0.018 & 0.898 & -0.355 & -0.138 & 0.004 \\
        0.010 & 0.366 & -0.051 & -0.157 & 0.898 & -0.456 & -0.360 & 0.916 & 0.692 \\
        0.072 & -0.349 & 0.783 & 0.171 & -0.355 & -0.360 & -0.369 & 0.182 & 0.244 \\
        0.438 & -0.207 & -0.436 & 0.343 & -0.138 & 0.916 & 0.182 & 0.616 & -0.245 \\
        -0.261 & -0.948 & 0.379 & -0.012 & 0.004 & 0.692 & 0.244 & -0.245 & -0.765 \\
        \end{array}
    \right],\\    
    {\bf r} &= \begin{bmatrix}
      0.332&
 0.351&
 0.357&
 2.728&
 2.723&
 2.775&
 0.353&
 2.980&
 3.676
    \end{bmatrix}.
\end{align*}
One can check that the corresponding GKP code has distance $3.556$, which is better than $3^{1/2}\sqrt{\pi}\approx3.070$, the distance for the nine-mode surface-GKP code. 
}
\end{widetext}

\bibliography{GKP_facts.bib}

\end{document}